\pdfoutput=1
\RequirePackage{ifpdf}
\ifpdf 
\documentclass[pdftex]{sigma}
\else
\documentclass{sigma}
\fi

\usepackage{enumerate}
\usepackage{tikz}

\newtheorem{thm}{Theorem}[section]

{\theoremstyle{definition}

\newtheorem{rem}[thm]{Remark}
}
\numberwithin{equation}{section}

\begin{document}

\newcommand{\arXivNumber}{1408.5643}

\allowdisplaybreaks

\renewcommand{\PaperNumber}{056}

\FirstPageHeading

\ShortArticleName{From Polygons to Ultradiscrete Painlev\'e Equations}

\ArticleName{From Polygons to Ultradiscrete Painlev\'e Equations}

\Author{Christopher Michael ORMEROD~$^\dag$ and Yasuhiko YAMADA~$^\ddag$}

\AuthorNameForHeading{C.M.~Ormerod and Y.~Yamada}

\Address{$^\dag$~Department of Mathematics, California Institute of Technology,\\
\hphantom{$^\dag$}~1200 E California Blvd, Pasadena, CA, 91125, USA}
\EmailD{\href{mailto:cormerod@caltech.edu}{cormerod@caltech.edu}}

\Address{$^\ddag$~Department of Mathematics, Kobe University, Rokko, 657--8501, Japan}
\EmailD{\href{mailto:yamaday@math.kobe-u.ac.jp}{yamaday@math.kobe-u.ac.jp}}

\ArticleDates{Received January 29, 2015, in f\/inal form July 10, 2015; Published online July 23, 2015}

\Abstract{The rays of tropical genus one curves are constrained in a way that def\/ines a bounded polygon. When we relax this constraint, the resulting curves do not close, giving rise to a system of spiraling polygons. The piecewise linear transformations that preserve the forms of those rays form tropical rational presentations of groups of af\/f\/ine Weyl type. We present a selection of spiraling polygons with three to eleven sides whose groups of piecewise linear transformations coincide with the B\"acklund transformations and the evolution equations for the ultradiscrete Painlev\'e equations.}


\Classification{14T05; 14H70; 39A13}

\vspace{-3mm}

\section{Introduction}

A signif\/icant contribution to our understanding of the Painlev\'e equations, both discrete and continuous, has been their characterization in terms of their rational surfaces of initial conditions \cite{Okamoto:SurfaceInitialConds, Sakai:rational}. These works related the symmetries of the Painlev\'e equations to Cremona isometries of rational surfaces \cite{Looijenga:Rational, Nagata:Rational1, Nagata:Rational2}, which are groups of af\/f\/ine Weyl type \cite{Dolgachev, duVal, KMNOY:Cremona}. This provided a geometric setting for many previous studies that were based purely on the symmetries of the Painlev\'e equations \cite{KNY:qP4, KNY:discretePAmAn, NY:AffinedPs}. In the autonomous limit, the Painlev\'e equations degenerate to elliptic equations or QRT maps \cite{QRT:QRTmap1, QRT:QRTmap2} and their associated surfaces of initial conditions are rational elliptic surfaces \cite{Duistermaat:QRT, Tsuda:QRT}.

Given a subtraction free discrete Painlev\'e equation, one may obtain an ultradiscrete Painlev\'e equation by applying the ultradiscretization procedure \cite{TTMJ:ultradiscretization}. The ultradiscretization procedure famously related integrable dif\/ference equations with integrable cellular automata \cite{TM:BBSudmKdV, TTM:BBS, TTMJ:ultradiscretization}, hence, the process is thought to preserve integrability \cite{KMT:ConservedudKdV, RTGO:ultimatediscretePs}. The ultradiscrete Painlev\'e equations are second order non-linear dif\/ference equations def\/ined over the max-plus semif\/ield that are integrable in the sense that they possess many of same properties of the continuous and discrete Painlev\'e equations that are associated with integrability, albeit, in some tropical form. These properties include tropical Lax representations \cite{JNO:uP3, Ormerod:qP6reduction} and tropical singularity conf\/inement \cite{JL:TropSC, Ormerod:TropicalSC}. They also admit symmetry groups of af\/f\/ine Weyl type \cite{KMNY:Ereps, KNY:qP4} and special solutions of rational and hypergeometric type \cite{Murata:ExactsolsuP2, Ormerod:uhypergeometric, TTGOR:udPSolutions}. The ultradiscrete QRT maps may also be obtained as autonomous limits of the ultradiscrete Painlev\'e equations \cite{Nobe:QRT, QCS:UDLaxpair}.

\begin{figure}[t]
\begin{tikzpicture}[scale=.7]
\clip (-5, -5) rectangle (5.5, 5.5);
\foreach \lam in {0,.5,...,10.5}
{
	\draw[blue!20,thick] (-1,-1-\lam) -- (-1-\lam,-1) -- (-1-\lam,1) -- (0,2+\lam) -- (2,2+\lam)-- (2+\lam,2)--(2+\lam,0) -- (1,-1-\lam) -- cycle;
}

\foreach \lam in {-1,-.5,...,0}
{
	\draw[blue!20,thick] (-1-\lam,-1-\lam) -- (-1-\lam,1) -- (0,2+\lam) -- (2+\lam,2+\lam)--(2+\lam,0) -- (1,-1-\lam) -- cycle;
}
\draw[very thick,blue] (-1,-1-1) -- (-1-1,-1) -- (-1-1,1) -- (0,2+1) -- (2,2+1)-- (2+1,2)--(2+1,0) -- (1,-1-1) -- cycle;

\end{tikzpicture}
\qquad
\begin{tikzpicture}[scale=.7]
\clip (-5, -5) rectangle (5.5, 5.5);
\foreach \lam in {0,.5,...,10.5}
{
	\draw[blue!20,thick] (-1,-1-\lam) -- (-1-\lam,-1) -- (-1-\lam,1) -- (0,2+\lam) -- (3,2+\lam)-- (3+\lam,2)--(3+\lam,0) -- (1,-2-\lam) -- (-1,-2-\lam);
}
\foreach \lam in {-1,-.5,...,0}
{
	\draw[blue!20,thick] (-1-\lam,-1-\lam) -- (-1-\lam,1) -- (0,2+\lam) -- (3+\lam,2+\lam)--(3+\lam,0) -- (1,-2-\lam) -- (-1,-2-\lam) ;
}
\draw[blue!20,thick] (0,0) -- (1,0);
\draw[blue!20,thick] (-.5,-.5) -- (1,-.5)--(1.5,0) -- (1.5,.5) -- (.5,.5);
\draw[blue,very thick] (-1,-1-1) -- (-1-1,-1) -- (-1-1,1) -- (0,2+1) -- (3,2+1)-- (3+1,2)--(3+1,0) -- (1,-2-1) -- (-1,-2-1);

\end{tikzpicture}
\caption{A f\/ibration of closed tropical curves (left) corresponds to ultradiscrete QRT maps. Breaking this closure condition results in spiraling polygons (right), which corresponds to ultradiscrete Painlev\'e equations.\label{Spirals}}
\vspace{-2mm}
\end{figure}
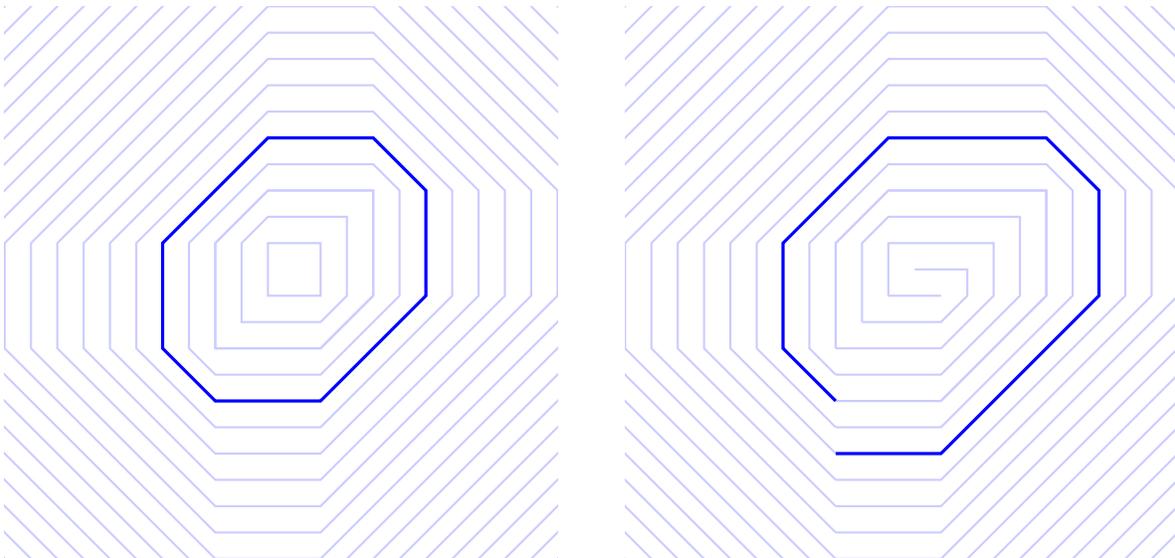

\looseness=-1
The ultradiscrete QRT maps preserve a pencil of curves arising as the level sets of tropical biquadratic functions \cite{Nobe:QRT, QCS:UDLaxpair}. Since every non-degenerate level set of a tropical biquadratic function is a tropical genus one curve, one may say that the ultradiscrete QRT maps can be lifted to automorphisms of tropical elliptic surfaces. Given the geometric interpretation of tro\-pi\-cal singularity conf\/inement \cite{Ormerod:TropicalSC}, the positions of the rays in any pencil of tropical genus one curves play the same role as the positions of the base points in a pencil of genus one curves. In this way, there is an analogous constraint on the positions of the rays of any pencil of tropical genus one curves, which when removed, results in curves that are no longer closed. We refer to the resulting set of piecewise linear curves as spiraling polygons, which are depicted in Fig.~\ref{Spirals}. This situation mimics the generalization of elliptic surfaces to surfaces of initial conditions for discrete Painlev\'e equations.

This article is concerned with groups of piecewise linear transformations of the plane which preserve the forms of the spiraling polygons. We specify a selection spiraling polygons with between three and eleven sides whose groups of transformations form representations of af\/f\/ine Weyl groups with types that coincide with those of the B\"acklund transformations for the multiplicative Painlev\'e equations~\cite{Sakai:rational}. The piecewise linear transformations corresponding to translations in the af\/f\/ine Weyl group are shown to be ultradiscrete Painlev\'e equations. A list of the correspondences between polygons, symmetry groups and ultradiscrete Painlev\'e equations, along with where these systems f\/irst appeared, is provided in Table~\ref{polgonalcorrespondence}. This work provides a geometric interpretation for the group of B\"acklund transformations of the ultradiscrete QRT maps and ultradiscrete Painlev\'e equations.

\begin{table}[t]
\centering
\caption{A labelling of the various polygons and the af\/f\/ine Weyl groups of symmetries they possess. The references refer to the f\/irst known appearence of the ultradiscrete Painlev\'e equation in the literature.}\label{polgonalcorrespondence}

\vspace{1mm}

\begin{tabular}{c c c  c } \hline\hline
Sides & Polygon & Af\/f\/ine Weyl group & Painlev\'e equation\\ \hline
3 & Triangle & $A_0^{(1)}$ &  \\
4 & Quadrilateral &$A_1^{(1)}$, $A_1^{(1)} + D_8$ & u-$\mathrm{P}_{\rm I}$, u-$\mathrm{P}_{\rm I}'$ \cite{RTGO:ultimatediscretePs}\\
5 & Pentagon &$\left(A_1+A_1\right)^{(1)}$ & u-$\mathrm{P}_{\rm II}$ \cite{RTGO:ultimatediscretePs}\\
6 & Hexagon &$\left(A_2+A_1\right)^{(1)}$ & u-$\mathrm{P}_{\rm III}$/ u-$\mathrm{P}_{\rm IV}$ \cite{KNY:qP4}\\
7 & Heptagon&$A_4^{(1)}$ & u-$\mathrm{P}_{\rm V}$ \cite{RTGO:ultimatediscretePs}\\
8 & Octagon& $D_5^{(1)}$ & u-$\mathrm{P}_{\rm VI}$ \cite{RTGO:ultimatediscretePs}\\
9 & Enneagon& $E_6^{(1)}$ & u-$\mathrm{P}\big(A_2^{(1)}\big)$  \cite{KMNY:Ereps}\\
10 & Decagon& $E_7^{(1)}$ & u-$\mathrm{P}\big(A_1^{(1)}\big)$  \cite{KMNY:Ereps}\\
11 & Undecagon & $E_8^{(1)}$ & u-$\mathrm{P}\big(A_0^{(1)*}\big)$  \cite{KMNY:Ereps} \\\hline\hline
\end{tabular}

\end{table}

Our construction replicates the ultradiscretization of known subtraction-free af\/f\/ine Weyl representations in the unpublished work of Kajiwara et al.~\cite{KMNY:Ereps}, however, our derivation does not use or require the ultradiscretization procedure. Finding generators for the representations is reduced to combinatorial properties of the underlying polygons. By considering genus one tropical plane cubic, quartic and sextic curves, we treat polygons with up to eleven sides. We mention that the case of octagons arising as level sets of tropical biquadratic functions also appeared in this context in the work of Rojas~\cite{OR2}, Nobe~\cite{Nobe:QRT} and Scully~\cite{Scully}, as do a very small collection of the symmetries we list in~\cite{OR2}.

We set out this paper as follows: we f\/irst brief\/ly review a geometric setting for QRT maps and the discrete Painlev\'e equations in Section~\ref{sec:geometry}, then we review the ultradiscretization procedure with some relevant tools from tropical geometry in Section~\ref{sec:trop}. A~description of the canonical classes of transformations that preserve given spiral structures is presented in Section~\ref{sec:classes}, which we use in Section~\ref{sec:reps} to give explicit presentations of the piecewise linear transformations that may be used to construct the ultradiscrete Painlev\'e equations. We have a brief discussion of the dif\/f\/iculties in extending this to polygons with greater than eleven sides in Section~\ref{sec:dodecagons}.

\section{The geometry of QRT maps and discrete Painlev\'e equations}\label{sec:geometry}

The QRT maps are integrable second order autonomous dif\/ference equations \cite{QRT:QRTmap1, QRT:QRTmap2}. They are Lax integrable, measure preserving and possess the singularity conf\/inement property. The QRT maps may broadly be considered discrete analogue of elliptic equations \cite{Tsuda:QRT}. To construct a~QRT map, one takes two linearly independent biquadratics, $h_0(x,y)$ and $h_1(x,y)$, and a generic point, $p = (x,y)$, to which we associate an element, $z = [z_0:z_1] \in \mathbb{P}_1$, by the relation
\begin{gather}\label{curve}
z_0 h_0(x,y) + z_1 h_1(x,y) = 0.
\end{gather}
That is to say that $h_0(x,y)$ and $h_1(x,y)$ def\/ine a pencil of biquadratic curves. If we let $h(x,y) = h_0(x,y)/h_1(x,y)$, then the QRT map, $\phi\colon (x,y) \to (\tilde{x},\tilde{y})$, is def\/ined by the condition that~$\tilde{x}$ and~$\tilde{y}$ are related to~$x$ and~$y$ by
\begin{subequations}\label{QRTdef}
\begin{gather}
h(x,y) = h(x,\tilde{y}),\\
h(x,\tilde{y}) = h(\tilde{x},\tilde{y}),
\end{gather}
\end{subequations}
where the trivial solutions, $x = \tilde{x}$ and $y = \tilde{y}$, are discarded \cite{QRT:QRTmap1, QRT:QRTmap2}. In this way, the map is an endomorphism of the curve def\/ined by \eqref{curve} for each value of~$z$.

\begin{figure}[t]
\centering
\begin{tikzpicture}[scale = 1.3]
\clip (-3,-3) rectangle (3,3);
\draw (-3,1) -- (3,1);
\draw (-3,-1) -- (3,-1);
\draw (-3,2) -- (3,2);
\draw (-3,-2) -- (3,-2);
\draw (1,-3) -- (1,3);
\draw (-1,-3) -- (-1,3);
\draw (2,-3) -- (2,3);
\draw (-2,-3) -- (-2,3);
\draw [red, thick,  domain=-3:-0.1, samples=20] plot ({\x}, {2/\x});
\draw [red, thick,  domain=-3:-0.1, samples=20] plot ({\x}, {-2/\x});
\draw [red, thick,  domain=0.1:3, samples=20] plot ({\x}, {2/\x});
\draw [red, thick,  domain=0.1:3, samples=20] plot ({\x}, {-2/\x});
\draw [blue, thick,  domain=0:360, samples=40] plot ({sqrt(5)*cos(\x)}, {sqrt(5)*sin(\x)});
\draw [purple, thick,  domain=-1.9:1.9, samples=40] plot ({\x}, {1-3/(\x*\x-4)});
\draw [purple, thick,  domain=-1.9:1.9, samples=40] plot ({\x}, {-1+3/(\x*\x-4)});
\draw [purple, thick,  domain=-1.9:1.9, samples=40] plot ({1-3/(\x*\x-4)},{\x});
\draw [purple, thick,  domain=-1.9:1.9, samples=40] plot ({-1+3/(\x*\x-4)},{\x});
\draw [purple, thick,  domain=-3:-2.1, samples=40] plot ({2+1/(\x*\x-4)},{\x});
\draw [purple, thick,  domain=2.1:3, samples=40] plot ({-2-1/(\x*\x-4)},{\x});
\draw [purple, thick,  domain=-3:-2.1, samples=40] plot ({\x},{-2-1/(\x*\x-4)});
\draw [purple, thick,  domain=2.1:3, samples=40] plot ({\x},{2+1/(\x*\x-4)});
\draw [purple, thick,  domain=-1.9:1.9, samples=40] plot ({1+9/((\x*\x-4)*(\x*\x-4))},{\x});
\draw [purple, thick,  domain=-1.9:1.9, samples=40] plot ({-1-9/((\x*\x-4)*(\x*\x-4))},{\x});
\draw [purple, thick,  domain=-1.9:1.9, samples=40] plot ({\x},{1+9/((\x*\x-4)*(\x*\x-4))});
\draw [purple, thick,  domain=-1.9:1.9, samples=40] plot ({\x},{-1-9/((\x*\x-4)*(\x*\x-4))});
\draw [blue, thick,  domain=0.1:89.9, samples=40] plot ({((17)^(.25))*cos(\x)^(.5)}, {((17)^(.25))*sin(\x)^(.5)});
\draw [blue, thick,  domain=0.1:89.9, samples=40] plot ({-((17)^(.25))*cos(\x)^(.5)}, {((17)^(.25))*sin(\x)^(.5)});
\draw [blue, thick,  domain=0.1:89.9, samples=40] plot ({-((17)^(.25))*cos(\x)^(.5)}, {-((17)^(.25))*sin(\x)^(.5)});
\draw [blue, thick,  domain=0.1:89.9, samples=40] plot ({((17)^(.25))*cos(\x)^(.5)}, {-((17)^(.25))*sin(\x)^(.5)});
\draw [blue, thick] (-.1,-2.03)--(.1,-2.03);
\draw [blue, thick] (-.1,2.03)--(.1,2.03);
\draw [blue, thick] (-2.03,-.1)--(-2.03,.1);
\draw [blue, thick] (2.03,.1)--(2.03,-.1);
\end{tikzpicture}

\caption{A collection of elements of the pencil of biquadratic curves with eight distinct base-points in~$\mathbb{R}^2$.\label{QRTfig}}
\end{figure}
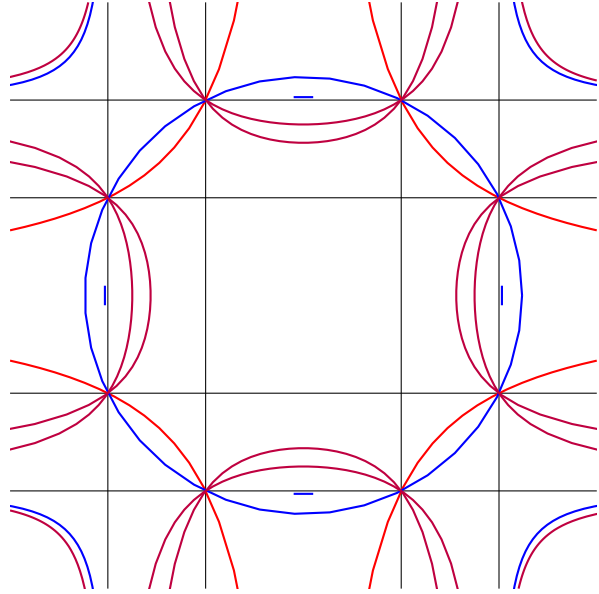

If we take a point in the intersection of the curves $h_0(x,y) = 0$ and $h_1(x,y) = 0$, then~$z_0$ and~$z_1$ may be chosen arbitrarily, hence, an entire pencil of curves intersect at these points. These points are called base-points and the number of base points for any pencil of biquadratics is 8, counting multiplicities. A case in which there are eight distinct base points in~$\mathbb{R}^2$ is depicted in Fig.~\ref{QRTfig}. By blowing up these base points, possibly multiple times in the case of higher multiplicities, we obtain a surface admitting a f\/ibration by smooth biquadratic curves (i.e., elliptic curves). Lifting the QRT map to this surface gives an automorphism of an elliptic surface~\cite{Duistermaat:QRT, Tsuda:QRT}.

A classic example is the QRT map def\/ined by the invariant
\begin{align}\label{QRTinvariant}
h(x,y) = \dfrac{y}{a_3} + \dfrac{y}{a_4} + \dfrac{(a_1+a_2)b_1b_2}{ya_1a_2} + \dfrac{(y+b_1)(y+b_2)}{x y} + \dfrac{x (y+b_3)(y+b_4)}{y a_3 a_4},
\end{align}
where we require the condition
\begin{gather}\label{QRTconst}
a_1a_2b_3b_4 = b_1b_2a_3a_4.
\end{gather}
The map, $(x,y) \to (\tilde{x}, \tilde{y})$, is specif\/ied by relations
\begin{subequations}\label{QRT}
\begin{gather}
\tilde{x} x = \dfrac{a_3 a_4(\tilde{y}+b_1 )(\tilde{y}+b_2)}{(\tilde{y}+b_3)(\tilde{y}+b_4)},\\
\tilde{y} y = \dfrac{b_3 b_4(x + a_1)(x+a_2)}{(x+a_3)(x+a_4)}.
\end{gather}
\end{subequations}
The base points of \eqref{QRT} lie on the lines $x,y = 0,\infty$ in $\mathbb{P}_1^2$. The blow-up at these points, with~\eqref{QRTconst} as a constraint, is an elliptic surface~\cite{Duistermaat:QRT}.

The discrete Painlev\'e equations are integrable second order dif\/ference equations that admit the continuous Painlev\'e equations as a continuum limit \cite{RGH:discretePs} and QRT maps in an autonomous limit. The discrete Painlev\'e equations and QRT maps are integrable by many of the same criteria; Lax integrability~\cite{JS:qP6, PNGR:Isomonodromic}, vanishing algebraic entropy~\cite{BV:Entropy} and singularity conf\/inement~\cite{RGH:discretePs}.

One way to obtain a non-autonomous second order dif\/ference equation from a QRT map is by assuming the parameters vary in a manner that preserves the singularity conf\/inement property~\cite{RGH:discretePs}. Given the autonomous system def\/ined by~\eqref{QRT}, we may deautonomize to the system to obtain the nonlinear $q$-dif\/ference equation
\begin{subequations}\label{qP6}
\begin{gather}
\tilde{y} y  = \dfrac{b_3 b_4(x + a_1t)(x+a_2t)}{(x+a_3)(x+a_4)},\\
\tilde{x} x  = \dfrac{a_3 a_4(\tilde{y}+qb_1 t)(\tilde{y}+qb_2t)}{(\tilde{y}+b_3)(\tilde{y}+b_4)},
\end{gather}
\end{subequations}
where $x = x(t)$, $y = y(t)$, $\tilde{x} = x(q t)$ and $\tilde{y} = y(qt)$. If we think of this as a dif\/ference equation for $y = y_n$ and $x= x_n$, with independent parameter, $n$, this is equivalent to $n$ appearing in an exponent as $t = t_0 q^n$. The parameter $q \in \mathbb{C} \setminus \{0\}$ is a constant def\/ined by the relation
\begin{gather}\label{qdef}
q = \dfrac{a_1a_2b_3b_4}{b_1b_2a_3a_4}.
\end{gather}
This system was f\/irst derived as a connection preserving deformation \cite{JS:qP6}. While these are often thought of as nonlinear $q$-dif\/ference equations in $t$, from the viewpoint of symmetries, it is more conducive to think of \eqref{qP6} as a~map
\begin{gather}\label{mapping}
\phi\colon \  \left(\begin{matrix} a_1,a_2, a_3, a_4\\ b_1, b_2, b_3,b_4 \end{matrix} ; x, y \right) \to \left(\begin{matrix} qa_1,qa_2, a_3, a_4\\ qb_1, qb_2, b_3,b_4 \end{matrix} ; \tilde{x}, \tilde{y} \right),
\end{gather}
where $\tilde{x}$ and $\tilde{y}$ are related by \eqref{qP6} and we absorb $t$ into the def\/initions of $a_1$, $a_2$, $b_1$ and $b_2$ (equivalent to setting $t =1$ in \eqref{qP6}). When we blow up the eight points, $P = \{p_1 ,\ldots, p_8 \} \subset \mathbb{P}_1^2$, given by
\begin{alignat*}{5}
& p_1 = (-a_1,0), \qquad &&  p_2 = (-a_2,0), \qquad &&  p_3 = (-a_3,\infty), \qquad && p_4 = (-a_4,\infty), & \\
& p_5 = (0,-b_1), \qquad && p_6 = (0,-b_2), \qquad && p_7 = (\infty, -b_3), \qquad && p_8 = (\infty,-b_4),&
\end{alignat*}
the resulting surface, $X_P$, has been called a generalized Halphen surface \cite{Sakai:rational}. Lifting the map def\/ined by~\eqref{qP6} is not an automorphism of $X_P$, but rather an isomorphism, $\varphi\colon  X_P \to X_{\tilde{P}}$, where $\tilde{P}$ is the set of points def\/ined by the image of~\eqref{mapping}. This map is bijective for the same reasons as for the QRT case. In the autonomous limit as $q = 1$, \eqref{qdef} coincides with \eqref{QRTconst}, $P = \tilde{P}$ and $\varphi$ is an automorphism of an elliptic surface that coincides with the lift of~\eqref{QRT}.

\begin{figure}[t]
\centering
\begin{tikzpicture}[scale = 3]
\draw[blue] (-.2,-.2) grid (1.2,1.2);
\filldraw[red] (.3,1) circle (.03);
\filldraw[red] (.6,1) circle (.03);
\filldraw[red] (.4,0) circle (.03);
\filldraw[red] (.7,0) circle (.03);
\filldraw[red] (1,.3) circle (.03);
\filldraw[red] (1,.6) circle (.03);
\filldraw[red] (0,.4) circle (.03);
\filldraw[red] (0,.7) circle (.03);
\node at (-.5,0) {$y = 0$};
\node at (-.5,1) {$y = \infty$};
\node at (0,-.3) {$x = 0$};
\node at (1,-.3) {$x = \infty$};
\end{tikzpicture}

\caption{The positions of the blow-up points for \eqref{QRT} and \eqref{qP6}.\label{QRTqPVIconfig}}
\end{figure}
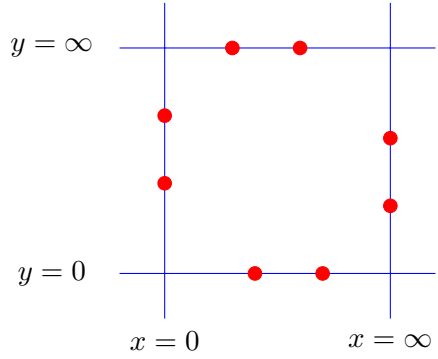

In the same way as \eqref{QRT}, the blow-up points for \eqref{qP6} lie on the lines $x,y = 0,\infty$, as shown in Fig.~\ref{QRTqPVIconfig}. We can identify the af\/f\/ine coordinates, $x$ and $y$, with projective coordinates, $[x_0:x_1]$ and $[y_0:y_1]$, via the relations $x= x_1/x_0$ and $y= y_1/y_0$ in which the points, $P$, lie on the decomposable curve def\/ined by $x_0x_1y_0y_1 = 0$.

If we were to follow up the construction of the surface, one notices that if we were to interchange the blow-up points, we obtain a surface that is isomorphic. We notice that the blow-up co-ordinates, $(z_0^1:z_1^1)$ and $(z_0^3:z_1^3)$, for the points, $p_1$ and $p_3$ respectively, satisfy the relations
\begin{gather*}
z_1^1(x+a_1)=z_0^1y, \qquad z_1^3(x + a_3) = \dfrac{z_0^3}{y},
\end{gather*}
then if we def\/ine the transformation $(x,y) \to (\hat{x},\hat{y})$, by
\begin{gather*}
\hat{x}=x, \qquad \hat{y} = y \dfrac{x+a_3}{x+a_1},
\end{gather*}
then the blow-up co-ordinates in $\hat{x}$ and $\hat{y}$ satisfy the relations
\begin{gather*}
z_1^1(\hat{x}+a_3) = z_0^1 \hat{y}, \qquad z_1^3(\hat{x} + a_1) = \dfrac{z_0^3}{\hat{y}}.
\end{gather*}
This transformation also has a scaling ef\/fect on the positions of $p_5$ and $p_6$.
\begin{gather}\label{w3qP6}
\left(\begin{matrix} a_1,a_2, a_3, a_4\\ b_1, b_2, b_3,b_4 \end{matrix} ; x, y \right) \to \left(\begin{matrix} a_3,a_2, a_1, a_4\\ b_1\frac{a_3}{a_1}, b_2\frac{a_3}{a_1}, b_3,b_4 \end{matrix} ; \hat{x}, \hat{y}\right).
\end{gather}
Both the constraint, \eqref{QRTconst}, and the variable~$q$, def\/ined by \eqref{qdef}, remain valid on the new surface, hence, the transformation $(x,y) \to (\hat{x},\hat{y})$ may be lifted to an isomorphism of surfaces.

Let $\sigma_{i,j}$ denote the isomorphism identifying the surfaces in which the blowups at points $p_i$ and $p_j$ are interchanged, then we have a natural set of elements, $w_0 = \sigma_{7,8}$, $w_1 = \sigma_{5,6}$, $w_4 = \sigma_{1,2}$ and $w_5 = \sigma_{3,4}$. We label the transformation from~\eqref{w3qP6} by~$w_3$ and the corresponding operation using points $p_5$ and $p_7$ by $w_2$. These transformations and two natural symmetries, $\rho_1$ and~$\rho_2$, form a~representation of an af\/f\/ine Weyl group of type $D_5^{(1)}$  (see \cite[Section~2]{Sakai:rational} for more details). Furthermore, as an inf\/inite order isomorphism, both~\eqref{QRT} and~\eqref{qP6} may be represented as a~product of these involutions as
\begin{gather*}
T = \rho_2 \circ w_2 \circ w_0 \circ w_1 \circ w_2 \circ \rho_1 \circ w_3 \circ w_5 \circ w_4 \circ w_3.
\end{gather*}
In many cases, such birational representations were studied independently.

While we have been considering biquadratics over $\mathbb{P}_1^2$, we may extend these arguments to plane curves in $\mathbb{P}_2$ via the birational map, $\pi\colon  \mathbb{P}_1^2 \to \mathbb{P}_2$, def\/ined by
\begin{gather*}
\pi \colon \  ([x_0:x_1],[y_0:y_1]) = [x_0y_0:x_1y_0:x_0y_1],
\end{gather*}
which is not def\/ined when $x_0 = y_0 = 0$ (corresponding to $(\infty,\infty)$). The inverse,
\begin{gather*}
\pi^{-1} ([u_0:u_1:u_2]) = ([u_0:u_1], [u_0:u_2]),
\end{gather*}
is not def\/ined at $[0:0:1]$ and $[0:1:0]$. These maps are isomorphisms when restricted to the copies of $\mathbb{C}^2$ def\/ined by $x_0 = y_0 = 1$ and $u_0 = 1$ respectively (or more precisely, $x_0$, $y_0$ and $u_0$ are not $0$). Any biquadratic curve,
\begin{gather*}
b(x,y) = \sum b_{i,j} x_0^ix_1^{2-i} y_0^jy_1^{2-j} = 0,
\end{gather*}
going through $(\infty,\infty)$ (i.e., $b_{2,2} = 0$) is mapped, via $\pi$, to a cubic plane curve
\begin{gather*}
c(u) = \sum_{0\leq i, j \leq 2, i+j > 0} c_{i,j} u_0^{i+j-1} u_1^{2-i} u_2^{2-j},
\end{gather*}
which goes through $[0:0:1]$ and $[0:1:0]$. In this way, our two generating biquadratics, $h_0$ and $h_1$ from \eqref{curve}, map to two cubic planar curves which generally intersect at $9$ points (also constrained). In this way, we can naturally pass from a pencil of biquadratics on $\mathbb{P}_1^2$, which is resolved by blowing up eight points to a pencil of cubic plane curves, and a surface obtained by blowing up $\mathbb{P}_2$ at nine points.

In passing from the QRT maps to discrete Painlev\'e equations via singularity conf\/inement, where the base points are allowed to move, the resulting systems are one of three types of nonautonomous dif\/ference equations; $h$-dif\/ference, $q$-dif\/ference or elliptic dif\/ference equations. The points can still lie in non-generic positions, but the additional constraint associated with the QRT maps is relaxed. The positions and multiplicities of these nine points determine the symmetries of the surface and of the equation. All the equations admitting ultradiscretization (or tropicalization) are special cases of $q$-dif\/ference equations, where all the parameters are assumed to be positive. The class of surfaces giving rise to $q$-dif\/ference equations was studied by Looijenga \cite{Looijenga:Rational}.

\begin{figure}[t]
\tikzstyle{block} = [font=\scriptsize,rectangle, draw, fill=blue!5, text width=1.6em, text centered, rounded corners, minimum height=2em]
\tikzstyle{line} = [draw, -latex']
\tikzstyle{eblock} = [font=\scriptsize,circle,draw, fill=blue!5, text width=1.5em, text centered, rounded corners, minimum height=2em]
\tikzstyle{beblock} = [font=\scriptsize,circle,draw, fill=blue!5, text width=2.2em, text centered, rounded corners, minimum height=2em]
\tikzstyle{qblock} = [font=\scriptsize,rectangle, draw, fill=blue!5, text width=4.6em, text centered, rounded corners, minimum height=2em]
\tikzstyle{dblock} = [font=\scriptsize,rectangle, draw, fill=blue!5, text width=2em, text centered, minimum height=2em]
\tikzstyle{bdblock} = [font=\scriptsize,rectangle, draw, fill=blue!5, text width=5em, text centered, minimum height=2em]
\centering
\begin{tikzpicture}[scale=1.7]
    \node [block] (E8) {$\dfrac{E_8^{(1)}}{A_0^{(1)}}$};
    \node [block, right of=E8, node distance=1.5cm] (E7) {$\dfrac{E_7^{(1)}}{A_1^{(1)}}$};
    \node [block, right of=E7, node distance=1.5cm] (E6) {$\dfrac{E_6^{(1)}}{A_2^{(1)}}$};
    \node [block, right of=E6, node distance=1.5cm] (D5) {$\dfrac{D_5^{(1)}}{A_3^{(1)}}$};
    \node [block, right of=D5, node distance=1.5cm] (A4) {$\dfrac{A_4^{(1)}}{A_4^{(1)}}$};
    \node [qblock, right of=A4, node distance=2.1cm] (A21) {$\dfrac{(A_2+A_1)^{(1)}}{A_5^{(1)}}$};
    \node [qblock, right of=A21, node distance=2.8cm] (A11) {$\dfrac{(A_1+A_1)^{(1)}}{A_6^{(1)}}$};
    \node [block, right of=A11, node distance=2cm] (A1) {$\dfrac{A_1^{(1)}}{A_7^{(1)}}$};
    \node [block, right of=A1, node distance=1.5cm] (A0) {$\dfrac{A_0^{(1)}}{A_8^{(1)}}$};
    \node [block, above of=A1, node distance=1.8cm] (A1s) {$\dfrac{\tilde{A}_1^{(1)}}{{A}_7^{(1)}}$};
    \draw [->] (E8) -- (E7);
    \draw [->] (E7) -- (E6);
    \draw [->] (E6) -- (D5);
    \draw [->] (D5) -- (A4);
    \draw [->] (A4) -- (A21);
    \draw [->] (A21) -- (A11);
    \draw [->] (A11) -- (A1s);
    \draw [->] (A11) -- (A1);
    \draw [->] (A1) -- (A0);
\end{tikzpicture}

\caption{The coalescence diagram for $q$/u-Painlev\'e equations. The symmetry of the equation appears on top and the surface type appears below.\label{dPs}}
\end{figure}
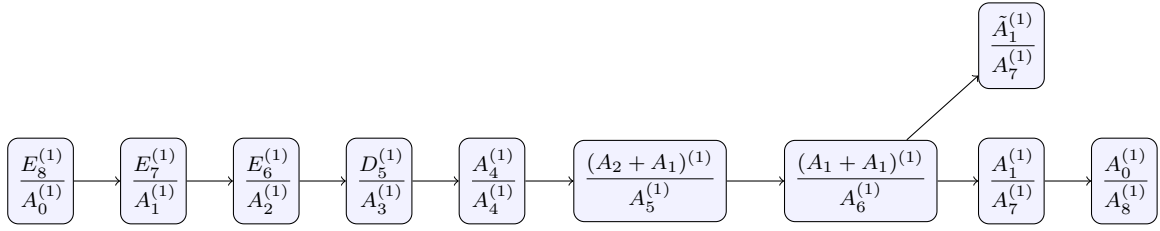

When the nine points are in any non-generic position and appear with dif\/ferent multiplicities, one can not interchange blow-up points in any ad-hoc manner. For example, in the case of \eqref{QRT}, the points lie on four distinct lines with an intersection form of type $A_3^{(1)}$, and the positions of those points are subject to the constraint \eqref{QRTconst}. The type of surface is characterized by this intersection form, and we may only interchange blow-up points in a way that preserves the intersection form. In this way we obtain two root systems, one describing the symmetry group of the equation, the other describing the surface type. A degeneration diagram which lists the surface type and the symmetries of the corresponding $q$-Painlev\'e equations is given in Fig.~\ref{dPs}.

By identifying the Picard lattices of isomorphic surfaces, we have an alternative interpretation of these maps and their symmetries \cite{Sakai:rational}. From the theory of rational surfaces (as blow-ups of the minimal surfaces $\Sigma_0 = \mathbb{P}_1^2$ or $\Sigma_1 = \mathbb{P}_2$), we have the isomorphism $\mathrm{Pic}(X) = \mathbb{H}^1(X,\mathcal{O}^*) \cong H_2(X,\mathbb{Z})$, with an endowed intersection form \cite{Nagata:Rational1, Nagata:Rational2}. The interchange of blow-up and blow-down structures \cite{WeakFactor} preserves this intersection form and leaves the canonical class f\/ixed \cite{Looijenga:Rational}, so we may interpret these as ref\/lections in $\mathrm{Pic}(X)$. This def\/ines a group of Cremona isometries, which are of af\/f\/ine Weyl type. The work of Sakai extended \cite{Looijenga:Rational} and realized the action of the translational Cremona isometries as discrete Painlev\'e equations \cite{Sakai:rational}.

\section{Tropicalization}\label{sec:trop}

Tropicalization can be thought of as the pointwise application of a nonarchimedean valuation to geometric structures. Tropicalization sends curves to lines, surfaces to polygons and more generally, smooth structures to piecewise linear ones \cite{BG:geometryvaluations, RST:TropicalGeometry}. In the integrable community a non-analytic limit known as ultradiscretization is used as a way of obtaining new and interesting piecewise linear integrable systems \cite{TTMJ:ultradiscretization}. Relating tropicalization with ultradiscretization gives us a way of understanding the geometry of ultradiscrete systems \cite{Ormerod:TropicalSC}.

Let us f\/irst consider the ultradiscretization procedure as it was originally considered in \cite{TTMJ:ultradiscretization}. Given a subtraction free rational function in a number of strictly positive variables, $f(x_1,\ldots, x_n)$, we introduce ultradiscrete variables, $X_1$, \ldots, $X_n$, related by $x_i = e^{X_i/\epsilon}$. The ultradiscretization of $f$, denoted $F$, is obtained by the limit
\begin{gather}
F(X_1, \ldots, X_n) := \lim_{\epsilon \to 0^+} \epsilon \ln f(x_1, \ldots, x_n).
\end{gather}
The subtraction free nature of the function is required so that we need not consider the logarithm of a negative number. Roughly speaking, the ultradiscretization procedure replaces variables and binary operations as follows:
\begin{gather*}
x_1 x_2   \to X_1 + X_2,\qquad
x_1 + x_2   \to \max(X_1,X_2),\qquad
x_1/x_2   \to X_1 - X_2,
\end{gather*}
where there is no (natural) replacement of subtraction.

Given a dif\/ference equation, such as \eqref{qP6}, we may apply the ultradiscretization procedure to obtain a system known as u-$\mathrm{P}_{\rm VI}$ \cite{RTGO:ultimatediscretePs}, given by
\begin{subequations}\label{uP6}
\begin{gather}
X + \tilde{X} =  A_3 +  A_4 + \max(Q + T+ B_1,\tilde{Y}) + \max(Q +T+ B_2,\tilde{Y}) \nonumber\\
\hphantom{X + \tilde{X} =}{} - \max(B_3,\tilde{Y}) - \max(B_4,\tilde{Y}),\\
Y + \tilde{Y} =  B_3 +  B_4 + \max(A_1+T,\tilde{X}) + \max(A_2+T,X)\nonumber\\
\hphantom{Y + \tilde{Y} =}{}- \max(B_3,X) - \max(B_4,X),
\end{gather}
\end{subequations}
where the variable $Q$ is specif\/ied by the relation
\begin{gather*}
Q = A_1 + A_2 -A_3 - A_4 - B_1 -B_2 + B_3 + B_4.
\end{gather*}
A special case of this system was shown to arise as an ultradiscrete connection preserving deformation \cite{Ormerod:qP6reduction}. In the same way as \eqref{qP6}, we may think of this as a map
\begin{gather*}
\Phi \colon \  \left(\begin{matrix} A_1,A_2, A_3, A_4\\ B_1, B_2, B_3,B_4 \end{matrix} ; X, Y \right) \to \left(\begin{matrix} Q+A_1,Q+ A_2, A_3, A_4\\ Q+B_1, Q+B_2, B_3,B_4 \end{matrix} ; \tilde{X}, \tilde{Y} \right).
\end{gather*}
In the autonomous limit, when we let $Q = 0$, the above ultradiscrete Painlev\'e equation becomes an ultradiscrete QRT map (i.e., the ultradiscretization of \eqref{QRT}), which was introduced in \cite{QCS:UDLaxpair} and studied from a tropical geometric viewpoint by Nobe~\cite{Nobe:QRT}. The ultradiscretization of \eqref{QRTinvariant} gives the following piecewise linear function
\begin{gather}
H(X,Y) = \max \big( Y-A_3,Y-A_4, B_1 + B_2 \max(-A_1,-A_2) - Y, \nonumber\\
\hphantom{H(X,Y) = \max \big(}{} \max(Y,B_1) + \max(Y,B_2) - X - Y,\nonumber \\
\hphantom{H(X,Y) = \max \big(}{}  X - Y + \max(Y,B_3) + \max(Y,B_4)- A_3 - A_4\big), \label{uinvariant}
\end{gather}
which is also an invariant of the ultradiscrete QRT map, i.e., $H(X,Y) = H(\tilde{X},\tilde{Y})$ \cite{Nobe:QRT}. Furthermore, the evolution of the ultradiscrete QRT map def\/ines a linear evolution on the Jacobian of the invariant, hence, the ultradiscrete QRT map may be expressed in terms of the addition law on a tropical elliptic curve~\cite{Vigeland:grouplaw, Nobe:QRT}.

While we may be able to solve \eqref{QRTdef} in a subtraction free manner, given an invariant such as~\eqref{uinvariant}, the equation $H(X,Y) = H(\tilde{X},\tilde{Y})$ involves a $\max$ on both the left and right, hence, cannot generally be solved within the limited framework of tropical arithmetic. Our approach is dif\/ferent in that we only consider transformations that preserve the structure of the tropical curves of the form $H(X,Y) = H_0$ where $H_0$ is some constant. Any automorphism of tropical curves of this form can be expressed in terms of compositions of more fundamental operations. We need to consider these curves more carefully, hence, we will brief\/ly review some tropical geometry~\cite{RST:TropicalGeometry}.

The discrete dynamical system, \eqref{uP6}, is most naturally def\/ined over a tropical semif\/ield \cite{Pin:tropicalsemirings}, more precisely, the max-plus semif\/ield, which is the set $\mathbb{T} = \mathbb{R} \cup \{-\infty\}$, equipped with the binary operations
\begin{gather*}
X_1 \oplus X_2  := \max(X_1,X_2),\qquad
X_1 \otimes X_2  := X_1+X_2,
\end{gather*}
which are known as tropical addition and tropical multiplication respectively. The element $-\infty$ plays the role of the tropical additive identity and $0$ plays the role of the tropical multiplicative identity~\cite{Pin:tropicalsemirings}.

The geometry of objects over the tropical semif\/ields is the subject of tropical geometry \cite{RST:TropicalGeometry}. A~tropical polynomial, $F \in \mathbb{T}[X_1,\ldots, X_n]$ def\/ines a piecewise linear function from $\mathbb{T}^n \to \mathbb{T}$, given by
\begin{gather}\label{tropfun}
F(X_1,\ldots, X_n) = \max_{j}  ( C_j + A_{j,1}X_1 + \cdots + A_{j,n}X_n ),
\end{gather}
where $\{A_{j,i}\}$ is a set of integers and $\{C_j\}$ is a set of elements of $\mathbb{T}$. The tropical variety associated with $F \in \mathbb{T}[X_1,\ldots, X_n]$, denoted $\mathcal{V}(F)$, is def\/ined to be
\begin{gather*}
\mathcal{V}(F) = \big\{ X = (X_1, \ldots, X_n) \in \mathbb{T}^n \textrm{ such that $F$ is not dif\/ferentiable at $X$}  \big\},
\end{gather*}
which occurs precisely when one argument of the $\max$-expression becomes dominant over another argument \cite{RST:TropicalGeometry}.

Another equivalent algebraic characterization of tropical varieties relies on nonarchimedean valuations. Every non-zero algebraic function, $f \in \mathbb{C}(t)$, admits a representation as a Puiseux series,
\begin{gather*}
f(t) = c_1 t^{q_1} + c_2 t^{q_2} + \cdots,
\end{gather*}
where $c_1 \neq 0$ and $\{q_i\}$ are rational and ordered such that $q_i < q_{i+1}$. The function, $\nu\colon \mathbb{C}(t)\to \mathbb{T}$, given by
\begin{gather*}
\nu(f) = -q_1,
\end{gather*}
is a nonarchimedean valuation. This may be extended to an algebraically and topologically closed f\/ield with a valuation ring of~$\mathbb{R}$, which we simply denote $\mathbb{K} = \overline{\mathbb{C}(t)}$ \cite{Markwig:FieldforTG}. If $I \subset \mathbb{K}[x_1^{\pm 1}, \ldots, x_n^{\pm 1}]$ is an ideal, then we def\/ine $V(I) \subset \mathbb{K}^n$ as
\begin{gather*}
V(I) = \{ (x_1, \ldots, x_n)\colon \, f(x_1, \ldots, x_n) = 0 \textrm{ for all } f \in I \}.
\end{gather*}
The tropical variety associated with $I$ is the topological closure of the point-wise application of~$\nu$ to~$V(I)$, i.e., $\mathcal{V}(I) = \overline{\nu(V(I))} \subset \mathbb{T}^n$. For every tropical variety~$\mathcal{V}(F)$, there exists a function,~$f$, such that $\mathcal{V}(F) = \mathcal{V}(\langle f \rangle)$ where $\langle f \rangle \subset \mathbb{K}[x_1^{\pm 1}, \ldots, x_n^{\pm 1}]$ denotes the ideal generated by $f$. This means that we may def\/ine a tropical variety in terms of either piecewise linear functions or ideals of $\mathbb{K}[x_1^{\pm 1}, \ldots, x_n^{\pm 1}]$. The equivalence of the set of points of non-dif\/ferentiability and the image of the valuations is outlined in \cite{RST:TropicalGeometry}. Each tropical curve is a collection vertices, f\/inite line segments, called edges, and a collection of semi-inf\/inite line segments, called rays.

In the same way as af\/f\/ine $n$-space may be considered to be embedded in projective space, we may naturally consider $\mathbb{T}^n$ as being embedded in tropical projective space. Def\/ine the equivalence relation, $\sim$, on $\mathbb{T}^{n+1}$ so that
\begin{gather*}
V \sim U \quad \textrm{if and only if} \quad V = U +  \lambda(1,1,\ldots, 1),
\end{gather*}
for some $\lambda$, then tropical projective $n$-space is the set
\begin{gather*}
\mathbb{TP}_n = \mathbb{T}^{n+1} / \sim.
\end{gather*}
A tropical function of the form \eqref{tropfun} is said to be homogeneous if there exists a~$d$ such that for every~$j$
\begin{gather*}
\sum_{i} A_{j,i} = d.
\end{gather*}
The set of non-dif\/ferentiable points of a tropically homogeneous polynomial def\/ines a tropical projective variety.

Given a rational function in a number of variables, $f(x_1, \ldots, x_n)$, we can lift the function up to the f\/ield of algebraic functions by letting $x_i = t^{X_i}$ for some $X_i$, then the ultradiscretization procedure is known to coincide with
\begin{gather}\label{udnu}
F(X_1, \ldots, X_n) = \nu (f(x_1,\ldots,x_n)  ),
\end{gather}
for \looseness=-1 all subtraction free functions \cite{Ormerod:uhypergeometric, Ormerod:TropicalSC}. The above extension, given by~\eqref{udnu}, is one of a~number of ways to incorporate a version of subtraction into the ultradiscretization procedure \cite{IGRS:s-ultradiscretization, KL:UDnegativity, Kondo:UDSGSymmMPAlg, ON:inversible}.

The most immediate consequence from the viewpoint of the geometry is that singularities of a map manifest themselves as points of non-dif\/ferentiability \cite{BG:geometryvaluations, Ormerod:TropicalSC, RST:TropicalGeometry}. This interpretation was also present in the work of Joshi and Lafortune who elucidated what the analogue of singularity conf\/inement should be for tropical integrable dif\/ference equations \cite{JL:TropSC}.

\begin{figure}[t]
\centering
\begin{tikzpicture}[scale=.7]
\clip (-4, -4) rectangle (4.5, 4.5);
\draw[very thick,blue] (-1,-1-1) -- (-1-1,-1) -- (-1-1,1) -- (0,2+1) -- (2,2+1)-- (2+1,2)--(2+1,0) -- (1,-1-1) -- cycle;
\draw[red, thick] (-1,-2) -- (-1,-6);
\draw[red, thick] (1,-2) -- (1,-6);
\draw[red, thick] (3,0) -- (7,0);
\draw[red, thick] (3,2) -- (7,2);
\draw[red, thick] (2,3) -- (2,7);
\draw[red, thick] (0,3) -- (0,7);
\draw[red, thick] (-2,1) -- (-7,1);
\draw[red, thick] (-2,-1) -- (-7,-1) ;
\end{tikzpicture}

\caption{A tropical biquadratic with the rays labeled in red.\label{tent}}
\end{figure}
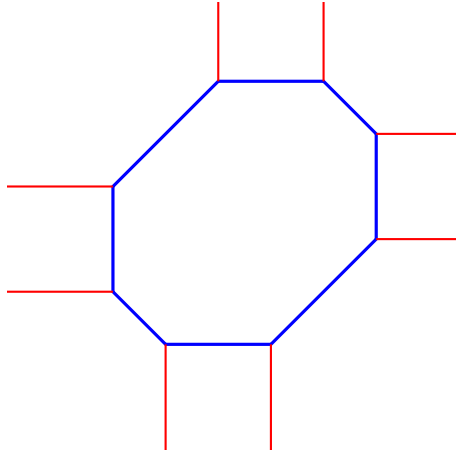

One of the characteristic features of the QRT map is that the invariant curves all intersect at the base points. From looking at the invariant curves of \eqref{QRT}, depicted in Fig.~\ref{Spirals}, this feature is not apparent in the tropical setting. When we consider the extension of the ultradiscretization via \eqref{udnu}, another way of looking at the invariant is that the level set is a subset of the tropical variety associated the ideal
\begin{gather*}
I_{H_0} =  \big\langle h(x,y)  - t^{H_0} \big\rangle,
\end{gather*}
in $\mathbb{K}[x,y]$, which is the set
\begin{align}\label{varinvar}
\mathcal{V}(I_{H_0}) := \overline{\nu (V(I_{H_0}) )}.
\end{align}
For each $x = t^X$ where $X \in \mathbb{Q}$, the equation
\begin{gather*}
h\big(t^X,y\big) - t^{H_0} = 0,
\end{gather*}
is quadratic in~$y$, and as $\mathbb{K}$ is algebraically closed, we have two algebraic solutions,~$y_1$ and~$y_2$ over~$\mathbb{K}$. That is for each $X$, we obtain values $Y_1 = \nu(y_1)$ and $Y_2 = \nu(y_2)$ in $\mathbb{T}$, which form inf\/inite rays (also called tentacles in~\cite{Vigeland:grouplaw,Nobe:QRT}). These form points of $\mathcal{V}(I_{H_0})$ that do not appear in the level set of~$H(X,Y)$. Notice that each of the rays intersect on the lines at $X = \pm \infty$ and $Y = \pm \infty$, and positions of the rays def\/ine where on that line they intersect. The inclusion of the rays to the level sets, as seen in Fig.~\ref{tent}, makes them smooth tropical curves in the sense of~\cite{RST:TropicalGeometry}.

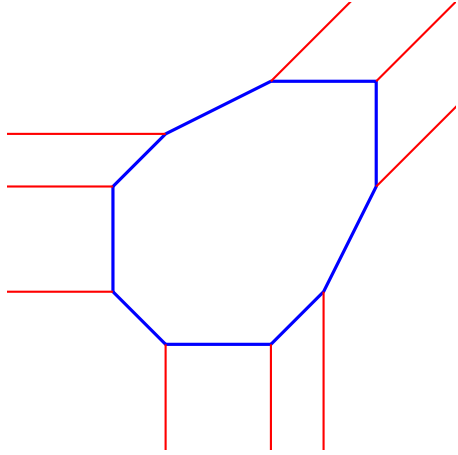
\begin{figure}[t]
\centering
\begin{tikzpicture}[scale=.7]
\clip (-4, -4) rectangle (4.5, 4.5);
\draw[very thick,blue] (-1,-2) -- (-2,-1) -- (-2,1) -- (-1,2) -- (1,3)-- (2+1,3)--(2+1,1) -- (2,-1)-- (1,-1-1) -- cycle;
\draw[red, thick] (-1,-2) -- (-1,-6);
\draw[red, thick] (1,-2) -- (1,-6);
\draw[red, thick] (2,-1) -- (2,-6);
\draw[red, thick] (3,1) -- (7,5);
\draw[red, thick] (3,3) -- (7,7);
\draw[red, thick] (1,3) -- (5,7);
\draw[red, thick] (-1,2) -- (-7,2);
\draw[red, thick] (-2,1) -- (-7,1);
\draw[red, thick] (-2,-1) -- (-7,-1) ;
\end{tikzpicture}

\caption{A tropical cubic plane curve with rays labeled in red. \label{cubicplane}}
\end{figure}

We may extend these tropical biquadratics to $\mathbb{TP}_1^2$ by using homogeneous co-ordinates $X = [X_0:X_1]$ and $Y = [Y_0:Y_1]$. The maps $\pi$ and $\pi^{-1}$ possess tropical analogues, $\Pi\colon \mathbb{TP}_1^2 \to \mathbb{TP}_2$ and $\Pi^{-1}\colon  \mathbb{TP}_2 \to \mathbb{TP}_1^2$, given by
\begin{alignat*}{3}
& \Pi\colon \ &&  ([X_0,X_1], [Y_0,Y_1]) \to [X_0 + Y_0 : X_1 + Y_0:X_0 + Y_1],& \\
& \Pi^{-1} \colon \ &&  [U_0:U_1:U_2] \to ([U_0:U_1],[U_0:U_2]).&
\end{alignat*}
These are isomorphisms between the copies of $\mathbb{T}^2$ specif\/ied by $X_0 = Y_0 = 0$ and $U_0 = 0$ respectively. The map~$\Pi$ is not def\/ined when~$X_0 = Y_0 = -\infty$ and the inverse is not def\/ined at $[-\infty:0:-\infty]$ and $[-\infty:-\infty:0]$. The level set of a tropical biquadratic function
\begin{gather*}
H(X,Y) = \max_{i,j = 0,1,2} \big( B_{i,j} + i X_0 + (2-i)X_1 + j Y_0 + (2-j)Y_1 \big),
\end{gather*}
in which $B_{2,2} = -\infty$ maps via~$\Pi$ to a tropical cubic plane curve, specif\/ied by the level set of some cubic,
\begin{gather*}
H(U) = \max_{0\leq i, j \leq 2,\, i+j > 0} \big(  C_{i,j} + (i+j-1) U_0 + (2-i)U_1 + (2-j) U_2\big).
\end{gather*}
Since $\Pi$ maps the rays and edges over $\mathbb{TP}_1^2$ to rays and edges in~$\mathbb{TP}_2$, we expect the image of the level set of a biquadratic to be at most an octagon, however, the most general cubic plane curve is an enneagon. If one considers the enneagon as the image of the variety over $\mathbb{K}[x,y,z]$, one recovers nine rays counting multiplicities. The case of nine distinct rays is depicted in Fig.~\ref{cubicplane}. In this way, the information we have on rays in~$\mathbb{P}_1^2$ applies equally well to the rays in~$\mathbb{TP}_2$.

As the rays def\/ine the positions of the vertices of each polygon, they will play an important role in the description of the symmetries. In Figs.~\ref{tent} and~\ref{cubicplane}, all the rays are asymptotic to one of three forms;
\begin{gather*}
L_i\colon \  X -A_i = 0, \qquad L_j\colon \ Y - A_j=0, \qquad L_k\colon \  Y-X -A_k = 0.
\end{gather*}
Since the rays in Figs.~\ref{tent} and~\ref{cubicplane} are part of every variety of the form \eqref{varinvar}, this is equivalent to each variety intersecting in $\mathbb{TP}_2$ at points
\begin{gather*}
[A_i:-\infty:0], \qquad [-\infty:A_j:0], \qquad [0:-\infty: A_k],
\end{gather*}
respectively. For the level set to close, there is a constraint on the positions of the rays, which when relaxed gives a spiral diagram. For smooth biquadratics, we obtain spiraling octagons (see Fig.~\ref{Spirals}). In the smooth cubic case we obtain spiraling enneagons (see Fig.~\ref{E6fig}). Given a polygon arising as a tropical curve, there are two types of degenerations:
\begin{itemize}\itemsep=0pt
\item{We may make two parallel rays coincide.}
\item{We may take two rays that are not parallel and merge them.}
\end{itemize}
The latter corresponds to setting a coef\/f\/icient of $H(X,Y)$ to $-\infty$.

This construction may be generalized to tropical genus one curves of higher degrees, which allows us to consider decagons and undecagons as level sets of tropical quartic and tropical sextic plane curves respectively. In these cases, one f\/inds twelve and thirteen rays, counting multiplicities (when rays coincide). The decagon used will be a tropical quartic with four rays of order one of the form $L_i\colon X -A_i$, four rays of order one of the form $L_j\colon Y - A_j$ and two rays of order two of the form $L_k\colon  Y-X -A_k = 0$. This would be the ultradiscretization of a curve of degree four with eight singularities of order one and two of order two, which gives a genus of one curve by the degree-genus formula,
\begin{gather}\label{degreegenus}
g = \dfrac{(d-1)(d-2)}{2} - \sum_{k} \dfrac{r_k(r_k-1)}{2},
\end{gather}
where $d$ is the degree of the curve and the $r_i$ is the order of the $k$-th singularity. In a similar way, our undecagon is a the ultradiscretization of a genus one curve of degree six curve with six rays of order one, two of order three and three of order two. This formula remains valid for tropical varieties \cite{Gathman:TropicalAlgebraicGeometry}.

\section{Piecewise linear transformations of polygons and spirals}\label{sec:classes}

Cremona transformations of the plane, and their subgroups, are a topic of classical and modern interest \cite{ Dolgachev,Hudson, KMNOY:Cremona}. The classical result of Noether \cite{Noether} (see also \cite{Hudson}) states that Cremona transformations are generated by the quadratic transformations, the simplest being the standard Cremona transformation
\begin{gather*}
\tau\colon \  [x:y:z] \to [yz:xz:xy],
\end{gather*}
which may be interpreted as the blow-up of the points $[1:0:0]$, $[0:1:0]$ and $[0:0:1]$ combined with a blow-down on the co-ordinate lines given by $xyz = 0$. In a similar vein, our aim is to specify a generating set of tropical Cremona transformations from which all the other transformations may be obtained. Our aim is to specify subgroups of these that preserve a given spiral diagram.

To specify any spiral diagram, we begin with a parameterization of the asymptotic form of the rays,
\begin{gather*}
\mathcal{X} = \{ L_i \textrm{ where } L_i\colon \, a_iX + b_i Y + c_i = 0,\textrm{ and } a_i, b_i , c_i \in \mathbb{Z} \}.
\end{gather*}
The shape of the spirals are determined by the invariants obtained in the autonomous limit. We seek a group of transformations that preserve the forms of these rays, more specif\/ically, we seek transformations, $\sigma$, such that
\begin{enumerate} \itemsep=0pt
\item[1)] $\sigma$ is a bijection of the plane;
\item[2)] for every ray, $L_j$, there is a ray, $L_i$, such that $\sigma\colon  L_i = \tilde{L}_j$, where $\tilde{L}_j$ dif\/fers only by some translation.
\end{enumerate}
These may be thought of as tropical Cremona isometries, as these conditions replicate conditions that require the canonical class and intersection form of the surface be f\/ixed.

Since the Cremona isometries are products of the interchange of blow-up and blow-down structures \cite{WeakFactor}, and the positions of these blow-up points are encoded in the positions of the rays, it is suf\/f\/icient to consider the shearing transformations that create and smooth out polygons whose vertices lie along these rays. Analagously to the results of Noether \cite{Hudson}, we propose the following two generators:
\begin{alignat}{3}
\label{generators1} & \iota_A \colon \ &&  (X,Y) \to (X,Y+\max(0,X-A)),& \\
\label{generators2} & \Xi \colon \ && (X,Y) \to (aX + cY, bX + dY), &
\end{alignat}
where $|ad-bc|= 1$ and $A \in \mathbb{T}$. The action of $\iota_A$ can be seen as an analogous to the interchange of blow-ups in the following way: if the vertices of the level sets of a polygon trace out the rays, then $\iota_A$ can smooth out all the vertices along a ray asymptotic to, $L\colon X = A$, while simultaneously creating a kink along all the level sets along a ray of the same form, but in the opposite direction. This means that if all the rays intersected at a point $P = (A,-\infty)$, the transformed polygon has rays that intersect at~$(A,\infty)$, or vise versa.

\begin{figure}[t]
\centering
\begin{tikzpicture}[scale=.9]
\clip (-2,-2) rectangle (6,6);
\draw[green,thick] (0,0) -- (-1,1) -- (-1,3)-- (1,5) -- (4,5)-- (5,4) -- (5,2) -- (3,0) -- cycle;
\draw[blue,thick] (0,0) -- (-1,1) -- (-1,3)-- (0,4) -- (4,4)-- (5,3) -- (5,1) -- (3,-1)--(1,-1) -- cycle;
\draw[red,thick] (0,-2)--(0,6);
\node[font=\scriptsize] at (1.35,-1.6) {$L_1$};
\node[font=\scriptsize] at (-0.35,4.9) {$L_2$};
\draw[red,thick] (1,-2)--(1,6);
\foreach \y in {-4,-3.5,...,7}
{
	\draw[blue,opacity=.5] (-2,\y) -- (0,\y) -- (1,\y+1) --(6,\y+1);
}
\end{tikzpicture}

\caption{A depiction of action of $\sigma$, described above on an octagon with rays~$L_1$ and~$L_2$. The blue octagon is the preimage and the green octagon is the image.\label{trans}}
\end{figure}
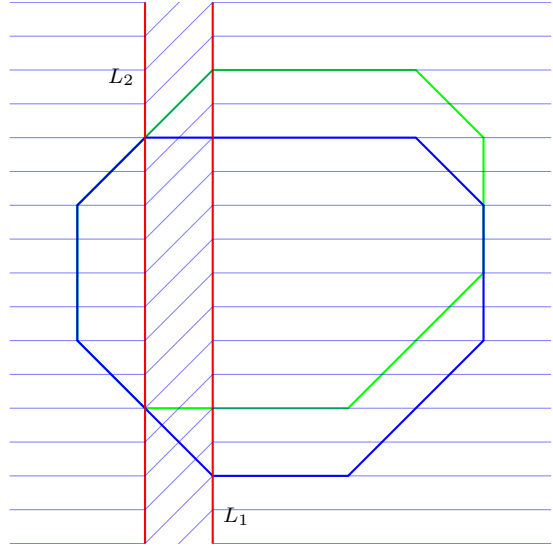

Let us use $\iota_A$ to interchange rays that are of the same form in asymptotically opposite directions. Suppose we have two rays,~$L_i$ and~$L_j$, which satisfy
\begin{gather*}
L_i\colon \  X - A = 0 \qquad \textrm{and} \qquad L_j\colon \ X-B = 0,
\end{gather*}
as $Y \to -\infty$ and $Y \to \infty$ respectively. In the simplest case, these rays are order one, in that the change in derivative is just one, in which case the transformation
\begin{gather*}
\sigma = \iota_B^{-1} \circ \iota_A\colon \  (X,Y) \to (X, Y+\max(0,X-A) - \max(0,X-B)),
\end{gather*}
has the ef\/fect of creating a ray along the line $X-A$ as $Y\to \infty$ and smooting out a set of kinks along $L_i$, and conservely doing the same for $L_j$. If we think of the surface as being parameterized by $A$ and $B$, then this action has the ef\/fect of swapping $A$ and $B$. The overall shape of the resulting polygon does not change by this transformation and the action is an isomorphism of polygons. The action of $\iota_A$ and $\sigma$ on the plane is depicted in Fig.~\ref{sABP12} and the action on the level set of the form in Fig.~\ref{tent} is depicted in Fig.~\ref{trans}.

\begin{figure}[t]
\centering
\begin{tikzpicture}[scale=.7]
\clip (-4.3,-4.3) rectangle (3.3,3.3);
\draw[very thick](-1,5) -- (-1,-4);
\foreach \x in {-5,-4,...,5}
{	
	\foreach \y in {-10,-9,..., 5}
	{
		\draw[blue,opacity=.5,->,thick] ({\x},{\y + max(0,\x+1)})--({\x+1},{\y + max(0,\x+1+1)});
	  \draw[green,opacity=.5,->,thick] ({\x},{\y + max(0,\x+1)})--({\x},{\y+1 + max(0,\x+1)});
	}
}
\end{tikzpicture}
\qquad
\begin{tikzpicture}[scale=.7]
\clip (-4.3,-4.3) rectangle (3.3,3.3);
\draw[very thick](-1,5) -- (-1,-4);
\draw[very thick](1,5) -- (1,-4);
\foreach \x in {-5,-4,...,5}
{	
	\foreach \y in {-7,-6,..., 5}
	{
		\draw[blue,opacity=.5,->,thick] ({\x},{\y + max(0,\x+1)- max(0,\x-1)})--({\x+1},{\y + max(0,\x+1+1)- max(0,\x+1-1)});
	  \draw[green,opacity=.5,->,thick] ({\x},{\y + max(0,\x+1)- max(0,\x+1)})--({\x},{\y+1 + max(0,\x+1)- max(0,\x-1)});
	}
}
\end{tikzpicture}

\caption{Assuming $B < A$, the ef\/fect of $\iota_B$ is depicted on the left, and $\sigma = \iota_A^{-1} \circ \iota_B$ on the right. \label{sABP12}}
\end{figure}
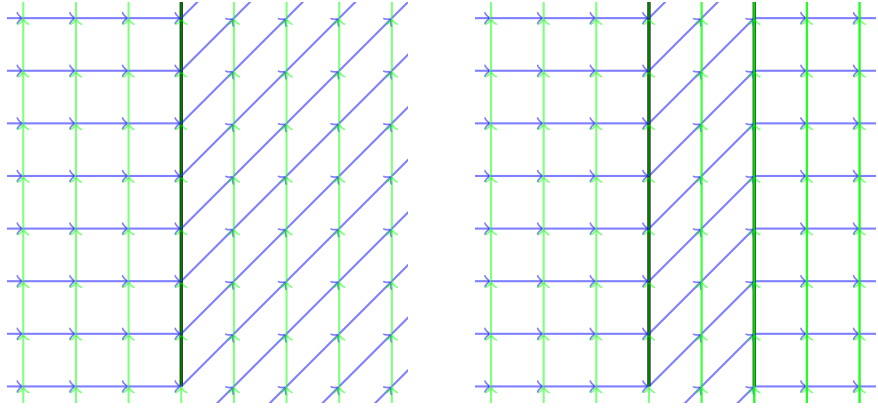

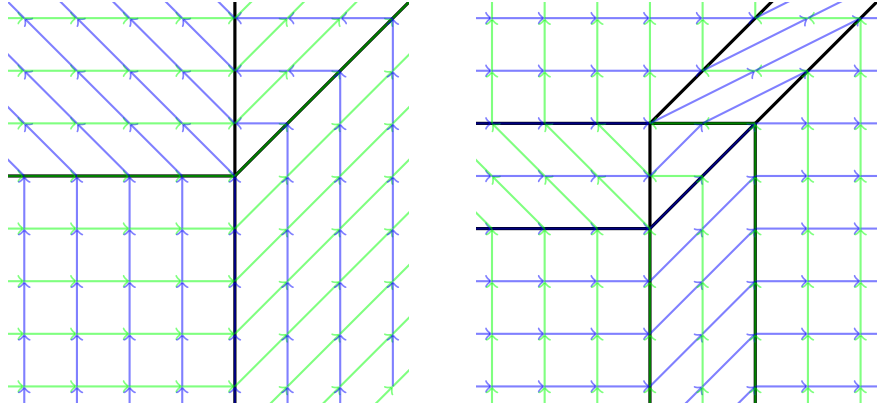
\begin{figure}[t]
\centering
\begin{tikzpicture}[scale=.7]
\clip (-4.3,-4.3) rectangle (3.3,3.3);
\draw[very thick](-5,0) -- (0,0);
\draw[very thick](0,-5) -- (0,0);
\draw[very thick](0,5) -- (0,0);
\draw[very thick](0,0) -- (4,4);
\foreach \x in {-5,-4,...,5}
{	
	\foreach \y in {-7,-6,..., 5}
	{
		\draw[blue,opacity=.5,->,thick] ({\x-max(0,\y)},{\y+max(0,\x,\y)-max(0,\y)})--({\x-max(0,\y+1)},{\y+1+max(0,\x,\y+1)-max(0,\y+1)});
	    \draw[green,opacity=.5,->,thick] ({\x-max(0,\y)},{\y+max(0,\x,\y)-max(0,\y)})--({\x+1-max(0,\y)},{\y+max(0,\x+1,\y)-max(0,\y)});
	}
}
\end{tikzpicture}
\qquad
\begin{tikzpicture}[scale=.7]
\clip (-4.3,-4.3) rectangle (3.3,3.3);
\draw[very thick](-5,-1) -- (-1,-1);
\draw[very thick](-5,1) -- (-1,1)--(4,6);
\draw[very thick] (1,1)--(1,-5);
\draw[very thick](-1,-1) -- (-1,1) -- (1,1);
\draw[very thick](-1,-5) -- (-1,-1);
\draw[very thick](-1,-1) -- (4,4);
\foreach \x in {-5,-4,...,5}
{	
	\foreach \y in {-7,-6,..., 5}
	{
		\draw[blue,opacity=.5,->,thick] ({max (0,\x-1,\y-1)-max (0,\x-1,\y+1)+\x},{-max(0,\x-1,\y+1)+max (0,\x+1,\y+1)+\y})--({max (0,\x+1-1,\y-1)-max (0,\x+1-1,\y+1)+\x+1},{-max(0,\x+1-1,\y+1)+max (0,\x+1+1,\y+1)+\y});
	  \draw[green,opacity=.5,->,thick] ({max (0,\x-1,\y-1)-max (0,\x-1,\y+1)+\x},{-max(0,\x-1,\y+1)+max (0,\x+1,\y+1)+\y})--({max (0,\x-1,\y+1-1)-max (0,\x-1,\y+1+1)+\x},{-max(0,\x-1,\y+1+1)+max (0,\x+1,\y+1+1)+\y+1});
	}
}
\end{tikzpicture}

\caption{The action of $\eta_B$ and $\sigma$ from \eqref{sigmaX} on $\mathbb{TP}_2$.\label{tAB}}
\end{figure}

\begin{figure}[t]
\centering
\begin{tikzpicture}[scale=.6]
\clip (-5.5, -4.3) rectangle (8.3, 7.3);
\draw[red,thick] (-5,0) -- (0,0) -- (0,-5);
\draw[red,thick] (-5,1) -- (1,1) -- (1,-5);
\draw[red,thick]  (1,1) -- (8,8);
\draw[red,thick]  (1,0) -- (8,7);
\draw[red,thick]  (0,0) -- (1,0);
\foreach \x in {-7,-6,...,8}
{
	\foreach \y in {-7,-6,...,8}
	{
			\draw[blue,opacity=.5] ({1+\x+max(0,\x,\y)-max(1,1+\x,\y)},{\y+max(1,\x,\y)-max(1,1+\x,\y)}) --({1+\x+1+max(0,\x+1,\y)-max(1,1+\x+1,\y)},{\y+max(1,\x+1,\y)-max(1,1+\x+1,\y)});
			\draw[green,opacity=.5] ({1+\x+max(0,\x,\y)-max(1,1+\x,\y)},{\y+max(1,\x,\y)-max(1,1+\x,\y)}) -- ({1+\x+max(0,\x,\y+1)-max(1,1+\x,\y+1)},{\y+1+max(1,\x,\y+1)-max(1,1+\x,\y+1)});
	}
}
\draw[purple,thick] (0,-3)-- (1,-4)-- (2,-4)-- (4,-2)-- (6,2)-- (6,3)-- (6,4)-- (6,5)-- (6,5)-- (6,6)-- (5,6)-- (4,6)-- (3,6)-- (2,6)-- (1,6)-- (-1,5)-- (-3,3)-- (-3,2)-- (-3,1)-- (-3,0)-- (-2,-1)-- (-1,-2)-- (0,-3);
\draw[blue,thick] (0,-3)-- (1,-3)-- (2,-3)-- (4,-1)-- (6,3)-- (6,4)-- (6,5)-- (6,6)-- (6,6)-- (5,6)-- (4,6)-- (3,6)-- (2,6)-- (1,6)-- (0,6)-- (-2,5)-- (-4,3)-- (-4,2)-- (-4,1)-- (-3,0)-- (-2,-1)-- (-1,-2)-- (0,-3);
\end{tikzpicture}

\caption{The ef\/fect of the $\sigma$ from \eqref{sigmaX} on a tropical cubic plane curve with rays $L_i\colon X-A = 0$ and $L_j\colon Y-B = 0$. \label{tau12fig}}
\end{figure}
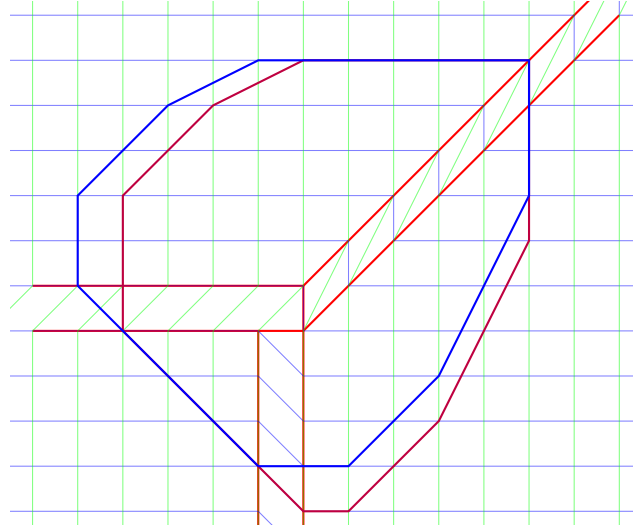

Let us now consider how to swap rays given by
\begin{gather*}
L_i\colon \  X-A=0, \qquad L_j\colon \ Y-B = 0,
\end{gather*}
as $Y \to -\infty$ and $X \to -\infty$ respectively. To describe this transformation, let us consider the transformation, $\rho\colon  \mathbb{T}^2 \to \mathbb{T}^2$, given by
\begin{gather*}
\rho\colon \  (X,Y) \to (X - \max(0,Y), X- \max(0,-Y)),
\end{gather*}
whose inverse is given by
\begin{gather*}
\rho^{-1}\colon \  (X,Y) \to (\max(X,Y), Y-X).
\end{gather*}
This transformation can be expressed as a composition of transformations of the form~\eqref{generators1} and~\eqref{generators2} as
\begin{gather*}
 \rho \colon \ \! (X,Y) \! \stackrel{\Xi}{\longrightarrow}(Y,X)  \stackrel{\iota_0}{\longrightarrow} (Y,X-\max(0,Y))  \stackrel{\Xi}{\longrightarrow} (X-\max(0,Y),Y)\\
\hphantom{\rho \colon \ \! (X,Y)}{} \! \stackrel{\Xi}{\longrightarrow} (X-\max(0,Y),Y+X-\max(0,Y)) = (X-\max(0,Y),X-\max(0,-Y)).\!
\end{gather*}
Roughly speaking, this sends every straight line of the form $X - A =0$ to one that is bent 90 degree along the line $Y= X$. The conjugation of $\iota_A$ by $\rho$, which we label $\eta_A = \rho \circ \iota_A \circ \rho^{-1}$, is given by the expression
\begin{gather*}
\eta_A(X,Y) = (X - \max(0,Y-A), Y + \max(A,X,Y)- \max(A,Y)).
\end{gather*}
It should be clear that this has the same ef\/fect as $\iota_A$ below the line $Y = X$, however, the ef\/fect of $\iota_A$ around $Y = \infty$ now occurs at $X = -\infty$. We may now state that the transformation swapping $L_i$ and $L_j$ is given by
\begin{gather*}
\sigma =  \eta_{B} \circ \eta_A^{-1},
\end{gather*}
whose max-plus expression may be simplif\/ied to
\begin{gather}
\sigma(X,Y) = \big(B+ X + \max(A,X,Y) - \max(A +B,B+X, A+Y),\nonumber\\
\hphantom{\sigma(X,Y) = \big(}{} A+Y+\max(B,X,Y)-\max(A+ B, B + X,A+Y)\big),\label{sigmaX}
\end{gather}
or equivalently, this is the tropical projective transformation
\begin{gather*}
\sigma([X:Y:Z]) = \big[B+ X + \max(A+Z,X,Y):A+Y+\max(B+Z,X,Y):\\
\hphantom{\sigma([X:Y:Z]) = \big[}{}  Z+\max(A+ B+Z, B + X,A+Y)\big].\nonumber
\end{gather*}
The ef\/fect of $\eta_B$ is shown in Fig.~\ref{tAB} and the ef\/fect on a cubic plane curve with these rays is depicted in Fig.~\ref{tau12fig}.

Lastly, for bookkeeping reasons, we include a set of transformations simply permute the roles of two rays that are of the same type. For example, if we have a tentacle, described by $L_i\colon X-A =0$ as $Y \to -\infty$ and another, described by $L_j\colon X-B = 0$ as $Y \to -\infty$, then one transformation simply swaps the roles of~$A$ and~$B$, which swaps~$L_i$ and~$L_j$. In this case, $\sigma$ acts as the identity map on the plane and as a simple transformation of the parameter space. This transformation can always be applied when there are two rays of the same form.

\begin{figure}[t]
\centering
\begin{tikzpicture}[scale=.42]
\begin{scope}
\clip (-3,-5) rectangle (5,5);
\draw[green,very thick] (-1,-1-1) -- (-1-1,-1) -- (-2,1) -- (-1,2) -- (3,2)-- (3+1,1)--(3+1,-1) -- (1,-2-2) -- (0,-2-2)--(-1,-3);
\draw[blue,very thick] (-1,-1-1) -- (-1-1,-1) -- (-1-1,1) -- (0,2+1) -- (3,2+1)-- (3+1,2)--(3+1,0) -- (1,-2-1) -- (-1,-2-1);
\draw[red!40] (-5.5,-1) -- (-1,-1);
\draw[red!40] (-5.5,1) -- (0,1);
\draw[red!40] (0,5.5) -- (0,1);
\draw[red!40] (3,5.5) -- (3,2);
\draw[red!40] (5.5,2) -- (3,2);
\draw[red!40] (5.5,0) -- (1,0);
\draw[red!40] (1,-5.5) -- (1,0);
\draw[red!40] (-1,-5.5) -- (-1,-1);
\draw[green!40] (0,5.5) -- (0,-5);
\draw[green!40] (-1,-5.5) -- (-1,5);
\end{scope}
\node at (0,-6) {$s_3$};
\begin{scope}[xshift=10cm]
\clip (-3,-5) rectangle (5,5);
\draw[green,very thick] (-1,-1-1) -- (-1-1,-1) -- (-2,1) -- (-1,2) -- (1,2)-- (3+1,-1)--(3+1,-3) -- (3,-4) --(0,-2-2)--(-1,-3);
\draw[blue,very thick] (-1,-1-1) -- (-1-1,-1) -- (-2,1) -- (-1,2) -- (3,2)-- (3+1,1)--(3+1,-1) -- (1,-2-2) -- (0,-2-2)--(-1,-3);
\draw[red!40] (-6,-1) -- (0,-1);
\draw[red!40] (-6,1) -- (-1,1);
\draw[red!40] (-1,5.5) -- (-1,1);
\draw[green!40] (3,5.5) -- (3,-6);
\draw[red!40] (5.5,1) -- (3,1);
\draw[red!40] (5.5,-1) -- (1,-1);
\draw[green!40] (1,-6) -- (1,5);
\draw[red!40] (0,-6) -- (0,-1);
\end{scope}
\node at (12,-6) {$s_5\circ s_4 \circ s_3$};
\begin{scope}[xshift=20cm]
\clip (-3,-5) rectangle (5,5);
\draw[green,very thick] (-1,2) -- (-1-1,1) -- (-2,-1) -- (-1,-2) -- (1,-2)-- (3+1,1)--(3+1,3) -- (3,4) -- (0,4) -- (-1,3);
\draw[blue,very thick] (-1,-1-1) -- (-1-1,-1) -- (-2,1) -- (-1,2) -- (1,2)-- (3+1,-1)--(3+1,-3) -- (3,-4) --(0,-2-2)--(-1,-3);
\draw[green!40] (-5,0) -- (5,0);
\draw[red!40] (-6,-1) -- (0,-1);
\draw[red!40] (-6,1) -- (-1,1);
\draw[red!40] (-1,5.5) -- (-1,1);
\draw[red!40] (3,-3) -- (3,-6);
\draw[red!40] (5.5,-1) -- (1,-1);
\draw[red!40] (5.5,-3) -- (3,-3);
\draw[red!40] (1,6) -- (1,-1);
\draw[red!40] (0,-6) -- (0,-1);
\end{scope}
\node at (23,-6) {$p_1$};

\begin{scope}[yshift=-12cm]
\clip (-4.5,-5) rectangle (5,5);
\draw[green,very thick] (-3,2) -- (-4,1) -- (-2,-1) -- (-1,-2) -- (1,-2)-- (2,-1)--(2,3) -- (1,4) -- (-2,4) -- (-3,3);
\draw[blue,very thick] (-1,2) -- (-1-1,1) -- (-2,-1) -- (-1,-2) -- (1,-2)-- (3+1,1)--(3+1,3) -- (3,4) -- (0,4) -- (-1,3);
\draw[green!40] (-5.5,-1) -- (6,-1);
\draw[red!40] (-5.5,1) -- (0,1);
\draw[red!40] (0,5.5) -- (0,1);
\draw[red!40] (3,5.5) -- (3,3);
\draw[red!40] (5.5,3) -- (3,3);
\draw[green!40] (5.5,1) -- (-6,1);
\draw[red!40] (1,-5.5) -- (1,1);
\draw[red!40] (-1,-5.5) -- (-1,-1);
\end{scope}
\node at (0,-18) {$s_2$};
\begin{scope}[xshift = 12cm,yshift=-12cm]
\clip (-6,-5) rectangle (4,5);
\draw[green,very thick] (-4,2) -- (-4,1) -- (-2,-1) -- (-1,-2) -- (1,-2)-- (2,-1)--(2,1) -- (-1,4) -- (-4,4) -- (-5,3);
\draw[blue,very thick] (-3,2) -- (-4,1) -- (-2,-1) -- (-1,-2) -- (1,-2)-- (2,-1)--(2,3) -- (1,4) -- (-2,4) -- (-3,3);
\draw[red!40] (-5.5,1) -- (0,1);
\draw[red!40] (-2,5.5) -- (-2,1);
\draw[red!40] (1,5.5) -- (1,3);
\draw[red!40] (5.5,3) -- (3,3);
\draw[green!40] (5.5,1) -- (-6,1);
\draw[red!40] (1,-5.5) -- (1,1);
\draw[red!40] (-1,-5.5) -- (-1,1);
\draw[green!40] (-5.5,3) -- (6,3);
\end{scope}
\node at (12,-18) {$s_2 \circ s_1 \circ s_0$};
\begin{scope}[xshift = 22cm,yshift=-12cm]
\clip (-5.5,-5) rectangle (5,5);
\draw[purple,very thick] (-1,-1-1) -- (-1-1,-1) -- (-1-1,1) -- (0,2+1) -- (3,2+1)-- (3+1,2)--(3+1,0) -- (1,-2-1) -- (-1,-2-1);
\draw[green,very thick] (4,2) -- (4,1) -- (2,-1) -- (1,-2) -- (-1,-2)-- (-2,-1)--(-2,1) -- (1,4) -- (4,4) -- (5,3);
\draw[blue,very thick] (-4,2) -- (-4,1) -- (-2,-1) -- (-1,-2) -- (1,-2)-- (2,-1)--(2,1) -- (-1,4) -- (-4,4) -- (-5,3);
\draw[red!40] (-4,5.5) -- (-4,3);
\draw[red!40] (-1,5.5) -- (-1,1);
\draw[red!40] (6,1) -- (-1,1);
\draw[red!40] (-1,1) -- (-6,1);
\draw[red!40] (1,-5.5) -- (1,-1);
\draw[red!40] (-1,-5.5) -- (-1,1);
\draw[red!40] (-5.5,3) -- (-4,3);
\draw[red!40] (5.5,-1) -- (1,-1);
\draw[green!40] (0,5) -- (0,-5);
\end{scope}
\node at (22,-18) {$p_2$};
\end{tikzpicture}

\caption{Starting with a single spiral, we show each signif\/icant step in the sequence~\eqref{up6comp}. In blue, we show the result of previous transformations, in green is the result of the transformations listed below. In the last step, we also show the original spiral (in red).\label{actions}}
\end{figure}
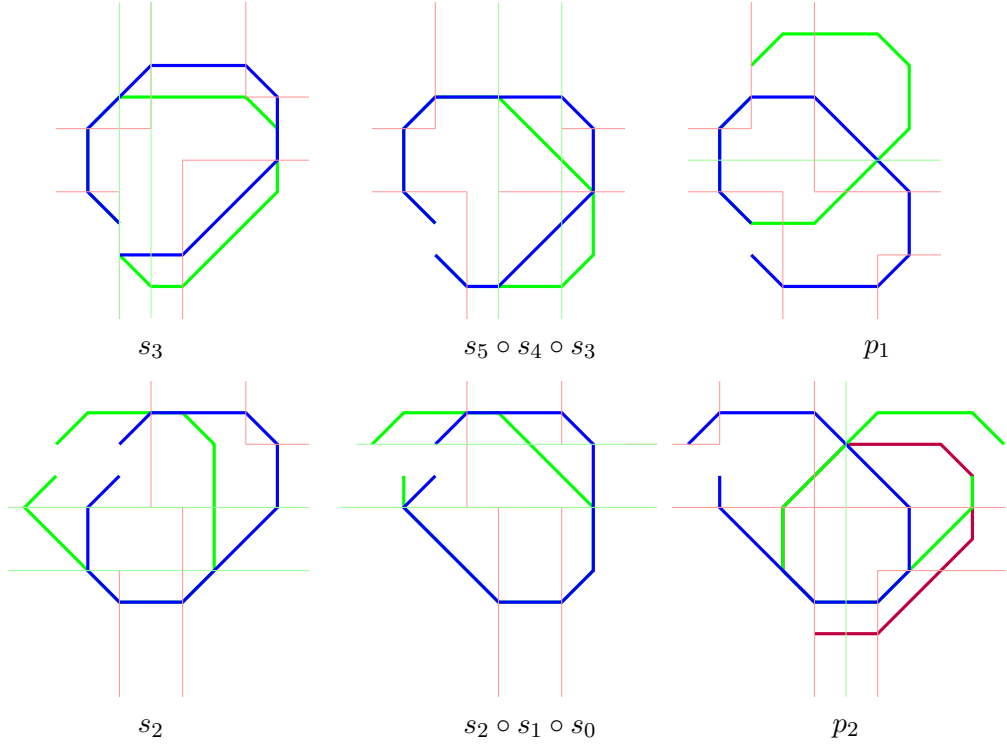

Each of these transformations is an isomorphism between either a collection of polygons or between some spiral diagrams of the same form. We can now specify that each of the ultradiscrete QRT maps and ultradiscrete Painlev\'e equations are inf\/inite order elements of the group of transformations that preserve a pencil of polygons def\/ined by tropical genus one curves or their corresponding spiral diagrams respectively. This means they may be expressed in terms of the simple transformations above. As an example, we consider \eqref{QRT} and \eqref{qP6}. We start by parameterizing the rays as follows:
\begin{alignat*}{5}
& L_1\colon \  X-A_1=0, \qquad && L_2\colon \  X-A_2=0, \qquad && L_3\colon \  X-A_3 = 0, \qquad && L_4\colon \  X-A_4 = 0, & \\
& L_5\colon \  Y-B_1=0, \qquad && L_6\colon \  Y-B_2=0, \qquad && L_7\colon \  Y-B_3 = 0, \qquad && L_8\colon \  Y-B_4 = 0, &
\end{alignat*}
where $L_1$ and $L_2$ extend downwords, $L_3$ and $L_4$ extend upwards, $L_5$ and $L_6$ extend to the left and $L_7$ and $L_8$ extend to the right. We now have a group of type $W(D_5^{(1)}) = \langle s_0, \ldots, s_5\rangle$ where
\begin{alignat*}{4}
& s_0 = \sigma_{7,8}, \qquad && s_1 = \sigma_{5,6}, \qquad && s_2 = \sigma_{5,7},& \\
& s_3 = \sigma_{1,3}, \qquad && s_4 = \sigma_{1,2}, \qquad && s_3 = \sigma_{3,4},&
\end{alignat*}
with two additional symmetries, $p_1$ and $p_2$, which are ref\/lections through the line $Y = (B_3 + B_4)/2$ and $X = (A_3 + A_4)/2$ respectively. We can now write the ultradiscrete QRT map and the ultradiscrete Painlev\'e equation as the composition
\begin{gather}\label{up6comp}
T = p_2 \circ s_2 \circ s_0 \circ s_1 \circ s_2 \circ s_1 \circ p_1  \circ s_3 \circ s_5 \circ s_4 \circ s_3.
\end{gather}
To show that each step is an isomorphism of spiral diagrams, we have depicted the nontrivial steps in $T$ on a typical spiral in Fig.~\ref{actions}.

We can present the nontrivial actions of these transformations as
\begin{gather*}
s_2 \colon \ X \to X + \max(Y,B_3) - \max(Y,B_1),\\
s_2 \colon \ A_1 \to A_1 + B_3 - B_1, \qquad s_2\colon \  A_2 \to A_2 + B_3- B_1,\\
s_3 \colon \  Y \to Y + \max(X,A_3) - \max(X,A_1),\\
s_3 \colon \  B_1 \to B_1 + A_3 - A_1, \qquad s_3\colon \  B_2 \to B_2 + A_3-A_1,\\
p_1 \colon \  Y \to B_3 + B_4 - Y, \qquad p_2\colon \ X \to A_3 + A_4 - X.
\end{gather*}
The composition in \eqref{up6comp} gives \eqref{uP6}.

\begin{rem}
The above constitutes the action on a tropical biquadratic that does not satisfy the requirement that the image under $\Pi$ is a tropical cubic plane curve. A cubic plane curve may be obtained by applying $\iota_{A_4}$, which has the ef\/fect of removing the ray given by $L_4$ and adding a~ray given by the same formula, but pointing downward instead of upwards. Up to translational invariance, this is equivalent to the polygon considered in Section~\ref{D5}. In particular, their groups of transformations are of the same af\/f\/ine Weyl type.
\end{rem}

An aspect of def\/ining the group of transformations for a polygon or spiral diagram that we have not introduced in the above example is that we may always remove two parameters by taking into account uniqueness of a group of transformations up to translational equivalence. We can take this into account by insisting that two rays, of dif\/ferent asymptotic forms, pass through the origin. This means that we will often compose one of the above types of transformation with a translation so that any ray which is supposed to pass through the origin does so after the transformation. This f\/ixes a representation based on which rays we choose to pass through the origin.

\section{Tropical representations of af\/f\/ine Weyl groups}\label{sec:reps}

While the task of f\/inding subtraction free versions of the Cremona transformations in Sakai's list was presented (but not published) by Kajiwara et al.~\cite{KMNY:Ereps}, what we wish to present is a dif\/ferent perspective. The derivation of the following list of af\/f\/ine Weyl representations will sometimes be a slightly dif\/ferent parameterization of the transformations of \cite{KMNY:Ereps} due to the manner in which they were derived. We will also provide some of the geometric motivation behind our choices of generators. To this end, we shall display a spiral diagram and a nontrivial translation for each of the cases in Table~\ref{polgonalcorrespondence}. When the Newton polygon is known, this will also accompany the spiral diagram on the right.

\subsection{Triangles}

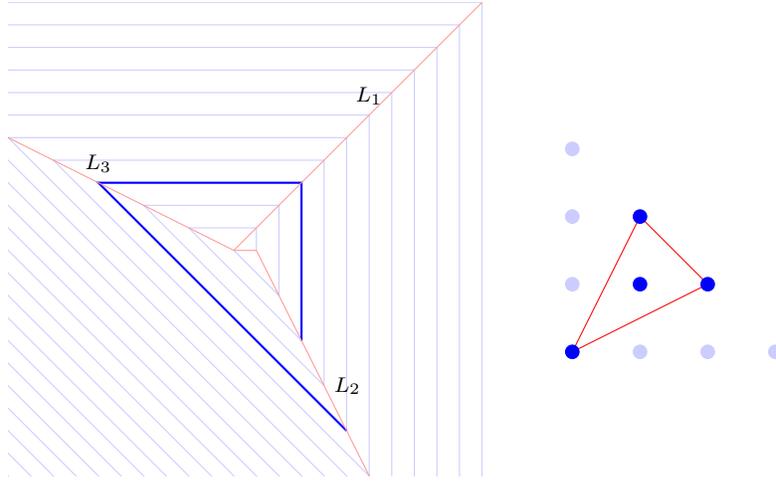
\begin{figure}[t]
\centering
\begin{tikzpicture}[scale=.6]
\begin{scope}
\clip (-5, -5) rectangle (5.5, 5.5);
\foreach \lam in {-.5,0,...,10.5}
{
	\draw[blue!20] (1+\lam,-1-2*\lam)  -- (1+\lam,1+\lam) -- (-2-2*\lam,1+\lam) -- (2+\lam,-3-2*\lam);
}
{
	\draw[blue,thick] (1+.5,-1-2*.5)  -- (1+.5,1+.5) -- (-2-2*.5,1+.5) -- (2+.5,-3-2*.5);
}
\draw[blue!20] (1,-1) -- (0,0);
\draw[red!40] (0,0) -- (6,6) node[black,fill=none,font=\scriptsize,midway,above] {$L_1$};
\draw[red!40] (0,0) -- (-6,3) node[black,fill=none,font=\scriptsize,midway,above] {$L_3$};
\draw[red!40] (.5,0) -- (3.5,-6) node[black,fill=none,font=\scriptsize,midway,right] {$L_2$};
\draw[red!40] (.5,0) -- (0,0);
\end{scope}

\begin{scope}[scale=1.5,xshift=-3cm,yshift=-1.5cm]
\draw[red] (8,0) -- (9,2)--(10,1)--cycle;
\filldraw[blue] (8,0) circle (.1);
\filldraw[blue!20] (9,0) circle (.1);
\filldraw[blue!20] (10,0) circle (.1);
\filldraw[blue!20] (11,0) circle (.1);
\filldraw[blue!20] (8,1) circle (.1);
\filldraw[blue] (9,1) circle (.1);
\filldraw[blue] (10,1) circle (.1);
\filldraw[blue!20] (8,2) circle (.1);
\filldraw[blue] (9,2) circle (.1);
\filldraw[blue!20] (8,3) circle (.1);
\end{scope}
\end{tikzpicture}\caption{The spiral diagram for the system with af\/f\/ine Weyl symmetry of type $A_0^{(1)}$. \label{A0spiral}}

\end{figure}

At the bottom of the hierarchy of multiplicative surfaces in \cite{Sakai:rational} is the system with a symmetry of the dihedral group of order $6$, which admits the presentation
\begin{gather*}
\mathcal{D}_6 = \big\langle p_1, p_2\colon \, p_1^3 = p_2^2 = (p_2p_1)^2 = 1 \big\rangle.
\end{gather*}
This is the group permuting the three rays in Fig.~\ref{A0spiral}. The rays may be parameterized by the equations
\begin{gather*}
L_1\colon \  Y-X-A=0, \qquad L_2\colon \ 2X+Y-B = 0, \qquad L_3\colon \ X+2Y-C = 0.
\end{gather*}
By exploiting scaling (i.e., $X \to X+ \lambda$ and $Y \to Y + \mu$), we can reduce this to the case where we f\/ix $B = C = 0$.

In this way, let $p_1$ permute the lines so that $p_1\colon (L_1,L_2,L_3) \to (L_3,L_1,L_2)$. Similarly, $p_2$ is the transformation that swaps $L_2$ and $L_3$ via a ref\/lection around the line $Y=X$. These are explicitly given by the piecewise  linear transformations
\begin{alignat*}{4}
& p_1\colon \  X \to -X-Y - \dfrac{A}{3}, \qquad && p_2\colon \  Y \to X + \dfrac{2A}{3}, \qquad  &&p_2\colon \  A \to A,& \\
& p_2\colon \  X \to Y, \qquad && p_2\colon \  X \to Y, \qquad && p_2\colon \  A \to -A.&
\end{alignat*}
As a very degenerate case, the transformations are simple given by (up to translations) a subgroup of actions of the type~\eqref{generators2}.

It is natural to see that $Q = A$. The limit which gives a f\/ibration by tropical biquadratics is the limit as $A = Q = 0$. The resulting polygons arise as level sets of
\begin{gather*}
H(X,Y) = \max(-X-Y, X,Y).
\end{gather*}
As the dihedral group, $\mathcal{D}_6$, contains no elements of inf\/inite order, there is no dif\/ference equation associated with this group.

\subsection{Rectangles} 
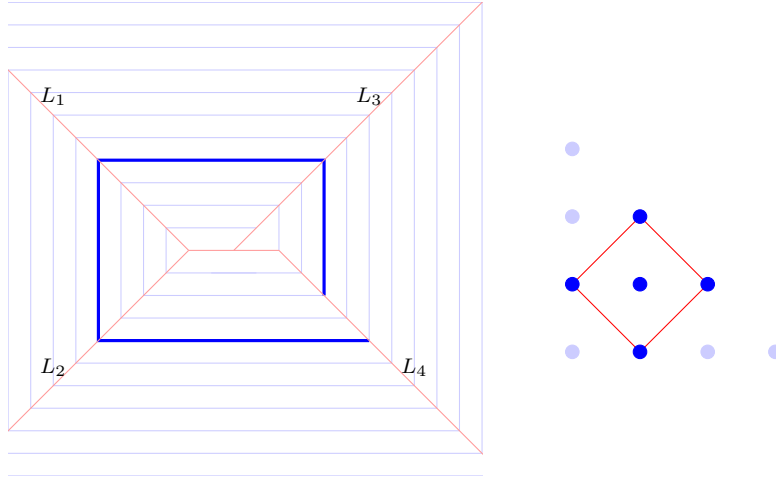
\begin{figure}[t]
\centering
\begin{tikzpicture}[scale=.6]
\begin{scope}
\clip (-5, -5) rectangle (5.5, 5.5);
\foreach \lam in {-1.5,-1,...,10.5}
{
	\draw[blue!20] (2+\lam,-1-\lam)  -- (2+\lam,2+\lam) -- (-3-\lam,2+\lam) -- (-3-\lam,-2-\lam)--(3+\lam,-2-\lam);
}
\draw[blue!20] (.5,-.5) -- (-.5,-0.5);
\draw[blue,very thick] (2+0,-1-0)  -- (2+0,2+0) -- (-3-0,2+0) -- (-3-0,-2-0)--(3+0,-2-0);
\draw[red!40] (1,0) -- (7,-6) node[black,fill=none,font=\scriptsize,midway,above] {$L_4$};
\draw[red!40] (0,0) -- (6,6) node[black,fill=none,font=\scriptsize,midway,above] {$L_3$};
\draw[red!40] (-1,0) -- (-7,6) node[black,fill=none,font=\scriptsize,midway,above] {$L_1$};
\draw[red!40] (-1,0) -- (-7,-6) node[black,fill=none,font=\scriptsize,midway,above] {$L_2$};
\draw[red!40] (-1,0) -- (1,0);
\end{scope}
\begin{scope}[scale=1.5,xshift=-3cm,yshift=-1.5cm]
\draw[red] (9,0)--(8,1) -- (9,2)--(10,1)--cycle;
\filldraw[blue!20] (8,0) circle (.1);
\filldraw[blue] (9,0) circle (.1);
\filldraw[blue!20] (10,0) circle (.1);
\filldraw[blue!20] (11,0) circle (.1);
\filldraw[blue] (8,1) circle (.1);
\filldraw[blue] (9,1) circle (.1);
\filldraw[blue] (10,1) circle (.1);
\filldraw[blue!20] (8,2) circle (.1);
\filldraw[blue] (9,2) circle (.1);
\filldraw[blue!20] (8,3) circle (.1);
\end{scope}
\end{tikzpicture}

\caption{The spiral diagram for the system with af\/f\/ine Weyl symmetry of type $A_1^{(1)}$ with an additional dihedral symmetry. \label{A1pspiral}}
\end{figure}

We consider a spiral diagram of quadralaterals which gives an af\/f\/ine Weyl group of type $W(A_1^{(1)})$ with an additional $\mathcal{D}_8$ symmetry. In the same way as above, we may exploit scaling so that the rays extending towards $X = -\infty$ pass through the origin. We paramaterize our rays as follows:
\begin{alignat*}{3}
& L_1\colon \  Y+X = 0, \qquad && L_3\colon \  Y - X - A =0, & \\
& L_2\colon \  Y-X = 0, \qquad && L_4\colon \  Y + X - B =0. &
\end{alignat*}
This is depicted in Fig.~\ref{A1pspiral}.

The symmetry group for this system is the semidirect product of $\mathcal{D}_8 = \langle p_1 , p_2 \rangle$ and $W\big(A_1^{(1)}\big)  = \langle s_0, s_1 \rangle$. A presentation is given by
\begin{gather*}
\mathcal{D}_8 \ltimes W\big(A_1^{(1)}\big) = \big\langle p_1 , p_2, s_0, s_1\colon \,  p_1^4 = p_2^2 = (p_1p_2)^2 =  s_0^2 = s_1^2 = s_0 p_2 s_1 p_2 = 1 \big\rangle,
\end{gather*}
where the action of $\mathcal{D}_8$ is specif\/ied up to translation by a clockwise rotation of the four def\/ining lines, $p_1$, whereas $p_2$ swaps $L_3$ and $L_4$. We write these transformations as
\begin{alignat*}{5}
& p_1\colon \ X \to Y+\dfrac{B}{2},\qquad && p_1 \colon \ Y \to -X - \dfrac{B}{2}, \qquad && p_1\colon \ A \to B, \qquad && p_1\colon \ B \to A,& \\
& p_2\colon \ X \to X, \qquad && p_2 \colon \ Y \to -Y ,  \qquad && p_2\colon \ A \to -B, \qquad && p_2\colon \ B \to -A.&
\end{alignat*}
Let $s_0$ be the conjugation of the transformation depicted in Fig.~\ref{sABP12} with the piecewise linear transformation that makes $L_2$ and $L_3$ parrellel to the $y$-axis (and $L_1$ and $L_4$ to the $x$-axis). The transformation $s_1$ may be obtained in a similar manner with $L_1$ and $L_4$, giving
\begin{gather*}
s_0 \colon \ X \to X + \max(0,Y-X+A) - \max(0,Y-X) - \dfrac{A}{2},\\
s_0 \colon \ Y \to Y + \max(0,Y-X+A) - \max(0,Y-X) + \dfrac{A}{2},\\
s_0 \colon \ A \to -A, \qquad s_0\colon \ B \to B - 2 A,
\end{gather*}
and $s_1 = p_2 \circ s_0 \circ p_2$, which we write as
\begin{gather*}
s_1 \colon \ X \to X + \max(0,-X-Y-B) - \max(0,-X-Y) + \dfrac{B}{2},\\
s_1 \colon \ Y \to Y + \max(0,-X-Y) - \max(0,-X-Y-B) + \dfrac{B}{2},\\
s_1 \colon \ A \to A-2B,\qquad s_1\colon \ B \to - B.
\end{gather*}
We f\/ind that $Q = A - B$ by tracing around the spiral. When $A = B$, we obtain the invariant
\begin{gather*}
H(X,Y) = \max(-X,-Y,Y,X - A).
\end{gather*}
For the element $T = s_1 \circ s_0$, we resort to co-ordinates $U$ and $V$, where $X = (U+V)/2$ and $Y = (U- V)/2$. The dynamical system in these variables is
\begin{gather*}
\tilde{U} - U =   B+ 2\max(A,A+V) - 2\max(0,V),\\
\tilde{V} - V =   3A + 2\max(0,\tilde{U}) - 2\max(B, 2A +\tilde{U}),\\
T\colon \  A  \to A + 2Q,  \qquad  T\colon \  A \to A - 2Q,
\end{gather*}
where $\tilde{U} = T(U)$ and $\tilde{V} = T(V)$.

\subsection{Quadralaterals} 
\begin{figure}[t]
\centering
\begin{tikzpicture}[scale=.6]
\begin{scope}
\clip (-5, -5) rectangle (5.5, 5.5);
\foreach \lam in {-1.5,-1,...,10.5}
{
	\draw[blue!20] (1+\lam,1+\lam)  -- (-2-\lam,1+\lam) -- (-2-\lam,-1) -- (2+\lam,-5-2*\lam) -- (2+\lam,2+\lam);
}
\draw[blue, very thick] (-1,-1) -- (1+0,-3) -- (1+0,1+0)  -- (-2-0,1+0) -- (-2-0,-1);
\draw[blue!20] (0,0)--(0,-1);
\draw[red!40] (-.5,-.5)--(7,7) node[black,fill=none,font=\scriptsize,midway,above] {$L_4$};
\draw[red!40] (-.5,-.5)--(-8,7) node[black,fill=none,font=\scriptsize,midway,above] {$L_1$};
\draw[red!40] (0,-1)--(-7.5,-1) node[black,fill=none,font=\scriptsize,midway,above] {$L_2$};
\draw[red!40] (0,-1)--(3,-7) node[black,fill=none,font=\scriptsize,midway,right] {$L_3$};
\end{scope}

\begin{scope}[scale=1.5,xshift=-3cm,yshift=-1.5cm]
\draw[red] (8,0)--(8,1) -- (9,2)--(10,1)--cycle;
\filldraw[blue] (8,0) circle (.1);
\filldraw[blue!20] (9,0) circle (.1);
\filldraw[blue!20] (10,0) circle (.1);
\filldraw[blue!20] (11,0) circle (.1);
\filldraw[blue] (8,1) circle (.1);
\filldraw[blue] (9,1) circle (.1);
\filldraw[blue] (10,1) circle (.1);
\filldraw[blue!20] (8,2) circle (.1);
\filldraw[blue] (9,2) circle (.1);
\filldraw[blue!20] (8,3) circle (.1);
\end{scope}
\end{tikzpicture}

\caption{The spiral diagram for the system with af\/f\/ine Weyl symmetry $A_1^{(1)}$. \label{A1spiral}}
\end{figure}
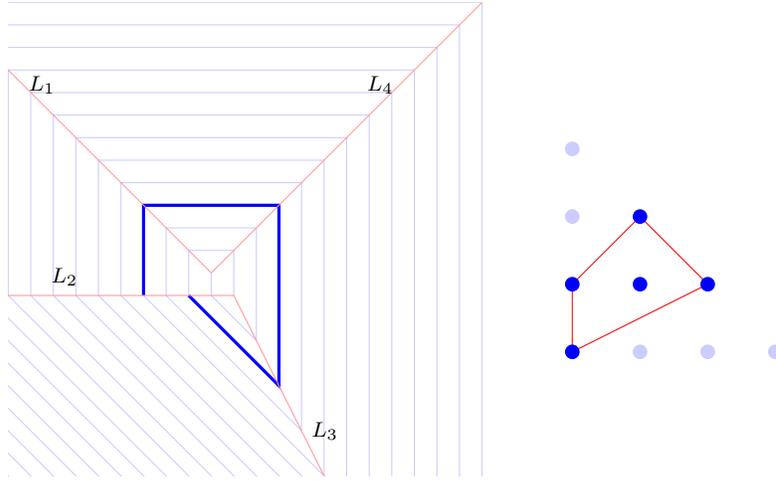

We have another quadralateral that does not possess an additional dihedral symmetry. We break the dihedral symmetry by f\/ixing the parameterization of the four rays in the following manner:
\begin{alignat*}{3}
&L_1\colon \  Y+X = 0, \qquad && L_3\colon \ Y+2X -A = 0,& \\
&L_2 \colon \  Y = 0, \qquad && L_4\colon \ Y-X-B = 0,
\end{alignat*}
as depicted in Fig.~\ref{A1spiral}.

The group of transformations that preserves this spiral diagram is of type
\begin{gather*}
W(A_1^{(1)}) = \big\langle s_0, s_1\colon \, (s_0)^2 = (s_1)^2 \big\rangle,
\end{gather*}
where $s_1 \circ s_0$ is the element of inf\/inite order. The f\/irst transformation, $s_0$, swaps the roles of $L_1$ with $L_2$ and $L_3$ with $L_4$, which we write as
\begin{alignat*}{3}
& s_0 \colon \  Y \to -Y-X, \qquad && s_0 \colon \  A \to -B,& \\
& s_0 \colon \ X \to X, \qquad && s_0 \colon \  B \to -A.&
\end{alignat*}
The other involution, $s_1$, is a ref\/lection around $X = B/2$ above $Y=0$ (so that $L_4$ is sent to~$L_1$) and a skewed ref\/lection below $Y = 0$, given by $X \to -Y-X-B/2$, which simplif\/ies to the following tropically rational transformation
\begin{alignat*}{3}
& s_1\colon \  X \to \max(0,-Y) - X - B, \qquad && s_1\colon \  A \to -2B - A,& \\
& s_1\colon \  Y \to Y,  \qquad && s_1\colon \  B \to B.&
\end{alignat*}
This is simply the conjugation of $\iota_0$ with a swap of $X$ and $Y$. We f\/ind $Q = A+B$ by tracing around one spiral. When $A = -B$, we obtain the invariant
\begin{gather*}
H(X,Y) = \max(-X-Y,-X,Y,X-A).
\end{gather*}
The composition, $T = s_1 \circ s_0$, gives the evolution equations
\begin{gather*}
\tilde{X} + X  = \max(0,Y+X) + A,\\
\tilde{Y} + Y  = -X,\\
T\colon \  A \to A + Q, \qquad T\colon \  B \to B - Q.
\end{gather*}
This element, $T$, is the generator for $\mathbb{Z}$ in the decomposition of $W(A_1^{(1)}) \cong \mathbb{Z} \ltimes \mathfrak{G}_2$ in~\cite{KMNY:Ereps, Sakai:rational}. Alternatively, we could write this system as a second order dif\/ference equation in $W =-Y$, where the resulting system becomes
\begin{gather*}
W + 2\tilde{W} + \tilde{\tilde{W}} = \max(0,\tilde{W}) + A,
\end{gather*}
which coincides with a more standard version of an ultradiscrete version of the f\/irst Painlev\'e equation~\cite{RTGO:ultimatediscretePs}.

\subsection{Pentagons} 
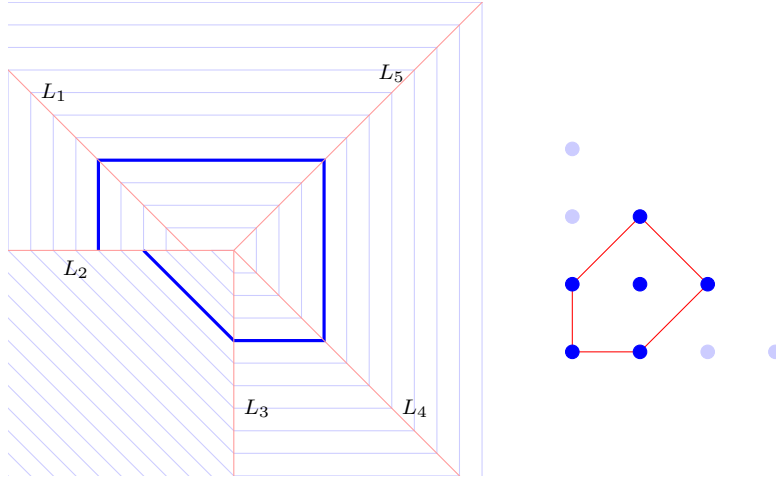
\begin{figure}[t]
\centering
\begin{tikzpicture}[scale=.6]
\begin{scope}
\clip (-5, -5) rectangle (5.5, 5.5);
\foreach \lam in {-.5,0,...,10.5}
{
	\draw[blue!20] (-1-\lam,0) -- (0,-1-\lam) -- (1+\lam,-1-\lam) -- (1+\lam,1+\lam) -- (-2-\lam,1+\lam) -- (-2-\lam,0);
}
\draw[blue!20] (-.5,0) -- (0,0);
\draw[blue,very thick] (-1-1,0) -- (0,-1-1) -- (1+1,-1-1) -- (1+1,1+1) -- (-2-1,1+1) -- (-2-1,0);
\draw[red!40] (0,0) -- (7,7)  node[black,fill=none,font=\scriptsize,midway,above] {$L_5$};
\draw[red!40] (0,0) -- (-7,0) node[black,fill=none,font=\scriptsize,midway,below] {$L_2$};
\draw[red!40] (0,0) -- (7,-7)  node[black,fill=none,font=\scriptsize,midway,right] {$L_4$};
\draw[red!40] (0,0) -- (0,-7) node[black,fill=none,font=\scriptsize,midway,right] {$L_3$};
\draw[red!40] (-1,0) -- (-8,7) node[black,fill=none,font=\scriptsize,midway,right] {$L_1$};
\end{scope}

\begin{scope}[scale=1.5,xshift=-3cm,yshift=-1.5cm]
\draw[red] (8,0)--(8,1) -- (9,2)--(10,1)--(9,0)--cycle;
\filldraw[blue] (8,0) circle (.1);
\filldraw[blue] (9,0) circle (.1);
\filldraw[blue!20] (10,0) circle (.1);
\filldraw[blue!20] (11,0) circle (.1);
\filldraw[blue] (8,1) circle (.1);
\filldraw[blue] (9,1) circle (.1);
\filldraw[blue] (10,1) circle (.1);
\filldraw[blue!20] (8,2) circle (.1);
\filldraw[blue] (9,2) circle (.1);
\filldraw[blue!20] (8,3) circle (.1);
\end{scope}
\end{tikzpicture}

\caption{The pentagon.\label{A1tA1spiral}}
\end{figure}

This case is associated with u-$\mathrm{P}_{\rm II}$. To preserve much of the structure of the two previous cases, we have parameterize the f\/ive rays as follows:
\begin{alignat*}{3}
&L_1 \colon \  Y+X - A = 0, \qquad && L_4\colon \  Y+X- B = 0,&\\
&L_2 \colon \  Y = 0, \qquad && L_5\colon \  Y-X-C = 0,&\\
& L_3 \colon \  X = 0, \qquad &&&
\end{alignat*}
as depicted in Fig.~\ref{A1tA1spiral}.

A presentation of the group of transformations is
\begin{gather*}
A_1^{(1)}\times A_1^{(1)} = \big\langle s_0,s_1, w_0, w_1\colon \, s_i^2 = w_i^2 = 1 \big\rangle,
\end{gather*}
where $s_0$ is a ref\/lection around the line $Y=X$, and $s_1$ is the same the action of $s_1$ in the previous section in that above the line $Y=0$, we have a ref\/lection, and below the line, we skew the plane. The generators are
\begin{alignat*}{4}
& s_0 \colon \  X \to Y, \qquad && p_0\colon \  A \to B, \qquad && p_0\colon \  C \to -C,& \\
& s_0 \colon \  Y \to X, \qquad && p_0\colon \  B \to A,\qquad &&&\\
& s_1 \colon \  X \to \max(0,-Y) -X + B, \qquad && p_1 \colon \  A \to B+C,\qquad && p_1\colon \  A_2 \to A-B,&\\
& s_1 \colon \  Y \to Y, \qquad && p_1 \colon \  B \to B.\qquad &&&
\end{alignat*}
As for the other part of the group, $w_0$ swaps $L_1$ and $L_4$ via a transformation that sheers between the lines~$L_1$ and~$L_4$, which can be written as
\begin{gather*}
w_0 \colon \  X \to X +\max(0,X+Y-A) - \max(0,X+Y-B), \\
w_0 \colon \  Y \to Y +\max(0,X+Y-B) - \max(0,X+Y-A),\\
w_0 \colon \  A \to B, \qquad  w_0\colon \  B \to A, \qquad  w_0\colon \  C \to C + 2A - 2B,
\end{gather*}
while $w_1$ is a piecewise linear sheering transformation swapping $L_2$ and $L_3$, which we write as
\begin{gather*}
w_1 \colon \  X \to X +C + \max(0,X,Y-C) - \max(0,X,Y), \\
w_1 \colon \  X \to Y + \max(C,X,Y-C) - \max(0,X,Y),\\
w_1 \colon \  A_0 \to A + C, \qquad  w_1 \colon \  B \to B + C, \qquad  s_1 \colon \  C \to -C.
\end{gather*}
Tracing around the f\/igure reveals that $Q$ is given by
\begin{gather*}
Q = B + C - A.
\end{gather*}
When $C = A-B$, we obtain the invariant
\begin{gather*}
H(X,Y) = \max(-X-Y,-X,-Y,X-B,Y-A).
\end{gather*}
One simple translation is the composition, $T = (s_0 \circ s_1)^2$, which can be written as
\begin{subequations}\label{uP2}
\begin{gather}
\tilde{X} + X  = \max(0,-Y) + B,\\
\tilde{Y} + Y  = \max(0,-\tilde{X}) + B + C,\\
T\colon \  A \to  A + Q, \qquad  T\colon \  B \to B + Q,
\end{gather}
\end{subequations}
where the other obvious translation, $(w_0 \circ w_1)$, commutes with $T$. This system is called u-$\mathrm{P}_{\rm II}$. Exact solutions of~\eqref{uP2} were studied in~\cite{Murata:ExactsolsuP2}.

\subsection{Hexagons} 
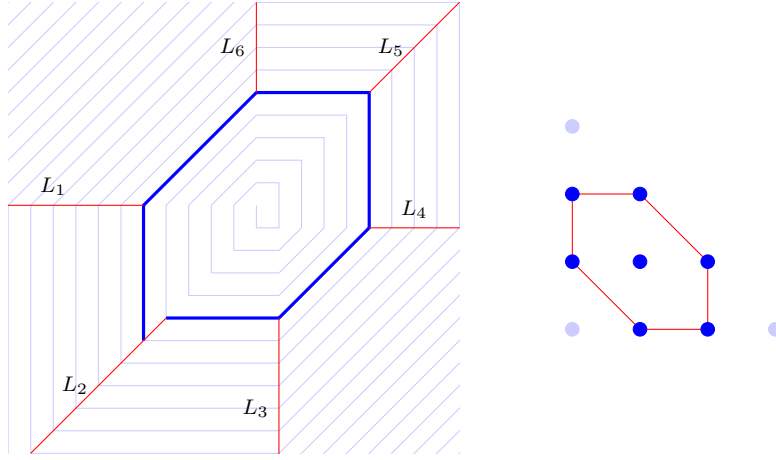
\begin{figure}[t]
\centering
\begin{tikzpicture}[scale=.6]
\begin{scope}
\clip (-5,-5) rectangle (5,5);
\foreach \m in {-.5,0,...,13}
{
	\draw[blue!20] (-.5-\m,-1-\m) -- (1,-1-\m)  -- (2+\m,0) -- (2+\m,2+\m)-- (.5,2+\m) -- (-\m-1,.5) -- (-\m-1,-1.5-\m);
}
\draw[blue!20]  (0,-.5)--(0,.5) -- (.5,1) -- (1,1) -- (1,0)--(.5,0) -- (0.5,.5);
\draw[blue,very thick] (-.5-1,-1-1) -- (1,-1-1) -- (2+1,0) -- (2+1,2+1)-- (.5,2+1) -- (-1-1,.5) -- (-1-1,-1-1.5) ;
\draw[red] (.5,3)--(.5,5) node[black,fill=none,font=\scriptsize,midway,left] {$L_6$};
\draw[red] (3,3)--(5,5) node[black,fill=none,font=\scriptsize,midway,left] {$L_5$};
\draw[red] (1,-2)--(1,-6) node[black,fill=none,font=\scriptsize,midway,left] {$L_3$};
\draw[red] (-2,.5)--(-6,.5) node[black,fill=none,font=\scriptsize,midway,above] {$L_1$};
\draw[red] (3,0)--(5,0) node[black,fill=none,font=\scriptsize,midway,above] {$L_4$};
\draw[red] (-1.5,-2)--(-4.5,-5) node[black,fill=none,font=\scriptsize,midway,left] {$L_2$};
\end{scope}
\begin{scope}[scale=1.5,xshift=-3cm,yshift=-1.5cm]
\draw[red] (9,0) -- (10,0) -- (10,1) -- (9,2) -- (8,2)-- (8,1)--cycle;
\filldraw[blue!20] (8,0) circle (.1);
\filldraw[blue] (9,0) circle (.1);
\filldraw[blue] (10,0) circle (.1);
\filldraw[blue!20] (11,0) circle (.1);
\filldraw[blue] (8,1) circle (.1);
\filldraw[blue] (9,1) circle (.1);
\filldraw[blue] (10,1) circle (.1);
\filldraw[blue] (8,2) circle (.1);
\filldraw[blue] (9,2) circle (.1);
\filldraw[blue!20] (8,3) circle (.1);
\end{scope}
\end{tikzpicture}

\caption{The spiral diagram for the discrete Painlev\'e equation with $A_2^{(1)}+A_1^{(1)}$ symmetry.\label{A2A1fig}}
\end{figure}

The tropical representation for $W(A_2^{(1)} + A_1^{(1)})$ was one of the f\/irst to be written down \cite{KNY:qP4}. There are a number of equivalent ways of obtaining a hexagon as a cubic plane curve, we choose to parameterize our rays so that our presentation coincides with the presentation of Noumi et  al.~\cite{KNY:qP4}. In particular, our rays are parameterized as follows:
\begin{alignat*}{3}
&L_1\colon \ Y = 0, \qquad && L_4\colon \ Y - B_2 = 0,&\\
&L_2\colon \ Y -X -B_1=0, \qquad && L_5\colon \ Y-X+A_0-B_1 = 0,&\\
&L_3\colon \ X = 0, \qquad && L_6\colon \ X+A_1 = 0,&
\end{alignat*}
which is depicted in Fig.~\ref{A2A1fig}.

The group of transformations preserving these spiral diagrams is of the af\/f\/ine Weyl type
\begin{gather*}
W\big(A_2^{(1)}+A_1^{(1)}\big) = \big\langle s_0, s_1 , s_2 , r_0, r_1\colon \, s_i^2 = r_i^2 = (s_i s_{i+1})^3\big\rangle.
\end{gather*}
We have a natural $A_2^{(1)}$ group acting on the pairs of lines opposite to each other, in particular, if we denote the piecewise linear transformation that shears the space between two lines (as in Fig.~\ref{sABP12}), $L_i$ and $L_j$, by $\sigma_{i,j}$, then we let $s_0 = \sigma_{2,5}$, $s_1 = \sigma_{3,6}$ and $s_2 = \sigma_{1,4}$. The action of these elements may be written as
\begin{gather*}
s_0 \colon \ X \to X + \max(X+B_1, Y) - \max(B_1+X,A_0 + Y), \\
s_0 \colon \ Y \to A_0 + Y + \max(X+B_1, Y) - \max(B_1+X,A_0 + Y), \\
s_1 \colon \ X \to A_1 + X, \qquad s_1\colon \ Y \to Y+\max(0,A_1+X) - \max(0,X),\\
s_2 \colon \ X \to X+ \max(A_2,Y)-A_2 - \max(0,Y), \qquad s_2\colon \ Y \to Y-A_2,
\end{gather*}
where the action on the parameters is
\begin{gather*}
s_i \colon \ A_i \to -A_i, \qquad  s_i \colon \ A_j \to A_j - 2 A_i.
\end{gather*}
The action of the $W\big(A_1^{(1)}\big) = \langle r_0, r_1 \rangle$ component is as follows:
\begin{gather*}
r_0\colon \ X \to X  + \max( X, X + Y-A_2, Y-A_1-A_2)\\
\hphantom{r_0\colon \ X \to X}{} - \max(X,A_0-B_1 + Y,A_0 + A_1-B_1+ X+Y),\\
r_0\colon \ Y \to Y  + \max( X, X + Y-A_2, Y-A_1-A_2)\\
\hphantom{r_0\colon \ Y \to Y}{}
- \max(X,A_0-B_1 + Y,A_0 + A_1-B_1+ X+Y),\\
r_1\colon \ X \to X  + \max( X+B_1, Y, B_1)- \max(X,Y,0),\\
r_1\colon \ Y \to Y  + \max( X, Y-B_1, -B_1)- \max(X,Y,0).
\end{gather*}
The Dynkin diagram automorphisms comprise of a ref\/lection around $L_2$, which swaps~$L_1$ with~$L_3$ and another that swaps~$L_4$ with~$L_6$, and hence, are given by
\begin{alignat*}{3}
& p_1 \colon \  A_{0,1,2} \to A_{1,2,0}, \qquad && p_1 \colon \  B_{0,1} \to B_{1,0}, & \\
&p_1 \colon \ X \to Y-A_2, \qquad &&p_1 \colon \ Y \to A_0-B_1 + Y-X,&\\
&p_2 \colon \ A_{0,1,2} \to -A_{0,2,1},\qquad & & p_2 \colon \ B_{0,1} \to -B_{1,0}, &\\
&p_2 \colon \ X \to A_2-Y, \qquad & &p_2 \colon \ Y \to -A_1-X.&
\end{alignat*}
These generators have been chosen to coincide with the original presentation of Noumi et al.~\cite{KNY:qP4}. The transformations~$p_1$ and~$p_2$ satisfy the relations
\begin{gather*}
p_1^3 = p_2^2 = p_1^{-1} \circ s_{i+1} \circ p_1 \circ s_i = p_2 \circ r_{i+1} \circ p_1 \circ r_i = 1.
\end{gather*}
We f\/ind the value of $Q$ is
\begin{gather*}
Q = A_0 + A_1 + A_2 = B_0 + B_1,
\end{gather*}
which, when $Q = 0$, gives invariant curves arising as the level sets of
\begin{gather*}
H(X,Y) = \max( -A_1 - B_1 - X, X, A_2-Y,A_2 + X-Y,
  Y-B_1, Y-A_1-B_1-X).
\end{gather*}
We have two distinct evolution equations corresponding to dif\/ferent lattice directions. Firstly, we have the translation $T_1 = p_1 \circ s_2 \circ s_1$, which sends $(X,Y)$ to $(\tilde{X},\tilde{Y})$, related via
\begin{gather*}
\tilde{X} - Y =  B_0 + \max(B_1+X,Y) - \max(X+A_1 +A_2,Y+B_0),\\
\tilde{Y} - Y + X =  A_0 + A_2 - B_1 + \max(\tilde{X},0) - \max(\tilde{X},A_0+A_2),\\
T_1\colon \ A_0  \to A_0 + Q, \qquad  T_1\colon \ A_1 \to A_1 - Q,
\end{gather*}
which corresponds to a version of u-$\mathrm{P}_{\rm III}$ \cite{KNY:qP4}. Secondly, we have $T_2 = p_2 \circ r_0$, which sends $(X,Y)$ to $(\hat{X},\hat{Y})$, where
\begin{gather*}
\hat{X} + Y  = B_1 + A_2 + \max(0,X,Y) - \max(0,Y,X+B_1),\\
\hat{Y} + X  = A_0 - B_0 + \max(0,X,Y) - \max(B_1,Y,X+B_1),\\
T_2 \colon \ B_0 \to B_0 + Q, \qquad  T_2\colon \ B_1 \to B_1 - Q.
\end{gather*}
which corresponds to a version of u-$\mathrm{P}_{\rm IV}$ \cite{KNY:qP4}.

\subsection{Septagons}\label{A4}

\begin{figure}[t]
\centering
\begin{tikzpicture}[scale=.6]
\begin{scope}

\clip (-5,-5) rectangle (5,5);
\foreach \m in {-.5,0,...,13}
{
	\draw[blue!20] (0,-1-\m) -- (1,-1-\m)  -- (2+\m,0) -- (2+\m,2+\m)-- (.5,2+\m) -- (-\m-1,.5) -- (-\m-1,-.5) -- (0,-1.5-\m);
}
\draw[blue!20] (0,-.5) --(0,.5) -- (.5,1) -- (1,1) -- (1,0)--(.5,0) -- (0.5,.5);
\draw[blue,very thick] (0,-1-1) -- (1,-1-1) -- (2+1,0) -- (2+1,2+1)-- (.5,2+1) -- (-1-1,.5) -- (-1-1,-.5) -- (0,-1.5-1);
\draw[red] (.5,3)--(.5,5) node[black,fill=none,font=\scriptsize,midway,left] {$L_6$};
\draw[red] (3,3)--(5,5) node[black,fill=none,font=\scriptsize,midway,above] {$L_7$};
\draw[red] (1,-2)--(1,-6) node[black,fill=none,font=\scriptsize,midway,right] {$L_5$};
\draw[red] (0,-2)--(0,-6) node[black,fill=none,font=\scriptsize,midway,left] {$L_4$};
\draw[red] (-2,.5)--(-6,.5) node[black,fill=none,font=\scriptsize,midway,above] {$L_3$};
\draw[red] (-2,-.5)--(-6,-.5) node[black,fill=none,font=\scriptsize,midway,below] {$L_1$};
\draw[red] (3,0)--(6,0) node[black,fill=none,font=\scriptsize,midway,above] {$L_2$};
\end{scope}
\begin{scope}[scale=1.5,xshift=-3cm,yshift=-1.5cm]
\draw[red] (8,0) -- (10,0) -- (10,1) -- (9,2) -- (8,2)--cycle;
\filldraw[blue] (8,0) circle (.1);
\filldraw[blue] (9,0) circle (.1);
\filldraw[blue] (10,0) circle (.1);
\filldraw[blue!20] (11,0) circle (.1);
\filldraw[blue] (8,1) circle (.1);
\filldraw[blue] (9,1) circle (.1);
\filldraw[blue] (10,1) circle (.1);
\filldraw[blue] (8,2) circle (.1);
\filldraw[blue] (9,2) circle (.1);
\filldraw[blue!20] (8,3) circle (.1);
\end{scope}
\end{tikzpicture}

\caption{The spiral diagram for the case of the ultradiscrete Painlev\'e equation with $A_4^{(1)}$ symmetry. \label{A4fig}}
\end{figure}
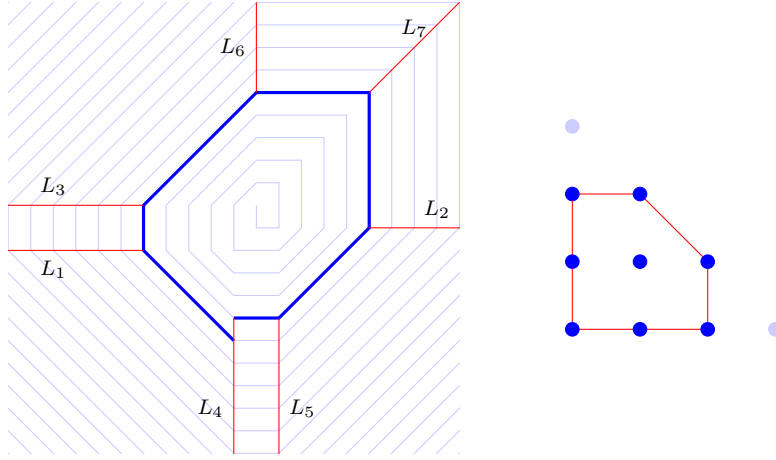

This case is associated with u-$\mathrm{P}_{\rm V}$ \cite{RTGO:ultimatediscretePs}. There are seven rays, specif\/ied as follows:
\begin{alignat*}{3}
&L_1\colon \ Y = 0, \qquad &&L_4\colon \ X=0, &\\
&L_2\colon \ Y - B_0 = 0, \qquad &&L_5\colon \ X-B_3=0,&\\
&L_3\colon \ Y + B_1 = 0,  \qquad &&L_6\colon \ X+B_4=0,&\\
& L_7\colon \ Y- X + B_1 + B_2 + B_3 + B_4 = 0,\qquad &&&
\end{alignat*}
which we depict in Fig.~\ref{A4fig}.

The group of transformations is of af\/f\/ine Weyl type
\begin{gather*}
W\big( A_4^{(1)} \big) = \langle s_0, \ldots, s_4 \rangle.
\end{gather*}
Rather than writing each relation, a presentation may be derived from the groups corresponding Dynkin diagram, which is shown below:
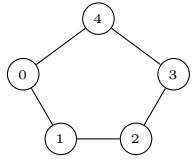
\begin{center}
\begin{tiny}
\begin{tikzpicture}
\node[draw, circle] (a41) at (0,0) {${}_{1}$};
\node[draw, circle] (a42) at (1,0) {${}_{2}$};
\node[draw, circle] (a43) at (1.5,.86) {${}_{3}$};
\node[draw, circle] (a44) at (.5,1.6) {${}_{4}$};
\node[draw, circle] (a40) at (-.5,.86) {${}_{0}$};
\draw (a41)--(a42);
\draw (a42)--(a43);
\draw (a43)--(a44);
\draw (a44)--(a40);
\draw (a40)--(a41);
\end{tikzpicture}
\end{tiny}
\end{center}

From the Dynkin diagram, the action of $s_i$ on is specif\/ied by
\begin{gather}
s_i\colon \  B_j =   \begin{cases}
-B_i & \text{if   $i = j$}, \\
B_j + B_i & \text{if $i \neq j$ and node $i$ is adjacent to node $j$}, \\
B_j & \text{otherwise}.
\end{cases}
\end{gather}
From this point, we will choose parameterizations of rays so that that the action of $s_i$ on the parameters determined by the Dynkin diagram in this way.

The f\/irst action is one that interchanges $L_1$ and $L_2$ by the piecewise linear shearing transformation
\begin{gather*}
s_0\colon \ X \to X + \max(Y,B_0) - B_0 - \max(0,Y), \qquad s_0\colon \ Y \to Y-B_0.
\end{gather*}
The second transformation is a simple translation,
\begin{gather*}
s_1\colon \ Y \to Y + B_1,
\end{gather*}
which has the ef\/fect of moving $L_3$ to $L_1$, hence, redef\/ining~$L_1$ and~$L_3$. The transformation~$s_2$ has the form
\begin{gather*}
 s_2\colon \ X \to X + \max(B_2,B_2+X,Y)-\max(0,X,Y),\\
 s_2\colon \ Y \to Y + \max(0,B_2+X,Y)- B_2 -\max(0,X,Y),
\end{gather*}
while $s_3$ simply is a translation in $X$ that redef\/ines $L_4$ and is given by
\begin{gather*}
 s_3\colon \ X \to X - B_3.
\end{gather*}
The last transformation is similar to $s_0$, but applied to the lines $L_4$ and $L_5$,
\begin{gather*}
s_4\colon \ X \to B_4 + X, \qquad s_4\colon \ Y \to \max(-X,B_4) - Y,
\end{gather*}
The Dynkin diagram automorphisms are generated by a rotation of the nodes
\begin{gather*}
p_1\colon \ B_i \to B_{i+1},\qquad
p_1\colon \ X \to \max(0,X,Y)-X-Y, \qquad
p_1\colon \ Y \to \max(0,X)-Y,
\end{gather*}
and a ref\/lection
\begin{gather*}
p_2 \colon \ B_{0,1,2,3,4} \to - B_{2,1,0,4,3},\qquad
p_2 \colon \ Y \to \max(0,X)-Y,
\end{gather*}
Tracing around the parameters provides the variable, $Q$, given by
\begin{gather*}
Q = B_0 + B_1 + B_2 + B_3 + B_4.
\end{gather*}
In the autonomous limit, when $Q =0$, the spiral diagram degenerates to a foliation by tropical cubic plane curves, specif\/ied by the level sets of
\begin{gather*}
H(X,Y) = \max\big( Y,\max(0,B_1)- B_1-B_4-X,Y-B_4-X,\\
\hphantom{H(X,Y) = \max\big(}{} \max(0,B_3)-B_1-B_3-B_4-Y,-B_1-B_4-X - Y,\\
\hphantom{H(X,Y) = \max\big(}{} X-B_1-B_3-B_4-Y\big).
\end{gather*}
The translation expressed as the composition
\begin{gather*}
T = s_4 \circ s_3 \circ s_2 \circ s_1 \circ p_1,
\end{gather*}
corresponds to the evolution equation
\begin{gather*}
\tilde{X} + X = B_3 + \max(0,\tilde{Y}) + \max(0,\tilde{Y}+B_1) - \max(B_0,Q+\tilde{Y}),\\
\tilde{Y} + Y = -B_1 - B_3 + \max(0,X) + \max(A_3,X) - \max(0,X+B_4),\\
T \colon \ A_0 \to A_0 - Q, \qquad  T\colon \ A_4 \to A_4 + Q,
\end{gather*}
which is known as the ultradiscrete version of the f\/ifth Painlev\'e equation \cite{RTGO:ultimatediscretePs}.

\subsection{Octagons}\label{D5} 
\begin{figure}[t]
\centering
\begin{tikzpicture}[scale=.6]
\begin{scope}
\clip (-5,-5) rectangle (5,5);
\foreach \m in {-.5,0,...,13}
{
	\draw[blue!20] (0,-1-\m) -- (1,-1-\m) -- (1.5,-.5-\m) -- (2+\m,1+\m) -- (2+\m,2+\m)-- (.5,2+\m) -- (-\m-1,.5) -- (-\m-1,-.5) -- (0,-1.5-\m);
}
\draw[blue!20] (0,-.5) --(0,.5) -- (.5,1) -- (1,1) -- (1,0)--(.5,0) -- (0.5,.5);
\draw[blue,very thick] (0,-1-1) -- (1,-1-1) -- (1.5,-.5-1) -- (2+1,1+1) -- (2+1,2+1)-- (.5,2+1) -- (-1-1,.5) -- (-1-1,-.5) -- (0,-1.5-1);
\draw[red] (.5,3)--(.5,5) node[black,fill=none,font=\scriptsize,midway,left] {$L_4$};
\draw[red] (3,3)--(5,5) node[black,fill=none,font=\scriptsize,midway,above] {$L_5$};
\draw[red] (3,2)--(6,5) node[black,fill=none,font=\scriptsize,midway,below] {$L_4$};
\draw[red] (1,-2)--(1,-6) node[black,fill=none,font=\scriptsize,midway,left] {$L_2$};
\draw[red] (0,-2)--(0,-6) node[black,fill=none,font=\scriptsize,midway,left] {$L_1$};
\draw[red] (1.5,-1.5)--(1.5,-6) node[black,fill=none,font=\scriptsize,midway,right] {$L_3$};
\draw[red] (-2,.5)--(-6,.5) node[black,fill=none,font=\scriptsize,midway,above] {$L_7$};
\draw[red] (-2,-.5)--(-6,-.5)  node[black,fill=none,font=\scriptsize,midway,below] {$L_8$};
\end{scope}
\begin{scope}[scale=1.5,xshift=-3cm,yshift=-1.5cm]
\draw[red] (8,0) -- (11,0) -- (9,2) -- (8,2)--cycle;
\filldraw[blue] (8,0) circle (.1);
\filldraw[blue] (9,0) circle (.1);
\filldraw[blue] (10,0) circle (.1);
\filldraw[blue] (11,0) circle (.1);
\filldraw[blue] (8,1) circle (.1);
\filldraw[blue] (9,1) circle (.1);
\filldraw[blue] (10,1) circle (.1);
\filldraw[blue] (8,2) circle (.1);
\filldraw[blue] (9,2) circle (.1);
\filldraw[blue!20] (8,3) circle (.1);
\end{scope}
\end{tikzpicture}

\caption{The spiral diagram for the case of the ultradiscrete Painlev\'e equation with $D_5^{(1)}$ symmetry. \label{D5fig}}
\end{figure}
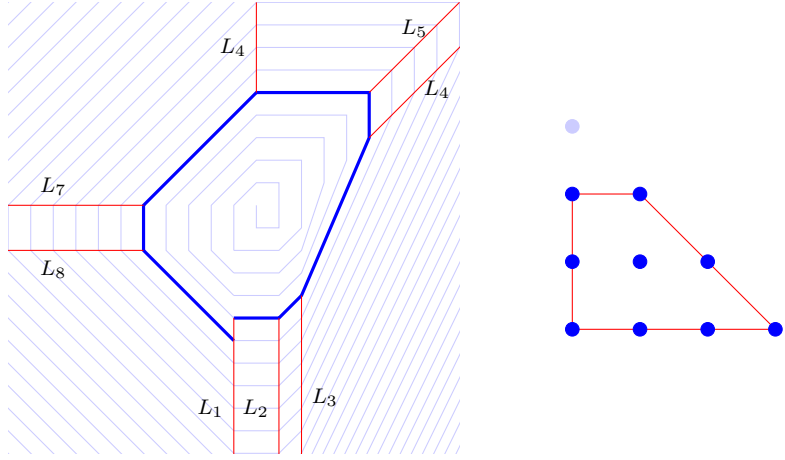

The biquadratic invariants obtained in the autonomous limit of $q$-$\mathrm{P}_{\rm VI}$ in Section~\ref{sec:trop} are not naturally mapped to cubic plane curves. However, under a simple transformation, we can present an equivalent system based on octagons arising as cubic plane curves whose rays are parameterized as follows:
\begin{alignat*}{3}
&L_1\colon \ X = 0, \qquad&&  L_5\colon \ Y=0, &\\
&L_2\colon \ X - B_2 =0, \qquad && L_6\colon \ Y+B_5 =0,&\\
&L_3\colon \ X - B_1-B_2 =0, \qquad && L_7\colon \ Y-X-B_3 =0,&\\
&L_4\colon \  X + B_0 =0 , \qquad && L_8\colon \ Y-X-B_3- B_4 =0,&
\end{alignat*}
which is depicted in Fig.~\ref{D5fig}

The group of transformations preserving these spiral diagrams is of af\/f\/ine Weyl type
\begin{gather*}
W\big( D_5^{(1)} \big) = \langle s_0, \ldots, s_5 \rangle.
\end{gather*}
A presentation may be derived from the Dynkin diagram below:
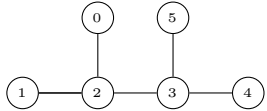
\begin{center}
\begin{tiny}
\begin{tikzpicture}
\node[draw, circle] (d51) at (0,0) {${}_{1}$};
\node[draw, circle] (d52) at (1,0) {${}_{2}$};
\node[draw, circle] (d53) at (2,0) {${}_{3}$};
\node[draw, circle] (d54) at (3,0) {${}_{4}$};
\node[draw, circle] (d50) at (1,1) {${}_{0}$};
\node[draw, circle] (d55) at (2,1) {${}_{5}$};
\draw (d51) -- (d52);
\draw (d51) -- (d52);
\draw (d52) -- (d53);
\draw (d53) -- (d54);
\draw (d50) -- (d52);
\draw (d55) -- (d53);
\end{tikzpicture}
\end{tiny}
\end{center}

We analogously specify the generators as we did before, where we denote the generators (in terms of $\sigma_{i,j}$ which swaps $L_i$ with $L_j$),
\begin{alignat*}{4}
& s_0  = \sigma_{1,4}, \qquad && s_1  = \sigma_{3,4}, \qquad && s_2 = \sigma_{1,3},& \\
& s_3  = \sigma_{5,7}, \qquad && s_4  = \sigma_{7,8}, \qquad && s_5 = \sigma_{5,6}.&
\end{alignat*}
These generators may be written
\begin{gather*}
s_0\colon \ X \to X + B_0, \qquad  s_0\colon \ Y \to Y+ \max(0,X+ B_0)-\max(0,X),\\
s_2\colon \ X \to X-B_2, \qquad  s_5\colon \ Y \to Y + B_5,\\
s_3\colon \ X\to X+ \max(B_3 + \max(0,X), Y) - \max(0,X,Y),\\
s_3\colon \ Y\to X+ \max(0,X + B_3, Y) - \max(0,X,Y) - B_3 ,
\end{gather*}
and the Dynkin diagram automorphisms, $p_1$ and $p_2$, are
\begin{alignat*}{3}
&p_1\colon \ B_{0,1,2,3,4,5} \to - B_{5,4,3,2,1,0},\qquad &&&\\
&p_1\colon \ X \to \max(0,X)-Y,\qquad  & &p_1\colon \ Y \to \max(0,X,Y)-X-Y,&  \\
&p_2\colon \ B_{0,1,2,3,4,5} \to - B_{0,1,2,3,5,4},\qquad &&& \\
&p_2\colon \ X \to -X, \qquad && p_2\colon \ Y \to Y-X-B_3,&
\end{alignat*}
Tracing around the particular values gives us the variable
\begin{gather*}
Q = B_0+B_1+2B_2+2B_3+B_4+B_5
\end{gather*}
In particular, when $Q=0$, we obtain the invariant
\begin{gather*}
H(X,Y) =  \max\big(\max(0,-B_5)-X, Y-X, \max(0,-B_1,-B_1-B_2)-B_5-Y,  \\
\hphantom{H(X,Y) =  \max\big(}{} B_0 + Y,\max(0,-B_1,-B_1-B_2) + X-B_2-B_5-Y,-b_5-X-Y,\\
\hphantom{H(X,Y) =  \max\big(}{} B_0+B_3+\max(0,B_4)+X , 2X-Y-B_1-2B_2-B_5\big).
\end{gather*}
The usual translation that is associated with the dynamics of u-$\mathrm{P}_{\rm VI}$ and the symmetry QRT equation is the action of
\begin{gather*}
T = p_2 \circ p_1 \circ p_2 \circ s_1 \circ s_2 \circ s_3\circ s_5 \circ s_4 \circ s_3 \circ s_2 \circ s_1.
\end{gather*}
To express the evolution of this system in a manner closer to that of \eqref{uP6}, we invert the transformation that was used to express the invariant as a tropical cubic curve. This is done by letting
\begin{gather*}
Z = \max(0,X) + \max(0,X-B_2) - \max(0,X+B_0) - Y,
\end{gather*}
which means the evolution in terms of $X$ and $Z$ is expressed as
\begin{gather*}
X+\tilde{X} = B_2  - B_5 + \max(0,Z) - \max(0, Z + B_0+B_2+B_3) \\
\hphantom{X+\tilde{X} =}{} + \max(B_5,Z) - \max(0, Z + B_0+B_2+B_3 +B_4), \\
Z+\tilde{Z} = B_5  - B_2 + \max(0,\tilde{X}) - \max(0, \tilde{X} + Q+B_0)\\
\hphantom{Z+\tilde{Z} =}{} +\max(B_2,\tilde{X}) - \max(0, \tilde{X} + Q-B_1-B_2),
\end{gather*}
which is equivalent to \eqref{uP6} above.

\subsection{Enneagons} 
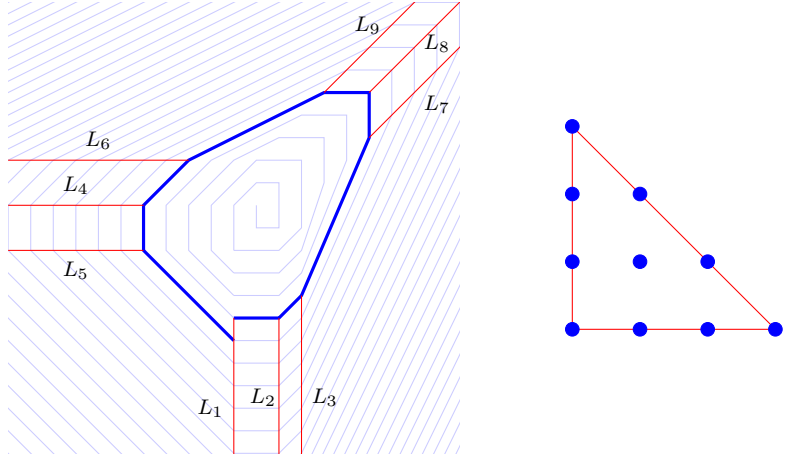
\begin{figure}[t]
\centering
\begin{tikzpicture}[scale=.6]
\begin{scope}
\clip (-5,-5) rectangle (5,5);
\foreach \m in {-.5,0,...,13}
{
	\draw[blue!20] (0,-1-\m) -- (1,-1-\m) -- (1.5,-.5-\m) -- (2+\m,1+\m) -- (2+\m,2+\m)--(1+\m,2+\m) -- (-\m,1.5) -- (-\m-1,.5) -- (-\m-1,-.5) -- (0,-1.5-\m);
}
\draw[blue!20] (0,-.5) --(0,.5) -- (.5,1) -- (1,1) -- (1,0)--(.5,0) -- (0.5,.5);
\draw[blue,very thick] (0,-1-1) -- (1,-1-1) -- (1.5,-.5-1) -- (2+1,1+1) -- (2+1,2+1)--(1+1,2+1) -- (-1,1.5) -- (-1-1,.5) -- (-1-1,-.5) -- (0,-1.5-1);
\draw[red] (2,3)--(5,6) node[black,fill=none,font=\scriptsize,midway,left=.01cm] {$L_9$};
\draw[red] (3,3)--(6,6)node[black,fill=none,font=\scriptsize,midway,below=-.01cm] {$L_8$};
\draw[red] (3,2)--(6,5) node[black,fill=none,font=\scriptsize,midway,below=.2cm] {$L_7$};
\draw[red] (1,-2)--(1,-6)  node[black,fill=none,font=\scriptsize,midway,above left=-.1cm] {$L_2$};
\draw[red] (0,-2)--(0,-6)  node[black,fill=none,font=\scriptsize,midway,left] {$L_1$};
\draw[red] (1.5,-1.5)--(1.5,-6) node[black,fill=none,font=\scriptsize,midway,right] {$L_3$};
\draw[red] (-2,.5)--(-5,.5)  node[black,fill=none,font=\scriptsize,midway,above=.01cm] {$L_4$};
\draw[red] (-2,-.5)--(-5,-.5) node[black,fill=none,font=\scriptsize,midway,below=.01cm] {$L_5$};
\draw[red] (-1,1.5)--(-5,1.5) node[black,fill=none,font=\scriptsize,midway,above=.01cm] {$L_6$};
\end{scope}
\begin{scope}[scale=1.5,xshift=-3cm,yshift=-1.5cm]
\draw[red] (8,0) -- (11,0) -- (8,3)--cycle;
\filldraw[blue] (8,0) circle (.1);
\filldraw[blue] (9,0) circle (.1);
\filldraw[blue] (10,0) circle (.1);
\filldraw[blue] (11,0) circle (.1);
\filldraw[blue] (8,1) circle (.1);
\filldraw[blue] (9,1) circle (.1);
\filldraw[blue] (10,1) circle (.1);
\filldraw[blue] (8,2) circle (.1);
\filldraw[blue] (9,2) circle (.1);
\filldraw[blue] (8,3) circle (.1);
\end{scope}
\end{tikzpicture}

\caption{The spiral diagram for the case of the ultradiscrete Painlev\'e equation with $E_6^{(1)}$ symmetry. \label{E6fig}}
\end{figure}

It is at this point we go beyond the QRT maps def\/ined by biquadratic cases \cite{Nobe:QRT, OR2}. We exploit the translational freedom to parameterize two rays coincide with the $y$-axis and $x$-axis respectively. The remaining ray are parameterized as follows:
\begin{alignat*}{3}
&L_1\colon \ X = 0, \qquad && L_6\colon \ Y + B_0+B_6=0, &\\
&L_2\colon \ X - B_2=0, \qquad && L_7\colon \ Y-X + B_3 =0,&\\
&L_3\colon \ X - B_1 - B_2=0,\qquad && L_8\colon \ Y-X +B_3 + B_4 =0,& \\
&L_4\colon \ Y = 0, \qquad &&  L_9\colon \ Y- X + B_3 + B_4 + B_5=0,&\\
&L_5\colon \ Y +B_6=0.\qquad &&&
\end{alignat*}
The relevant spiral diagram is of irregular enneagons, depicted in Fig.~\ref{E6fig}.

The group of transformations are of af\/f\/ine Weyl type
\begin{gather*}
W\big( E_6^{(1)} \big) = \langle s_0, \ldots, s_6 \rangle.
\end{gather*}
The presentation, and action on the parameters, is specif\/ied by the Dynkin diagram below:
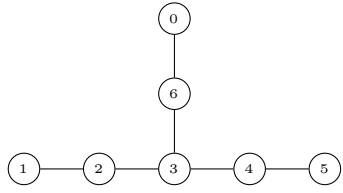
\begin{center}
\begin{tiny}
\begin{tikzpicture}
\node[draw, circle] (e61) at (0,0) {${}_{1}$};
\node[draw, circle] (e62) at (1,0) {${}_{2}$};
\node[draw, circle] (e63) at (2,0) {${}_{3}$};
\node[draw, circle] (e64) at (3,0) {${}_{4}$};
\node[draw, circle] (e65) at (4,0) {${}_{5}$};
\node[draw, circle] (e66) at (2,1) {${}_{6}$};
\node[draw, circle] (e60) at (2,2) {${}_{0}$};
\draw (e61) -- (e62);
\draw (e62) -- (e63);
\draw (e63) -- (e64);
\draw (e64) -- (e65);
\draw (e66) -- (e63);
\draw (e66) -- (e60);
\end{tikzpicture}
\end{tiny}
\end{center}

\begin{figure}[t]
\centering
\begin{tikzpicture}[scale=.8]
\begin{scope}
\clip (-4,-4) rectangle (4,4);
\draw[red] (0,-1-1.5) -- (.5,-1-1.5) -- (1.5,-.5-1) -- (2+1,1+1) -- (2+1,2+1)--(1+1,2+1) -- (-1,1.5) -- (-1-1,.5) -- (-1-1,-.5) -- (0,-1.5-1);
\draw[blue] (0,-1-1) -- (1,-1-1) -- (1.5,-.5-1) -- (2+1,1+1) -- (2+1,2+1)--(1+1,2+1) -- (-1,1.5) -- (-1-1,.5) -- (-1-1,-.5) -- (0,-1.5-1);
\draw[red] (2,3)--(5,6);
\draw[red] (3,3)--(6,6);
\draw[red] (3,2)--(6,5) ;
\draw[red] (1,-2)--(1,-4)  node[black,fill=none,font=\scriptsize,midway,above right=-.1cm] {${}_{L_i}$};
\draw[red,very thick] (.5,-2.5)--(.5,-5)  node[black,fill=none,font=\scriptsize,midway,above left=-.1cm] {${}_{T(L_i)}$};
\draw[red] (0,-2)--(0,-6);
\draw[red] (1.5,-1.5)--(1.5,-6) node[black,fill=none,font=\scriptsize,midway,right] {${}_{L_j}$};
\draw[red] (-2,.5)--(-5,.5);
\draw[red] (-2,-.5)--(-5,-.5);
\draw[red] (-1,1.5)--(-5,1.5);
\fill[blue] (-1.5,-1) circle (.1);
\draw[dashed] (.5,-6) -- (.5,1) -- (4.5,5);
\draw[dashed] (-.5,-6) -- (-.5,-1) -- (4.5,4);
\draw[dashed] (.5,1) -- (-4.5,1);
\draw[dashed] (-.5,-1) -- (-4.5,-1);
\fill[red] (-1.5,1) circle (.1);
\fill[red] (2.5,3) circle (.1);
\fill[red] (3,2.5) circle (.1);
\fill[red] (-.5,-2) circle (.1);
\node at (-1.6,-1.5) {${}_{P}$};
\end{scope}
\begin{scope}[xshift=10cm]
\clip (-4,-4) rectangle (4,4);
\draw[red] (0,-1-1.5) -- (.5,-1-1.5) -- (1.5,-.5-1) -- (2+1,1+1) -- (2+1,2+1)--(1+1,2+1) -- (-1,1.5) -- (-1-1,.5) -- (-1-1,-.5) -- (0,-1.5-1);
\draw[blue] (0,-1-1) -- (1,-1-1) -- (1.5,-.5-1) -- (2+1,1+1) -- (2+1,2+1)--(1+1,2+1) -- (-1,1.5) -- (-1-1,.5) -- (-1-1,-.5) -- (0,-1.5-1);
\draw[red] (2,3)--(5,6);
\draw[red] (3,3)--(6,6);
\draw[red] (3,2)--(6,5) ;
\draw[red] (1,-2)--(1,-4)  node[black,fill=none,font=\scriptsize,midway,above right=-.1cm] {${}_{L_i}$};
\draw[red,very thick] (.5,-2.5)--(.5,-5)  node[black,fill=none,font=\scriptsize,midway,above left=-.1cm] {${}_{T(L_i)}$};
\draw[red] (0,-2)--(0,-6);
\draw[red] (1.5,-1.5)--(1.5,-6) node[black,fill=none,font=\scriptsize,midway,right] {${}_{L_j}$};
\draw[red] (-2,.5)--(-5,.5);
\draw[red] (-2,-.5)--(-5,-.5);
\draw[red] (-1,1.5)--(-5,1.5);
\draw[dashed] (1.5,-6) -- (1.5,1) -- (4.5,4);
\draw[dashed] (-4.5,1) -- (1.5,1);
\draw[dashed] (-.5,-6) -- (-.5,0) -- (4.5,5);
\draw[dashed] (-4.5,0) -- (-.5,0);
\fill[blue] (-2,0) circle (.1);
\fill[red] (-1.5,1) circle (.1);
\fill[red] (2.5,3) circle (.1);
\fill[red] (3,2.5) circle (.1);
\fill[red] (-.5,-2) circle (.1);
\node at (-2.7,-.3) {${}_{T(P)}$};
\end{scope}
\end{tikzpicture}

\caption{This is a pictorial represection of \eqref{upE6} where the dashed lines intersect the polygon at the four f\/ixed points (in red) and $P$ and $T(L_i)$ on the left and $T(P)$ and $L_j$ on the right.\label{grouplaw}}
\end{figure}
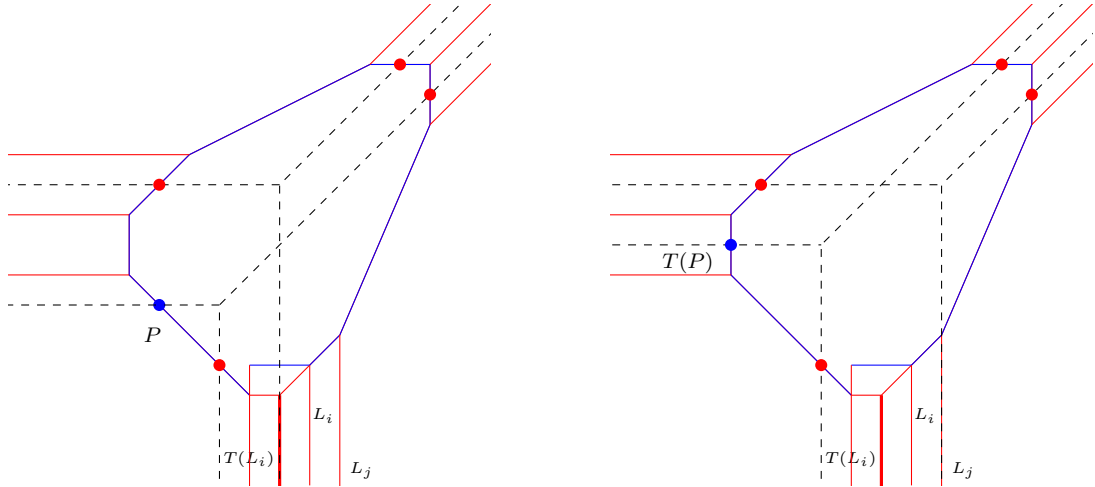

We may now parameterize the af\/f\/ine Weyl group actions by f\/irst specifying the generators that have a little ef\/fect on $X$ and $Y$, by letting $s_0 = \sigma_{5,6}$, $s_1 = \sigma_{2,3}$, $s_2 = \sigma_{1,2}$, $s_4 = \sigma_{7,8}$, $s_5 = \sigma_{8,9}$ and $s_6 = \sigma_{4,5}$. In these cases, the ef\/fect on $X$ and $Y$ are trivial, except for $s_2$ and $s_6$, which have the ef\/fect
\begin{gather*}
s_2\colon \ X \to X-B_2, \qquad  s_6\colon \ Y \to Y+ B_6.
\end{gather*}
The action of $s_3$ is given by
\begin{gather*}
 s_3\colon \ X \to X + \max(B_3,B_3+X,Y) - \max(0,X,Y),\\
 s_3\colon \ Y \to Y -B_3 + \max(0,B_3+X,Y) - \max(0,X,Y).
\end{gather*}
The Dynkin diagram automorphisms are given by
\begin{gather*}
p_1 \colon \ B_{0,1,2,3,4,5,6} \to -B_{5,1,2,3,6,0,4},\\
p_1 \colon \ X\to -X, \qquad  p_2\colon \ Y \to Y-X-B_3,\\
p_2 \colon \ B_{0,1,2,3,4,5,6} \to - B_{1,0,6,3,4,5,2},\\
p_2 \colon \ X \to Y, \qquad  p_2\colon \ Y \to X.
\end{gather*}
Tracing around the enneagon, we obtain the variable
\begin{gather*}
Q = B_0 + B_1 + 2B_2 + 3B_3+2B_4+B_5 + B_6.
\end{gather*}
In the autonomous limit, when $Q = 0$, this spiral diagram degenerates to give a f\/ibration by cubic plane curves, which may be expressed as the tropical curves that arise as the level sets of
\begin{gather*}
H(X,Y) = \max\big(2X-Y-B_1-B_2, \max(0,-B_2,-B_1-B_2)+X-Y-B_2,  \\
\hphantom{H(X,Y) ={}}{} \max(0,-B_5,-B_4-B_5)+X-B_1-2B_2-B_3, \max(0,-B_1,-B_2-B_2)-Y,\\
\hphantom{H(X,Y) ={}}{} Y + \max(0,-B_4,-B_4-B_5)-B_1-2B_2-2B_3-B_4, 2Y+B_0+2B_6-X, \\
\hphantom{H(X,Y) ={}}{} Y+B_6+ \max(0,B_0,B_0+ B_6) - X ,\max(0,B_6,B_0 + B_6)-X,-X - Y\big).
\end{gather*}
The translation that is associated with the dynamics of the discrete Painlev\'e equation in this case is given by
\begin{gather*}
T = p_1 \circ p_2\circ s_1 \circ s_2\circ s_3\circ s_4\circ s_6\circ s_0\circ s_3\circ s_2\circ s_1\circ s_6\circ s_3 \circ s_2\circ s_4\circ s_3\circ s_6\circ s_0.
\end{gather*}
Though the evolution equation is very complicated, the action can be evaluated quite easily by the geometric method in~\cite{10E9sol}. In the case of a tropical cubic genus one curve, the action of~$T$ can be describes as follows: we choose two two rays, say~$L_i$ and~$L_j$, and let~$T$ move~$L_i$ to the point in which the other rays and $T(L_i)$ def\/ine a pencil of tropical cubic genus one curves that foliate the plance, i.e., rather than spirals, we have closed curves. Any point, $P \in \mathbb{TP}_2$, is now on some closed genus one cubic curve,~$C$, in the pencil. We send~$P$ to~$T(P)$, so that~$T(P)$ satisf\/ies
\begin{gather}\label{upE6}
T(P)+T(L_i)=P + L_j,
\end{gather}
where we interpret $T(L_i)$ and $L_i$ in terms of the unique stable intersection of~$T(L_i)$ and~$L_j$ with~$C$ respectively and the addition is in accordance with the group law on~$C$ (see~\cite{Vigeland:grouplaw}). Finally we send~$L_j$ to a point in which~$T(L_j)$ satisf\/ies
\begin{gather*}
L_i + L_j = T(L_i) + T(L_j),
\end{gather*}
on $C$. We have illustrated this in Fig.~\ref{grouplaw}.

\subsection{Decagons} The rational surface of type $A_1^{(1)}$ was obtained by blowing up three points on a line and six on a quadratic curve \cite{Sakai:rational}. The resulting surface is rationally equivalent to the surface obtained by blowing up four points at lines at inf\/inity and two points at on the third line at inf\/inity. Hence, in the discrete setting, the underlying surface and the symmetries $W\big(E_7^{(1)}\big)$ obtained here are equivalent up to a rational transformation to those of \cite{Sakai:rational}.

\begin{figure}[t]
\centering
\begin{tikzpicture}[scale=.6]
\clip (-6,-2) rectangle (4,8);
\foreach \m in {-2,-1.5,...,13}
{
	\draw[blue!20] (-2,-1-\m) -- (0,-1-\m) -- (1,-\m) -- (2,2-\m) -- (2+\m/2+1,2+\m/2+3) -- (2+\m/2+1-2,2+\m/2+5) -- (-2-\m,6) -- (-4-\m,5)--(-5-\m,4) -- (-5-\m,1.5) -- (-2,-1.5-\m);
}
\draw[blue!20] (-2,1) -- (-2.5,1.5) -- (-2.5,4) -- (-1.5,5) -- (-0.5,5.5) -- (1.5,3.5) -- (1,2.5) -- (0,1.5) -- (-2,1.5) -- (-2,4) -- (-1,5) -- (1,3) -- (0,2) -- (-1.5,2) -- (-1.5,4) -- (0,2.5) -- (-1,2.5) -- (-1,3);
\def \m {-1};
\draw[blue,very thick] (-2,-1-\m) -- (0,-1-\m) -- (1,-\m) -- (2,2-\m) -- (2+\m/2+1,2+\m/2+3) -- (2+\m/2+1-2,2+\m/2+5) -- (-2-\m,6) -- (-4-\m,5)--(-5-\m,4) -- (-5-\m,1.5) -- (-2,-1.5-\m);
\draw [red] (-2,0) -- (-2,-6);
\draw [red] (0,0) -- (0,-6);
\draw [red] (1,1) -- (1,-6);
\draw [red] (2,3) -- (2,-6);
\draw [red] (2+\m/2+1-2,2+\m/2+5) --(2+\m/2+1-2+5,2+\m/2+5+5);
\draw [red] (2+\m/2+1,2+\m/2+3)-- (2+\m/2+1+5,2+\m/2+3+5);
\draw [red] (-4,1.5) -- (-4-4,1.5);
\draw [red] (-4,4) -- (-4-4,4);
\draw [red] (-3,5) -- (-4-4,5);
\draw [red] (-1,6) -- (-4-4,6);
\end{tikzpicture}

\caption{A spiral diagram for the $E_7^{(1)}$ case.\label{Fige7}}
\end{figure}
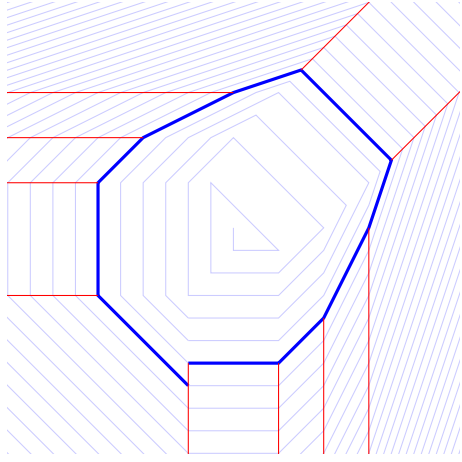

The most general tropical cubic plane curve is a enneagon, hence, to describe the decagon spirals, we need to consider spiral degenerations of quartic plane curves with two rays of order two. We choose to parameterize these rays as follows:
\begin{alignat*}{3}
&L_1 \colon \ X+B_4= 0, \qquad&& L_5\colon \ Y=0,&\\
&L_2 \colon \ X+B_4+B_5=0, \qquad&& L_6\colon \ Y-B_3=0,&\\
&L_3 \colon \ X+B_4+B_5+B_6=0, \qquad && L_7\colon \ Y-B_3-B_2=0,&\\
&L_4 \colon \ X+B_4+B_5+B_6+B_7=0,\qquad && L_8\colon \ Y-B_3-B_2-B_1= 0,&\\
&L_9 \colon \ Y-X - B_0=0, \qquad && L_{10}\colon \ Y-X = 0.&
\end{alignat*}
It should be noted that $L_9$ and $L_{10}$ are of order $2$. With these considerations, the resulting system of spiraling polygons is depicted in Fig.~\ref{Fige7}.

The resulting group of transformations is of af\/f\/ine Weyl type
\begin{gather*}
W\big(E_7^{(1)}\big) = \langle s_0, \ldots, s_7 \rangle.
\end{gather*}
This groups Dynkin diagram is below:
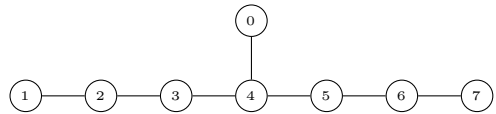
\begin{center}
\begin{tiny}
\begin{tikzpicture}
\node[draw, circle] (e71) at (0,0) {${}_{1}$};
\node[draw, circle] (e72) at (1,0) {${}_{2}$};
\node[draw, circle] (e73) at (2,0) {${}_{3}$};
\node[draw, circle] (e74) at (3,0) {${}_{4}$};
\node[draw, circle] (e75) at (4,0) {${}_{5}$};
\node[draw, circle] (e76) at (5,0) {${}_{6}$};
\node[draw, circle] (e77) at (6,0) {${}_{7}$};
\node[draw, circle] (e70) at (3,1) {${}_{0}$};
\draw (e71) -- (e72);
\draw (e72) -- (e73);
\draw (e73) -- (e74);
\draw (e74) -- (e75);
\draw (e75) -- (e76);
\draw (e76) -- (e77);
\draw (e74) -- (e70);
\end{tikzpicture}\end{tiny}
\end{center}

The ref\/lections are given by $s_0 = \sigma_{9,10}$, $s_1 = \sigma_{7,8}$, $s_2 = \sigma_{6,7}$, $s_3 = \sigma_{5,6}$, $s_5 = \sigma_{1,2}$, $s_6 = \sigma_{2,3}$ and $s_7 = \sigma_{3,4}$, hence, the nontrivial actions are given by
\begin{gather*}
s_0 \colon \ X \to X + B_0, \qquad s_3\colon \ Y \to Y+B_3, \qquad s_3\colon \ X \to X + B_3,\\
s_4\colon \ X \to X - B_4 + \max(B_4,X,Y) - \max(B_4,X,B_4+Y),\\
s_4\colon \ Y \to Y + \max(0,X,Y) - \max(B_4,X,B_4+Y).
\end{gather*}
We also have a single Dynkin diagram automorphism. This Dynkin diagram automorphism has the ef\/fect of sending~$X$ to~$-X$, and~$Y$ to~$-Y$, which swaps all the eight f\/irst order rays, however, it also has the ef\/fect of ref\/lecting the two rays,~$L_9$ and~$L_{10}$, in opposite direction, hence, we compose this a transformation of the form of~$\iota_A$ along~$L_9$ and~$L_{10}$, giving
\begin{gather*}
p_1\colon \ X \to Y+B_4 - \max(Y,X) - \max(Y,B_0+X),\\
p_1\colon \ Y \to X+B_0+B_4 - \max(Y,X) - \max(Y,B_0+X),\\
p_1\colon \ B_{0,1,2,3,4,5,6,7} \to B_{0,7,6,5,4,3,2,1}.
\end{gather*}
Tracing around the spirals reveals that
\begin{gather*}
Q = 2B_0 + B_1 + 2B_2 + 3B_3 + 4B_4 + 3B_5 + 2B_6+B_7,
\end{gather*}
which in the autonomous limit, when $Q= 0$, gives a closing of spirals to give the following invariant
\begin{gather*}
H(X,Y)=\max \big( 0,Y+\mu_1,2Y+\mu _2,X+\mu _5,2X+\mu _6,\\
\hphantom{H(X,Y)=\max \big(}{} \max(Y+\mu _3,X-B_0+\mu _7)+\max (Y,X+B_0)+\max (X,Y),\\
\hphantom{H(X,Y)=\max \big(}{} 2 \max (Y,X+B_0)+2 \max (X,Y)+\mu _4\big)
-X-Y,
\end{gather*}
where the values of $\mu_i$ are def\/ined by
\begin{gather*}
 \max(0,X) + \max(0,B_3+X)+ \max(0,B_3+B_2+X) \\
 \qquad {} + \max(0,B_3+B_2+B_1+X)  = \max(0,\mu_1 + X, \mu_2 + 2X, \mu_3 + 3X,\mu_4 + 4X),\\
\max(0,X-B_4) + \max(0,X-B_4-B_5)+ \max(0,X-B_4-B_5-B_6)\\
\qquad {} +\max(0,X-B_4-B_5-B_6-B_7)  = \max(0,\mu_5 + X, \mu_6 + 2X, \mu_7 + 3X,\mu_8 + 4X).
\end{gather*}
A translation associated with the ultradiscrete Painlev\'e equation is given by
\begin{gather*}
T = p_1 \circ s_1\circ s_2\circ s_3\circ s_4\circ s_0\circ s_5\circ s_4\circ s_3\circ s_2\circ s_1\circ s_6\circ s_5\circ s_4\circ s_0\\
\hphantom{T=}{} \circ s_3\circ s_2\circ s_4\circ s_3\circ s_5\circ s_4\circ s_0\circ s_6\circ s_5\circ s_4\circ s_3\circ s_2\circ s_1,
\end{gather*}
whose action on $\mathbb{T}^2$ is too complicated to be written here. However, the action of $T$ may be described by the theory of~\cite{10E9sol}, where the evolution takes the form
\begin{gather}\label{upE7}
T(P)+T(L_i)=P + L_j,
\end{gather}
where the addition here is def\/ined in terms of the group law on a tropical quartic genus one curve.

\subsection{Undecagon} The blow-up points in the original classif\/ication of Sakai~\cite{Sakai:rational} on~$\mathbb{P}_2$ lie on a single nodal cubic. This conf\/iguration is birationally equivalent (by a series of blow-ups and blow-downs) to a~conf\/iguration in~$\mathbb{P}_2$ in which there are three order two singularities on the line at $y = 0$, two order three singularities on the line $x = 0$ and six order one singularities on the line $z = 0$. Hence, in the discrete setting, the underlying surface and the symmetries of af\/f\/ine Weyl type~$E_8^{(1)}$ obtained here are equivalent up to a rational transformation to those of~\cite{Sakai:rational}.

\begin{figure}[t]
\centering
\begin{tikzpicture}[scale=.4]
\clip (-7,-5) rectangle (12,13);
\draw[blue!20] (1,1.5)--(1,2)-- (.75,2)--(.41666,1)--(.8,0.333)--(1.1666,0.5)--(1.66,1.66)--(1.66,2.66)--(1,2.66)--(.5,2.333)--(-0.1,1)--(.66,-.5)-- (1,-.1666)--(2,1.333)-- (2.5,2.5)--(2.5,3.5)-- (1.5,3.5)--(0.1666,2.6666)-- (-.6667,1)--(.333,-1)--(1,-.6667)--(2.5,1.5)--(3.5,3.5)--(3.5,4.5)--(2.5,4.5)-- (-.166666,3)--(-1.16666,1)--(0,-1.5) -- (1.,-1.16667) -- (1.5,-0.5) -- (3.5,2.5) -- (4.5,4.5) -- (4.5,5.5) -- (3.5,5.5) -- (1.5,4.5) -- (-0.75,3.) -- (-1.75,1) -- (-0.375,-1.75) -- (0,-2) -- (1,-1.66667) -- (3.,1.) -- (5.,4.) -- (6.,6.) -- (6.,7.) -- (5.,7.) -- (3.,6.) -- (-1.5,3.) -- (-2.5,1.) -- (-1.125,-1.75) -- (0.,-2.5) -- (1.,-2.16667) --(4.5,2.5) -- (6.5,5.5) -- (7.5,7.5) -- (7.5,8.5) -- (6.5,8.5) -- (4.5,7.5) -- (-2.25,3.) -- (-3.25,1.) -- (-1.875,-1.75) -- (0.,-3.) -- (1.,-2.66667) -- (6.,4.) -- (8.,7.) -- (9.,9.) -- (9.,10.) -- (8.,10.) -- (6.,9.) -- (-3.,3.) -- (-4.,1.) -- (-2.625,-1.75) -- (0.,-3.5) -- (1.,-3.16667) -- (7.5,5.5) -- (9.5,8.5) -- (10.5,10.5) -- (10.5,11.5) -- (9.5,11.5) -- (7.5,10.5) -- (-3.75,3.) -- (-4.75,1.) -- (-3.375,-1.75) -- (0.,-4.) -- (1.,-3.66667) -- (9.,7.) -- (11.,10.) -- (12.,12.) -- (12.,13.) -- (11.,13.) -- (9.,12.) -- (-4.5,3.) -- (-5.5,1.) -- (-4.125,-1.75) -- (0.,-4.5) -- (1.,-4.16667) -- (10.5,8.5) -- (12.5,11.5) -- (13.5,13.5) -- (13.5,14.5) -- (12.5,14.5) -- (10.5,13.5) -- (-5.25,3.) -- (-6.25,1.) --  (-4.875,-1.75) -- (0.,-5.) -- (1.,-4.66667) -- (12.,10.) -- (14.,13.) -- (15.,15.) -- (15.,16.) -- (14.,16.) -- (12.,15.) -- (-6.,3.) --  (-7.,1.) -- (-5.625,-1.75) -- (0.,-5.5) -- (1.,-5.16667) --  (13.5,11.5) -- (15.5,14.5) -- (16.5,16.5) -- (16.5,17.5) -- (15.5,17.5) -- (13.5,16.5) -- (-6.75,3.) -- (-7.75,1.) --  (-6.375,-1.75) -- (0.,-6.) -- (1.,-5.66667) -- (15.,13.) -- (17.,16.) -- (18.,18.) -- (18.,19.) -- (17.,19.) -- (15.,18.) -- (-7.5,3.) -- (-7.125,-1.75) -- (0.,-6.5) -- (1.,-6.16667) --  (16.5,14.5) -- (18.5,17.5) -- (19.5,19.5) -- (19.5,20.5) --  (18.5,20.5) -- (16.5,19.5) -- (-8.25,3.) -- (-7.875,-1.75) -- (0.,-7.) -- (1.,-6.66667) -- (18.,16.) -- (20.,19.) -- (21.,21.) -- (21.,22.) -- (20.,22.) -- (18.,21.) -- (-9.,3.) -- (-8.625,-1.75) -- (0.,-7.5) -- (1.,-7.16667) -- (19.5,17.5) -- (21.5,20.5) -- (22.5,22.5) -- (22.5,23.5) --(21.5,23.5) -- (19.5,22.5) -- (-9.75,3.)  -- (-9.375,-1.75) -- (0.,-8.) -- (1.,-7.66667) -- (21.,19.) -- (23.,22.) -- (24.,24.) -- (24.,25.) -- (23.,25.) -- (21.,24.) -- (-10.5,3.)  -- (-10.125,-1.75) -- (0.,-8.5) -- (1.,-8.16667) -- (22.5,20.5) -- (24.5,23.5) -- (25.5,25.5) -- (25.5,26.5) -- (24.5,26.5) -- (22.5,25.5) -- (-11.25,3.)  -- (-10.875,-1.75) -- (0.,-9.) -- (1.,-8.66667) -- (24.,22.) -- (26.,25.) -- (27.,27.) -- (27.,28.) -- (26.,28.) -- (24.,27.) -- (-12.,3.) -- (-11.625,-1.75) -- (0.,-9.5) -- (1.,-9.16667) -- (25.5,23.5) -- (27.5,26.5) -- (28.5,28.5) -- (28.5,29.5) -- (27.5,29.5) -- (25.5,28.5) -- (-12.75,3.) -- (-13.75,1.) -- (-12.375,-1.75) -- (0.,-10.) -- (1.,-9.66667) -- (27.,25.) -- (29.,28.) -- (30.,30.) -- (30.,31.) -- (29.,31.) -- (27.,30.) -- (-13.5,3.) -- (-14.5,1.) -- (-13.125,-1.75) -- (0.,-10.5) -- (1.,-10.1667) -- (28.5,26.5) -- (30.5,29.5) -- (31.5,31.5) -- (31.5,32.5) -- (30.5,32.5) -- (28.5,31.5) -- (-14.25,3.) -- (-15.25,1.) -- (-13.875,-1.75) -- (0.,-11.) -- (1.,-10.6667) -- (30.,28.) -- (32.,31.) -- (33.,33.) -- (33.,34.) -- (32.,34.) -- (30.,33.) -- (-15.,3.) -- (-16.,1.) -- (-14.625,-1.75) -- (0.,-11.5) -- (1.,-11.1667) -- (31.5,29.5) -- (33.5,32.5) -- (34.5,34.5) -- (34.5,35.5) -- (33.5,35.5) -- (31.5,34.5) -- (-15.75,3.) -- (-16.75,1.) -- (-15.375,-1.75) -- (0.,-12.) -- (1.,-11.6667) -- (33.,31.) -- (35.,34.) -- (36.,36.) -- (36.,37.) -- (35.,37.) -- (33.,36.) -- (-16.5,3.) -- (-17.5,1.) -- (-16.125,-1.75) -- (0.,-12.5) -- (1.,-12.1667) -- (34.5,32.5) -- (36.5,35.5) -- (37.5,37.5) -- (37.5,38.5) -- (36.5,38.5) -- (34.5,37.5) -- (-17.25,3.) -- (-18.25,1.) -- (-16.875,-1.75) -- (0.,-13.) -- (1.,-12.6667) -- (36.,34.) -- (38.,37.) --(39.,39.) -- (39.,40.) -- (38.,40.) -- (36.,39.) -- (-18.,3.) -- (-19.,1.) -- (-17.625,-1.75) -- (0.,-13.5) -- (1.,-13.1667) -- (37.5,35.5) -- (39.5,38.5) -- (40.5,40.5) -- (40.5,41.5) -- (39.5,41.5) -- (37.5,40.5) -- (-18.75,3.) -- (-19.75,1.) -- (-18.375,-1.75) -- (0.,-14.) -- (1.,-13.6667) -- (39.,37.) -- (41.,40.) -- (42.,42.) -- (42.,43.) -- (41.,43.) -- (39.,42.) -- (-19.5,3.) -- (-20.5,1.) -- (-19.125,-1.75) -- (0.,-14.5) -- (1.,-14.1667) -- (40.5,38.5) -- (42.5,41.5) -- (43.5,43.5) -- (43.5,44.5) -- (42.5,44.5) -- (40.5,43.5) -- (-20.25,3.) -- (-21.25,1.) -- (-19.875,-1.75) -- (0.,-15.) -- (1.,-14.6667) -- (42.,40.);
\draw[blue,very thick] (0.,-3.) -- (1.,-2.66667) -- (6.,4.) -- (8.,7.) -- (9.,9.) -- (9.,10.) -- (8.,10.) -- (6.,9.) -- (-3.,3.) -- (-4.,1.) -- (-2.625,-1.75) -- (0.,-3.5);
\draw[blue!20]  (-7,12.333)-- (-6,13);
\draw[blue!20]  (8.66,-5)-- (12,-0.33);
\draw[blue!20]  (9.0,-5)-- (12,-0.83);
\draw[blue!20]  (9.33,-5)-- (12,-1.33);
\draw[blue!20]  (9.66,-5)-- (12,-1.83);
\draw[blue!20]  (10,-5)-- (12,-2.33);
\draw[blue!20]  (10.33,-5)-- (12,-2.83);
\draw[blue!20]  (10.66,-5)-- (12,-3.33);
\draw[blue!20]  (11,-5)-- (12,-3.83);
\draw[blue!20]  (11.33,-5)-- (12,-4.25);
\draw[blue!20]  (11.66,-5)-- (12,-4.66);
\draw[red] (0,-5) -- (0,-3)  node[black,fill=none,font=\scriptsize,midway,left] {$L_1$};
\draw[red] (1,-5) -- (1,-2.6667)  node[black,fill=none,font=\scriptsize,midway,right] {$L_2$};
\draw[red] (-2.625,-1.75) -- (-7,-1.75)  node[black,fill=none,font=\scriptsize,midway,above] {$L_4$};
\draw[red] (-4,1) -- (-7,1)  node[black,fill=none,font=\scriptsize,midway,above] {$L_5$};
\draw[red] (-3,3) -- (-7,3)  node[black,fill=none,font=\scriptsize,midway,above] {$L_6$};
\draw[red] (6,4) -- (12,10)  node[black,fill=none,font=\scriptsize,midway,below] {$L_7$};
\draw[red] (8,7) -- (12,11)  node[black,fill=none,font=\scriptsize,midway,below] {$L_8$};
\draw[red] (9,9) -- (12,12)  node[black,fill=none,font=\scriptsize,midway,right] {$L_9$};
\draw[red] (9,10) -- (12,13)  node[black,fill=none,font=\scriptsize,midway,right] {$L_{10}$};
\draw[red] (8,10) -- (11,13)  node[black,fill=none,font=\scriptsize,midway,left] {$L_{11}$};
\draw[red] (6,9) -- (10,13)  node[black,fill=none,font=\scriptsize,midway,left] {$L_{12}$};
\end{tikzpicture}

\caption{A spiral diagram for the $E_8^{(1)}$ case.\label{fige8}}
\end{figure}
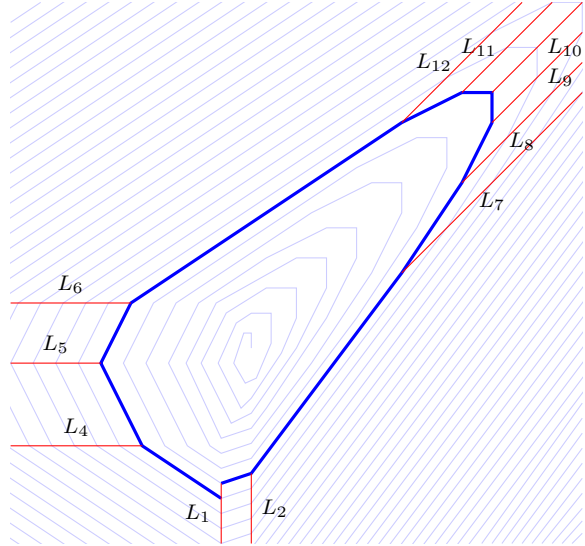

The tropical analogue requires we have a conf\/iguration of two, three and six rays, which may be parameterized as follows:
\begin{alignat*}{3}
&L_1\colon \ X = 0, \qquad && L_7\colon \ Y-X-B_3-B_4 =0,&\\
&L_2\colon \ X - B_0 = 0, \qquad &&L_8\colon \ Y-X-B_3-B_4-B_5 =0,&\\
&L_3\colon \ Y = 0, \qquad && L_9\colon \  Y-X-B_3-B_4-B_5-B_6 = 0,&\\
&L_4\colon \ Y+B_2= 0, \qquad && L_{10}\colon \ Y-X-B_3-B_4-B_5-B_6 -B_7= 0,&\\
&L_5\colon \  Y +B_1 + B_2=0, \qquad &&  L_{11}\colon \  Y-X-B_3-B_4-B_5-B_6 -B_7-B_8= 0,&\\
&L_6\colon \ Y - X - B_3= 0,\qquad &&&
\end{alignat*}
where $L_1$ and $L_2$ are of order $3$, $L_4$, $L_5$ and $L_6$ are of order $2$ and the remaining rays are order~$1$. Such a conf\/iguration is depicted in Fig.~\ref{fige8}.

The top case of the multiplicative type Painlev\'e equations of~\cite{Sakai:rational} is one that has a symmetry group of type
\begin{gather*}
W\big(E_8^{(1)}\big) = \langle s_0, \ldots, s_8 \rangle.
\end{gather*}
A presentation may be derived from the groups corresponding Dynkin diagram, which is shown below:
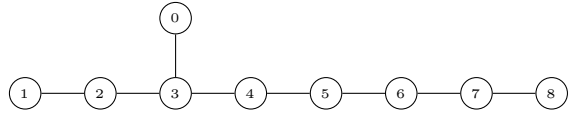
\begin{center}
\begin{tiny}
\begin{tikzpicture}
\node[draw, circle] (e81) at (0,0) {${}_{1}$};
\node[draw, circle] (e82) at (1,0) {${}_{2}$};
\node[draw, circle] (e83) at (2,0) {${}_{3}$};
\node[draw, circle] (e84) at (3,0) {${}_{4}$};
\node[draw, circle] (e85) at (4,0) {${}_{5}$};
\node[draw, circle] (e86) at (5,0) {${}_{6}$};
\node[draw, circle] (e87) at (6,0) {${}_{7}$};
\node[draw, circle] (e88) at (7,0) {${}_{8}$};
\node[draw, circle] (e80) at (2,1) {${}_{0}$};
\draw (e81) -- (e82);
\draw (e82) -- (e83);
\draw (e83) -- (e84);
\draw (e84) -- (e85);
\draw (e85) -- (e86);
\draw (e86) -- (e87);
\draw (e87) -- (e88);
\draw (e80) -- (e83);
\end{tikzpicture}
\end{tiny}
\end{center}

There are no Dynkin diagram automorphisms. We may specify the elemtents in terms of~$\sigma_{i,j}$ as~$s_0 = \sigma_{0,3}$, $s_1 = \sigma_{4,5}$, $s_2 = \sigma_{1,2}$, $s_4 = \sigma_{6,7}$, $s_5 = \sigma_{7,8}$ ,$s_6 = \sigma_{8,9}$, $s_7 = \sigma_{9,10}$ and $s_8 = \sigma_{10,11}$. The nontrivial actions are given by
\begin{gather*}
s_0\colon \ X \to X- B_0, \qquad s_2\colon \ Y \to Y+B_2, \\
s_3\colon \ X \to X + \max(B_3,B_3 + X,Y) - \max(0,X,Y),\\
s_3\colon \ Y \to Y -B_3 + \max(0,B_3+X,Y) - \max(0,X,Y).
\end{gather*}
By tracing around the f\/igure, we f\/ind that
\begin{gather*}
Q = 3B_0+2B_1+4B_2+6B_3+5B_4+4B_5 +3B_6 + 2B_7 +B_8.
\end{gather*}
In the autonomous limit, this becomes a foliation of tropical sextic curves, specif\/ied by the level sets of
\begin{gather*}
H(X,Y)=\max\{i X+j Y+c_{i,j} \, | \, 0\leq i, 0\leq j, i+j\leq 6\}-2X-3Y,
\end{gather*}
where
\begin{gather*}
 c_{0,0}= 2 \lambda_3, \qquad
c_{0,2}= 2 \lambda_2, \qquad
c_{0,4}= 2 \lambda_1,\\
 c_{1,1}= \max (2 \kappa_2+\mu_5,\kappa_1-\kappa_2+\lambda_2+\lambda_3 ),\\
c_{1,2}= \max  (\kappa_1-\kappa_2+2\lambda_2,2 \kappa_2+\mu_5+\lambda_2-\lambda_3,\mu_1+\lambda_3 ),\\
c_{1,3}= \max  (\mu_1+\lambda_2,\kappa_1-\kappa_2+\lambda_1+\lambda_2,2 \kappa_2+\mu_5+\lambda_1-\lambda_3 ),\\
c_{1,4}= \max  (\mu_1+\lambda_1,\kappa_1-\kappa_2+\lambda_2,2 \kappa_2+\mu_5-\lambda_3 ),\qquad
c_{1,5}= \mu_1,\\
c_{2,1}= \max  (\kappa_1+\kappa_2+\mu_5,2 \kappa_1-2 \kappa_2+\lambda_2+\lambda_3 ),\\
c_{2,2}= \max  (\kappa_2+\mu_4,2\lambda_2-\kappa_2,\kappa_1+\kappa_2+\mu_5+\lambda_2-\lambda_3,\kappa_1-\kappa_2+\mu_1+\lambda_3,  \\
\hphantom{c_{2,2}= \max  (}{} 2 \kappa_1-2 \kappa_2+\lambda_1+\lambda_3 ),\\
c_{2,4}= \mu_2, \qquad
c_{3,0}= 3 \kappa_1-3 \kappa_2+2\lambda_3, \qquad
c_{3,1}= \max  (2 \kappa_1+\mu_5,\kappa_1-2 \kappa_2+\lambda_2+\lambda_3 ),\\
c_{3,2}= \max  (\kappa_1+\mu_4,\kappa_2+\mu_5+\lambda_2-\lambda_3,-\kappa_2+\mu_1+\lambda_3,\kappa_1-2 \kappa_2+\lambda_1+\lambda_3 ),\\
c_{3,3}= \mu_3, \qquad
c_{4,1}= \max  (\kappa_1+\mu_5,-2 \kappa_2+\lambda_2+\lambda_3 ),\\
c_{4,2}= \mu_4,\qquad
c_{5,1}= \mu_5,\qquad
c_{6,0}= \mu_6, \qquad
c_{0,1}=c_{0,3}=c_{2,3}=-\infty,\\
 \max(0,X+\kappa_1,2X+\kappa_2)=\max(0,X)+\max(0,X+B_0),\\
 \max(0,X+\lambda_1,2X+\lambda_2,3X+\lambda_3)=\max(0,X)+\max(0,X-B_2)\\
\qquad {} +\max(0,X-B_1-B_2),\\
 \max(0,X+\mu_1,\dots,6X+\mu_6)=\max(0,X+B_3)\\
\qquad{} +\max(0,X+B_3+B_4)+ \cdots+\max(0,X+B_3+B_4+\cdots+B_8).
\end{gather*}
The translation is the composition
\begin{gather*}
\begin{split}
& T = s_3  \circ s_2 \circ s_4 \circ s_3 \circ s_1 \circ s_2 \circ s_5 \circ s_4 \circ s_3 \circ s_6\circ s_5\circ s_4\circ s_0\circ s_3\circ s_2\circ s_1\\
& \hphantom{T=}{} \circ s_7 \circ s_6\circ s_5\circ s_4\circ s_3\circ s_2\circ s_8\circ s_7\circ s_6\circ s_5\circ s_4\circ s_3\circ s_0\circ s_3\circ s_4\\
& \hphantom{T=}{} \circ s_5\circ s_6\circ s_7\circ s_8\circ s_2\circ s_3\circ s_4\circ s_5\circ s_6\circ s_7\circ s_1\circ s_2\circ s_3\circ s_0\circ s_4\\
& \hphantom{T=}{} \circ s_5\circ s_6\circ s_3\circ s_4\circ s_5\circ s_2\circ s_1\circ s_3\circ s_4\circ s_2\circ s_3\circ s_0.
\end{split}
\end{gather*}
Once again, the evolution is too complicated in its tropical form to give here. However, the geometric interpretation is that the evolution is def\/ined as
\begin{gather}\label{upE8}
T(P)+T(L_i)=P + L_j,
\end{gather}
where the addition is with respect to the group law on a tropical sextic genus one curve.

\section{Discussion of dodecagons, triskaidecagons and higher}\label{sec:dodecagons}

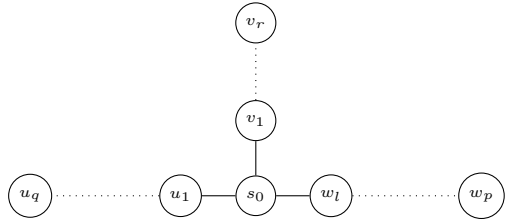
\begin{figure}[t]
\begin{tiny}
\centering
\begin{tikzpicture}
\node[draw, circle] (t000) at (0,0) {${s}_{0}$};
\node[draw, circle] (t001) at (1,0) {${w}_{l}$};
\node[draw, circle] (t002) at (3,0) {${w}_{p}$};
\node[draw, circle] (t010) at (-1,0) {${u}_{1}$};
\node[draw, circle] (t020) at (-3,0) {${u}_{q}$};
\node[draw, circle] (t100) at (0,1) {${v}_{1}$};
\node[draw, circle] (t200) at (0,2.3) {${v}_{r}$};
\draw (t000) -- (t001);
\draw (t000) -- (t010);
\draw (t000) -- (t100);
\draw[dotted] (t001) -- (t002);
\draw[dotted] (t010) -- (t020);
\draw[dotted] (t100) -- (t200);
\end{tikzpicture}

\end{tiny}
\caption{A Dynkin diagram of type $T_{p,q,r}$.\label{tree}}
\end{figure}

We wish to breif\/ly discuss some of the dif\/f\/iculties extending the above arguments to more than eleven sides. We have two constructions that we believed were related; tropical maps of the plane arising from polygons with greater than~11 sides and tropical birational representations of the Weyl group $W(T_{p,q,r})$ constructed in~\cite{Tsuda:Tropical2}, whose Dynkin diagram is given in Fig.~\ref{tree}.

\begin{figure}[t]
\centering
\begin{tikzpicture}[scale=.7]
\clip (-7,-5) rectangle (6,7);
\draw[blue!40] (-2.,-1.) -- (-2.5,-.5) -- (-2.5,2) -- (-.5,4) -- (1,4.75) -- (1.75,4.75) -- (2.25,4.25)--(2.25,3.25) -- (2,2.5) -- (1,.5) -- (0,-.5) -- (-2,-.5) -- (-2,2) -- (0,4)--(1,4.5) --(1.5,4.5)--(2,4)--(2,3) --(1,1)--(0,0) --(-1.5,0) --(-1.5,2) --(.5,4)--(1,4.25)--(1.25,4.25)--(1.75,3.75)--(1.75,3)--(1,1.5)--(0,.5)--(-1,.5)--(-1,2)--(1,4)--(1.5,3.5)--(1.5,3)--(1,2)--(0,1)--(-.5,1)--(-.5,2)--(1,3.5)--(1.25,3.25)--(1.25,3)--(1,2.5)--(0,1.5)--(0,2)--(1,3);
\draw[blue!40] (-2.,-1.) -- (0,-1.) -- (1.,0.) -- (2.,2.) -- (2.5,3.5) -- (2.5,4.5) -- (2.,5.) -- (1.,5.) -- (1.,5.) -- (-1.,4.) -- (-3.,2.) -- (-3.,-0.5) -- (-2.,-1.5) -- (-2.,-1.5) -- (0,-1.5) -- (1.,-0.5) -- (2.,1.5) -- (2.75,3.75) -- (2.75,4.75) -- (2.25,5.25) -- (1.25,5.25) -- (0.5,5.) -- (-1.5,4.) -- (-3.5,2.) -- (-3.5,-0.5) -- (-2.,-2.) -- (-2.,-2.) -- (0,-2.) -- (1.,-1.) -- (2.,1.) -- (3.,4.) -- (3.,5.) -- (2.5,5.5) -- (1.5,5.5) -- (0,5.) -- (-2.,4.) -- (-4.,2.) -- (-4.,-0.5) -- (-2.,-2.5) -- (-2.,-2.5) -- (0,-2.5) -- (1.,-1.5) -- (2.,0.5) -- (3.25,4.25) -- (3.25,5.25) -- (2.75,5.75) -- (1.75,5.75) -- (-0.5,5.) -- (-2.5,4.) -- (-4.5,2.) -- (-4.5,-0.5) -- (-2.,-3.) -- (-2.,-3.) -- (0,-3.) -- (1.,-2.) -- (2.,0) -- (3.5,4.5) -- (3.5,5.5) -- (3.,6.) -- (2.,6.) -- (-1.,5.) -- (-3.,4.) -- (-5.,2.) -- (-5.,-0.5) -- (-2.,-3.5) -- (-2.,-3.5) -- (0,-3.5) -- (1.,-2.5) -- (2.,-0.5) -- (3.75,4.75) -- (3.75,5.75) -- (3.25,6.25) -- (2.25,6.25) -- (-1.5,5.) -- (-3.5,4.) -- (-5.5,2.) -- (-5.5,-0.5) -- (-2.,-4.) -- (-2.,-4.) -- (0,-4.) -- (1.,-3.) -- (2.,-1.) -- (4.,5.) -- (4.,6.) -- (3.5,6.5) -- (2.5,6.5) -- (-2.,5.) -- (-4.,4.) -- (-6.,2.) -- (-6.,-0.5) -- (-2.,-4.5) -- (-2.,-4.5) -- (0,-4.5) -- (1.,-3.5) -- (2.,-1.5) -- (4.25,5.25) -- (4.25,6.25) -- (3.75,6.75) -- (2.75,6.75) -- (-2.5,5.) -- (-4.5,4.) -- (-6.5,2.) -- (-6.5,-0.5) -- (-2.,-5.) -- (-2.,-5.) -- (0,-5.) -- (1.,-4.) -- (2.,-2.) -- (4.5,5.5) -- (4.5,6.5) -- (4.,7.) -- (3.,7.) -- (-3.,5.) -- (-5.,4.) -- (-7.,2.) -- (-7.,-0.5) -- (-2.,-5.5) -- (-2.,-5.5) -- (0,-5.5) -- (1.,-4.5) -- (2.,-2.5) -- (4.75,5.75) -- (4.75,6.75) -- (4.25,7.25) -- (3.25,7.25) -- (-3.5,5.) -- (-5.5,4.) -- (-7.5,2.) -- (-7.5,-0.5) -- (-2.,-6.) -- (-2.,-6.) -- (0,-6.) -- (1.,-5.) -- (2.,-3.) -- (5.,6.) -- (5.,7.) -- (4.5,7.5) -- (3.5,7.5) -- (-4.,5.) -- (-6.,4.) -- (-8.,2.) -- (-8.,-0.5) -- (-2.,-6.5) -- (-2.,-6.5) -- (0,-6.5) -- (1.,-5.5) -- (2.,-3.5) -- (5.25,6.25) -- (5.25,7.25) -- (4.75,7.75) -- (3.75,7.75) -- (-4.5,5.) -- (-6.5,4.) -- (-8.5,2.) -- (-8.5,-0.5) -- (-2.,-7.) -- (-2.,-7.) -- (0,-7.) -- (1.,-6.) -- (2.,-4.) -- (5.5,6.5) -- (5.5,7.5) -- (5.,8.) -- (4.,8.) -- (-5.,5.) -- (-7.,4.) -- (-9.,2.) -- (-9.,-0.5) -- (-2.,-7.5) -- (-2.,-7.5) -- (0,-7.5) -- (1.,-6.5) -- (2.,-4.5) -- (5.75,6.75) -- (5.75,7.75) -- (5.25,8.25) -- (4.25,8.25) -- (-5.5,5.) -- (-7.5,4.) -- (-9.5,2.) -- (-9.5,-0.5) -- (-2.,-8.) -- (-2.,-8.) -- (0,-8.) -- (1.,-7.) -- (2.,-5.) -- (6.,7.) -- (6.,8.) -- (5.5,8.5) -- (4.5,8.5) -- (-6.,5.) -- (-8.,4.) -- (-10.,2.) -- (-10.,-0.5) -- (-2.,-8.5) -- (-2.,-8.5) -- (0,-8.5) -- (1.,-7.5) -- (2.,-5.5) -- (6.25,7.25) -- (6.25,8.25) -- (5.75,8.75) -- (4.75,8.75) -- (-6.5,5.) -- (-8.5,4.) -- (-10.5,2.) -- (-10.5,-0.5) -- (-2.,-9.) -- (-2.,-9.) -- (0,-9.) -- (1.,-8.) -- (2.,-6.) -- (6.5,7.5) -- (6.5,8.5) -- (6.,9.) -- (5.,9.) -- (-7.,5.) -- (-9.,4.) -- (-11.,2.) -- (-11.,-0.5) -- (-2.,-9.5) -- (-2.,-9.5) -- (0,-9.5) -- (1.,-8.5) -- (2.,-6.5) -- (6.75,7.75) -- (6.75,8.75) -- (6.25,9.25) -- (5.25,9.25) -- (-7.5,5.) -- (-9.5,4.) -- (-11.5,2.) -- (-11.5,-0.5) -- (-2.,-10.) -- (-2.,-10.) -- (0,-10.) -- (1.,-9.) -- (2.,-7.) -- (7.,8.) -- (7.,9.) -- (6.5,9.5) -- (5.5,9.5) -- (-8.,5.) -- (-10.,4.) -- (-12.,2.) -- (-12.,-0.5) -- (-2.,-10.5) -- (-2.,-10.5) -- (0,-10.5) -- (1.,-9.5) -- (2.,-7.5) -- (7.25,8.25) -- (7.25,9.25) -- (6.75,9.75) -- (5.75,9.75) -- (-8.5,5.) -- (-10.5,4.) -- (-12.5,2.) -- (-12.5,-0.5) -- (-2.,-11.);
\draw[blue,very thick] (-2.,-2.) -- (0,-2.) -- (1.,-1.) -- (2.,1.) -- (3.,4.) -- (3.,5.) -- (2.5,5.5) -- (1.5,5.5) -- (0,5.) -- (-2.,4.) -- (-4.,2.) -- (-4.,-0.5) -- (-2.,-2.5) -- (-2.,-2.5);
\foreach \x in {3,3.2,...,10}
{
	\draw[blue!40] (\x,-5) -- (6,13-3*\x);
}
\foreach \x in {0,.2,...,2}
{
	\draw[blue!40] (-7,5.7+\x) -- (-3-3*\x,7);
}
\draw [red] (-2,-2)--(-2,-5)  node[black,fill=none,font=\scriptsize,midway,left] {$L_1$};
\draw [red] (0,-2)--(0,-5)  node[black,fill=none,font=\scriptsize,midway,left] {$L_2$};
\draw [red] (1,-1)--(1,-5) node[black,fill=none,font=\scriptsize,midway,right] {$L_3$};
\draw [red] (2,1)--(2,-5) node[black,fill=none,font=\scriptsize,midway,right] {$L_4$};
\draw [red] (3,4)--(6,7) node[black,fill=none,font=\scriptsize,midway,below] {$L_9$};
\draw [red] (3,5)--(6,8) node[black,fill=none,font=\scriptsize,midway,below] {$L_{10}$};
\draw [red] (2.5,5.5)--(4,7) node[black,fill=none,font=\scriptsize,midway,left] {$L_{11}$};
\draw [red] (1.5,5.5)--(3,7) node[black,fill=none,font=\scriptsize,midway,left] {$L_{12}$};
\draw [red] (-4,2)--(-7,2) node[black,fill=none,font=\scriptsize,midway,below] {$L_{6}$};
\draw [red] (-4,-.5)--(-7,-.5) node[black,fill=none,font=\scriptsize,midway,below] {$L_{5}$};
\draw [red] (-2,4)--(-7,4)  node[black,fill=none,font=\scriptsize,midway,above] {$L_{7}$};
\draw [red] (0,5)--(-7,5) node[black,fill=none,font=\scriptsize,midway,above] {$L_{8}$};
\end{tikzpicture}

\caption{A model for the dodecagon.\label{trisk}}
\end{figure}
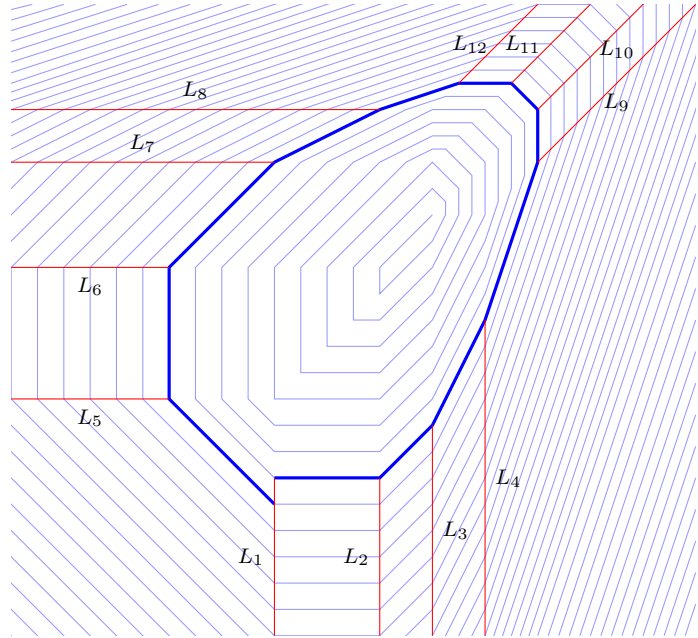

Each one of the polygons we have considered so far arise from tropical genus one curves. If we go to a higher number of sides, a simple combinatorial argument based on \eqref{degreegenus} shows us that the higher sided polygons must come from higher genus cases. The simplest example is the autonomous system def\/ined on a pencil of tropic quartics such that there are four distinct rays of order one in each direction. Suppose we parameterize these by
\begin{alignat*}{3}
&L_1\colon \  X = 0, \qquad && L_7\colon \  Y + B_7+B_8=0,&\\
&L_2\colon \  X - B_0 =0, \qquad && L_8\colon \  Y + B_7+B_8+B_9=0,&\\
&L_3\colon \  X - B_0-B_1=0, \qquad && L_9\colon \  Y-X + B_3 =0,&\\
&L_4\colon \  X - B_0-B_1 -B_2=0,\qquad && L_{10}\colon \  Y-X + B_3+B_4=0,&\\
&L_5\colon \  Y = 0, \qquad && L_{11}\colon \  Y-X + B_3+B_4+B_5=0,&\\
&L_6\colon \  Y + B_7=0, \qquad && L_{12}\colon \  Y-X + B_3+B_4+B_5+B_6=0,&
\end{alignat*}
as labelled in Fig.~\ref{trisk}.

The diagram in Fig.~\ref{trisk} has been obtained by following a path in which the rays have been f\/ixed, and follow the level curves of a biquartic invariant where three of the parameters, the coef\/f\/icients of $X+Y$, $2X+Y$ and $X +2Y$, have been set to $-\infty$. Tracing around the diagram, we f\/ind that $Q$ is given by
\begin{gather*}
Q = 3B_0+2B_1+B_2+4B_3+3B_4+2B_5+B_6+3B_7+2B_8+B_9.
\end{gather*}
Def\/ine a translation, $T$, by letting $T(L_i)$ move to a point that def\/ines a pencil of closed polygons. For any point $P$, we have a unique closed curve, $C$, intersecting with $P$. The evolution def\/ined by~\eqref{upE6},~\eqref{upE7} and \eqref{upE8} were in terms of a group law on genus one curves, however, these resulting closed curves in this more general setting are no longer of genus one, hence, describing the group structure on such curves is not so straightforward.

To obtain a higher number of sides (with the constraint that all rays of the same form are of the same order), we may realize 13-sided polygons as specializations of tropical curves of degree twelve and 14-sided polygons as specializations of tropical curves of degree six. A rudimentary search reveals $n$-agons for all $n$ up to 30 sides.

The problem is that in the autonomous limit, the naive extension to $W(T_{4,4,4})$ using canonical permutations and an analogue $s_3$ in the $E_8^{(1)}$ case does not preserve all the required quartic plane curves in the pencil constructed. The action of this generator generally gives a curve of degree f\/ive, hence, it is not $s_3$ invariant. It seems likely that this fails to preserve all the required degenerate curves when the closed piecewise linear curve become small. It seems that the dynamics we describe may have an interpretation in terms of the addition on some tropical hyper-elliptic curve, for example \cite{Inoue} where certain tropical dynamics was studied by using the tropical addition formulae on the spectral curve of hyper-elliptic type.

\section{Conclusion}

What has been presented is a way of naturally obtaining a group of transformations that preserve the structure of a spiral diagram. It is possible to extend this to cases that do not arise as ultradiscrete Painlev\'e equations, however the invariants seem more elusive. The autonomous limits do not necessarily result in foliations of genus one curves. Where the role of the addition law on cubic plane curves in the Painlev\'e equations and QRT maps is central~\cite{10E9sol}, perhaps similar integrable systems could be based on the addition laws for hyperelliptic curves, which are in general, much more complicated~\cite{CFADLNV:handbook, Inoue}.

Another possible direction is to explore the tropical Cremona transformations more thoroughly. Interesting tropical versions of del Pezzo surfaces have emerged with~$W(E_6)$ and~$W(E_7)$ symmetry during the write-up of this paper~\cite{TropDelPezzo}. A homological approach that follows \cite{duVal, Looijenga:Rational, Nagata:Rational1, Nagata:Rational2} more closely would also be of interest.

\subsection*{Acknowledgements}

Christopher M.~Ormerod would like to acknowledge Eric Rains for his helpful discussions.  Y.~Yamada is supported by JSPS KAKENHI Grant Number 26287018.

\pdfbookmark[1]{References}{ref}
\LastPageEnding


\begin{thebibliography}{99}
\footnotesize \itemsep=-1pt

\bibitem{WeakFactor}
Abramovich D., Karu K., Matsuki K., W{\l}odarczyk J., Torif\/ication and
  factorization of birational maps, \href{http://dx.doi.org/10.1090/S0894-0347-02-00396-X}{\textit{J.~Amer. Math. Soc.}} \textbf{15}
  (2002), 531--572, \href{http://arxiv.org/abs/math.AG/9904135}{math.AG/9904135}.

\bibitem{BV:Entropy}
Bellon M.P., Viallet C.M., Algebraic entropy, \href{http://dx.doi.org/10.1007/s002200050652}{\textit{Comm. Math. Phys.}}
  \textbf{204} (1999), 425--437, \href{http://arxiv.org/abs/chao-dyn/9805006}{chao-dyn/9805006}.

\bibitem{BG:geometryvaluations}
Bieri R., Groves J.R.J., The geometry of the set of characters induced by
  valuations, \href{http://dx.doi.org/10.1515/crll.1984.347.168}{\textit{J.~Reine Angew. Math.}} \textbf{347} (1984), 168--195.

\bibitem{CFADLNV:handbook}
Cohen H., Frey G., Avanzi R., Doche C., Lange T., Nguyen K., Vercauteren F.
  (Editors), Handbook of elliptic and hyperelliptic curve cryptography,
  \textit{Discrete Mathematics and Its Applications}, Chapman \& Hall/CRC, Boca Raton,
  FL, 2006.

\bibitem{Vigeland:grouplaw}
Dehli~Vigeland M., The group law on a tropical elliptic curve, \textit{Math.
  Scand.} \textbf{104} (2009), 188--204, \href{http://arxiv.org/abs/math.AG/0411485}{math.AG/0411485}.

\bibitem{Dolgachev}
Dolgachev I.V., Iskovskikh V.A., Finite subgroups of the plane {C}remona group,
  in Algebra, Arithmetic, and Geometry: in Honor of {Y}u.{I}.~{M}anin,
  {V}ol.~{I}, \href{http://dx.doi.org/10.1007/978-0-8176-4745-2_11}{\textit{Progr. Math.}}, Vol.~269, Birkh\"auser Boston, Inc.,
  Boston, MA, 2009, 443--548, \href{http://arxiv.org/abs/math.AG/0610595}{math.AG/0610595}.

\bibitem{duVal}
du~Val P., On the {K}antor group of a set of points in a plane, \href{http://dx.doi.org/10.1112/plms/s2-42.1.18}{\textit{Proc.
  London Math. Soc.}} \textbf{S2-42} (1937), 18--51.

\bibitem{Duistermaat:QRT}
Duistermaat J.J., Discrete integrable systems. QRT maps and elliptic surfaces,
  \href{http://dx.doi.org/10.1007/978-0-387-72923-7}{\textit{Springer Monographs in Mathematics}}, Springer, New York, 2010.

\bibitem{Gathman:TropicalAlgebraicGeometry}
Gathmann A., Tropical algebraic geometry, \textit{Jahresber. Deutsch.
  Math.-Verein.} \textbf{108} (2006), 3--32, \href{http://arxiv.org/abs/math.AG/0601322}{math.AG/0601322}.

\bibitem{Hudson}
Hudson H.P., Cremona transformations in plane and space, Cambridge University
  Press, Cambridge, 1927.

\bibitem{Inoue}
Inoue R., Takenawa T., Tropical spectral curves and integrable cellular
  automata, \href{http://dx.doi.org/10.1093/imrn/rnn019}{\textit{Int. Math. Res. Not.}} \textbf{2008} (2008), Art ID.~rnn019,
  27~pages, \href{http://arxiv.org/abs/0704.2471}{arXiv:0704.2471}.

\bibitem{IGRS:s-ultradiscretization}
Isojima S., Grammaticos B., Ramani A., Satsuma J., Ultradiscretization without
  positivity, \href{http://dx.doi.org/10.1088/0305-4470/39/14/011}{\textit{J.~Phys.~A: Math. Gen.}} \textbf{39} (2006), 3663--3672.

\bibitem{JS:qP6}
Jimbo M., Sakai H., A {$q$}-analog of the sixth {P}ainlev\'e equation,
  \href{http://dx.doi.org/10.1007/BF00398316}{\textit{Lett. Math. Phys.}} \textbf{38} (1996), 145--154,
  \href{http://arxiv.org/abs/chao-dyn/9507010}{arXiv:chao-dyn/9507010}.

\bibitem{JL:TropSC}
Joshi N., Lafortune S., Integrable ultra-discrete equations and singularity
  analysis, \href{http://dx.doi.org/10.1088/0951-7715/19/6/005}{\textit{Nonlinearity}} \textbf{19} (2006), 1295--1312.

\bibitem{JNO:uP3}
Joshi N., Nijhof\/f F.W., Ormerod C., Lax pairs for ultra-discrete {P}ainlev\'e
  cellular automata, \href{http://dx.doi.org/10.1088/0305-4470/37/44/L03}{\textit{J.~Phys.~A: Math. Gen.}} \textbf{37} (2004),
  L559--L565.

\bibitem{10E9sol}
Kajiwara K., Masuda T., Noumi M., Ohta Y., Yamada Y., {${}_{10}E_9$} solution
  to the elliptic {P}ainlev\'e equation, \href{http://dx.doi.org/10.1088/0305-4470/36/17/102}{\textit{J.~Phys.~A: Math. Gen.}}
  \textbf{36} (2003), L263--L272, \href{http://arxiv.org/abs/nlin.SI/0303032}{nlin.SI/0303032}.

\bibitem{KMNOY:Cremona}
Kajiwara K., Masuda T., Noumi M., Ohta Y., Yamada Y., Point conf\/igurations,
  {C}remona transformations and the elliptic dif\/ference {P}ainlev\'e equation,
  in Th\'eories asymptotiques et \'equations de {P}ainlev\'e, \textit{S\'emin.
  Congr.}, Vol.~14, Soc. Math. France, Paris, 2006, 169--198.

\bibitem{KMNY:Ereps}
Kajiwara K., Masuda T., Noumi M., Yamada Y., Tropical af\/f\/ine {W}eyl group
  representation of type ${E}^{(1)}_n$, 2004.

\bibitem{KNY:qP4}
Kajiwara K., Noumi M., Yamada Y., A study on the fourth {$q$}-{P}ainlev\'e
  equation, \href{http://dx.doi.org/10.1088/0305-4470/34/41/312}{\textit{J.~Phys.~A: Math. Gen.}} \textbf{34} (2001), 8563--8581,
  \href{http://arxiv.org/abs/nlin.SI/0012063}{nlin.SI/0012063}.

\bibitem{KNY:discretePAmAn}
Kajiwara K., Noumi M., Yamada Y., Discrete dynamical systems with
  {$W(A_{m-1}^{(1)}\times A_{n-1}^{(1)})$} symmetry, \href{http://dx.doi.org/10.1023/A:1016298925276}{\textit{Lett. Math. Phys.}}
  \textbf{60} (2002), 211--219, \href{http://arxiv.org/abs/nlin.SI/0106029}{nlin.SI/0106029}.

\bibitem{KMT:ConservedudKdV}
Kanki M., Mada J., Tokihiro T., Conserved quantities and generalized solutions
  of the ultradiscrete {K}d{V} equation, \href{http://dx.doi.org/10.1088/1751-8113/44/14/145202}{\textit{J.~Phys.~A: Math. Theor.}}
  \textbf{44} (2011), 145202, 13~pages, \href{http://arxiv.org/abs/1012.4061}{arXiv:1012.4061}.

\bibitem{KL:UDnegativity}
Kasman A., Lafortune S., When is negativity not a problem for the ultradiscrete
  limit?, \href{http://dx.doi.org/10.1063/1.2360394}{\textit{J.~Math. Phys.}} \textbf{47} (2006), 103510, 16~pages,
  \href{http://arxiv.org/abs/nlin.SI/0609034}{nlin.SI/0609034}.

\bibitem{Kondo:UDSGSymmMPAlg}
Kondo K., Ultradiscrete sine-{G}ordon equation over symmetrized max-plus
  algebra, and noncommutative discrete and ultradiscrete sine-{G}ordon
  equations, \href{http://dx.doi.org/10.3842/SIGMA.2013.068}{\textit{SIGMA}} \textbf{9} (2013), 068, 39~pages,
  \href{http://arxiv.org/abs/1311.2675}{arXiv:1311.2675}.

\bibitem{Looijenga:Rational}
Looijenga E., Rational surfaces with an anticanonical cycle, \href{http://dx.doi.org/10.2307/1971295}{\textit{Ann. of
  Math.}} \textbf{114} (1981), 267--322.

\bibitem{Markwig:FieldforTG}
Markwig T., A f\/ield of generalised {P}uiseux series for tropical geometry,
  \textit{Rend. Semin. Mat. Univ. Politec. Torino} \textbf{68} (2010), 79--92,
  \href{http://arxiv.org/abs/0709.3784}{arXiv:0709.3784}.

\bibitem{Murata:ExactsolsuP2}
Murata M., Exact solutions with two parameters for an ultradiscrete
  {P}ainlev\'e equation of type {$A^{(1)}_6$}, \href{http://dx.doi.org/10.3842/SIGMA.2011.059}{\textit{SIGMA}} \textbf{7} (2011), 059, 15~pages,
  \href{http://arxiv.org/abs/1106.3384}{arXiv:1106.3384}.

\bibitem{Nagata:Rational1}
Nagata M., On rational surfaces. {I}.~{I}rreducible curves of arithmetic
  genus~{$0$} or~{$1$}, \textit{Mem. Coll. Sci. Univ. Kyoto Ser.~A Math.}
  \textbf{32} (1960), 351--370.

\bibitem{Nagata:Rational2}
Nagata M., On rational surfaces.~{II}, \textit{Mem. Coll. Sci. Univ. Kyoto
  Ser.~A Math.} \textbf{33} (1960/1961), 271--293.

\bibitem{Nobe:QRT}
Nobe A., Ultradiscrete {QRT} maps and tropical elliptic curves,
  \href{http://dx.doi.org/10.1088/1751-8113/41/12/125205}{\textit{J.~Phys.~A: Math. Theor.}} \textbf{41} (2008), 125205, 12~pages.

\bibitem{Noether}
N\"other M., Ueber die auf Ebenen eindeutig abbildbaren algebraischen
  Fl\"achen, \textit{G\"ott. Nachr.}  (1870), 1--6.

\bibitem{NY:AffinedPs}
Noumi M., Yamada Y., Af\/f\/ine {W}eyl groups, discrete dynamical systems and
  {P}ainlev\'e equations, \href{http://dx.doi.org/10.1007/s002200050502}{\textit{Comm. Math. Phys.}} \textbf{199} (1998),
  281--295, \href{http://arxiv.org/abs/math.AG/9804132}{math.AG/9804132}.

\bibitem{ON:inversible}
Ochiai T., Nacher J.C., Inversible max-plus algebras and integrable systems,
  \href{http://dx.doi.org/10.1063/1.1925247}{\textit{J.~Math. Phys.}} \textbf{46} (2005), 063507, 17~pages,
  \href{http://arxiv.org/abs/nlin.SI/0405067}{nlin.SI/0405067}.

\bibitem{Okamoto:SurfaceInitialConds}
Okamoto K., Sur les feuilletages associ\'es aux \'equations du second ordre \`a
  points critiques f\/ixes de {P}.~{P}ainlev\'e. Espaces des conditions
  initiales, \href{http://doi.org/10.4099/math1924.5.1}{\textit{Japan.~J. Math.~(N.S.)}} \textbf{5} (1979), 1--79.

\bibitem{Ormerod:uhypergeometric}
Ormerod C.M., Hypergeometric solutions to an ultradiscrete {P}ainlev\'e
  equation, \href{http://dx.doi.org/10.1142/S140292511000060X}{\textit{J.~Nonlinear Math. Phys.}} \textbf{17} (2010), 87--102,
  \href{http://arxiv.org/abs/nlin.SI/0610048}{nlin.SI/0610048}.

\bibitem{Ormerod:qP6reduction}
Ormerod C.M., Reductions of lattice m{K}d{V} to {$q$}-{${\rm P}_{\rm VI}$},
  \href{http://dx.doi.org/10.1016/j.physleta.2012.09.008}{\textit{Phys. Lett.~A}} \textbf{376} (2012), 2855--2859, \href{http://arxiv.org/abs/1112.2419}{arXiv:1112.2419}.

\bibitem{Ormerod:TropicalSC}
Ormerod C.M., Tropical geometric interpretation of ultradiscrete singularity
  conf\/inement, \href{http://dx.doi.org/10.1088/1751-8113/46/30/305204}{\textit{J.~Phys.~A: Math. Theor.}} \textbf{46} (2013), 305204,
  15~pages, \href{http://arxiv.org/abs/0802.1959}{arXiv:0802.1959}.

\bibitem{PNGR:Isomonodromic}
Papageorgiou V.G., Nijhof\/f F.W., Grammaticos B., Ramani A., Isomonodromic
  deformation problems for discrete analogues of {P}ainlev\'e equations,
  \href{http://dx.doi.org/10.1016/0375-9601(92)90905-2}{\textit{Phys. Lett.~A}} \textbf{164} (1992), 57--64.

\bibitem{Pin:tropicalsemirings}
Pin J.E., Tropical semirings, in Idempotency ({B}ristol, 1994), \href{http://dx.doi.org/10.1017/CBO9780511662508.004}{\textit{Publ.
  Newton Inst.}}, Vol.~11, Cambridge University Press, Cambridge, 1998, 50--69.

\bibitem{QCS:UDLaxpair}
Quispel G.R.W., Capel H.W., Scully J., Piecewise-linear soliton equations and
  piecewise-linear integrable maps, \href{http://dx.doi.org/10.1088/0305-4470/34/11/337}{\textit{J.~Phys.~A: Math. Gen.}} \textbf{34}
  (2001), 2491--2503.

\bibitem{QRT:QRTmap1}
Quispel G.R.W., Roberts J.A.G., Thompson C.J., Integrable mappings and soliton
  equations, \href{http://dx.doi.org/10.1016/0375-9601(88)90803-1}{\textit{Phys. Lett.~A}} \textbf{126} (1988), 419--421.

\bibitem{QRT:QRTmap2}
Quispel G.R.W., Roberts J.A.G., Thompson C.J., Integrable mappings and soliton
  equations.~{II}, \href{http://dx.doi.org/10.1016/0167-2789(89)90233-9}{\textit{Phys.~D}} \textbf{34} (1989), 183--192.

\bibitem{RGH:discretePs}
Ramani A., Grammaticos B., Hietarinta J., Discrete versions of the {P}ainlev\'e
  equations, \href{http://dx.doi.org/10.1103/PhysRevLett.67.1829}{\textit{Phys. Rev. Lett.}} \textbf{67} (1991), 1829--1832.

\bibitem{RTGO:ultimatediscretePs}
Ramani A., Takahashi D., Grammaticos B., Ohta Y., The ultimate discretisation
  of the {P}ainlev\'e equations, \href{http://dx.doi.org/10.1016/S0167-2789(97)00192-9}{\textit{Phys.~D}} \textbf{114} (1998),
  185--196.

\bibitem{TropDelPezzo}
Ren Q., Shaw K., Sturmfels B., Tropicalization of del Pezzo surfaces,
  \href{http://arxiv.org/abs/1402.5651}{arXiv:1402.5651}.

\bibitem{RST:TropicalGeometry}
Richter-Gebert J., Sturmfels B., Theobald T., First steps in tropical geometry,
  in Idempotent mathematics and mathematical physics, \href{http://dx.doi.org/10.1090/conm/377/06998}{\textit{Contemp. Math.}},
  Vol.~377, Amer. Math. Soc., Providence, RI, 2005, 289--317,
  \href{http://arxiv.org/abs/math.AG/0306366}{math.AG/0306366}.

\bibitem{OR2}
Rojas O., From discrete integrable systems to cellular automata, Ph.D.~Thesis,
  La Trobe University, Australia, 2009.

\bibitem{Sakai:rational}
Sakai H., Rational surfaces associated with af\/f\/ine root systems and geometry of
  the {P}ainlev\'e equations, \href{http://dx.doi.org/10.1007/s002200100446}{\textit{Comm. Math. Phys.}} \textbf{220} (2001),
  165--229.

\bibitem{Scully}
Scully J., An exploration of integrable two-dimensional maps, {H}onours Thesis
  Mathematics Department, La Trobe University, Australia, 1999.

\bibitem{TM:BBSudmKdV}
Takahashi D., Matsukidaira J., Box and ball system with a carrier and
  ultradiscrete modif\/ied {K}d{V} equation, \href{http://dx.doi.org/10.1088/0305-4470/30/21/005}{\textit{J.~Phys.~A: Math. Gen.}}
  \textbf{30} (1997), L733--L739.


\bibitem{TTGOR:udPSolutions}
Takahashi D., Tokihiro T., Grammaticos B., Ohta Y., Ramani A., Constructing
  solutions to the ultradiscrete {P}ainlev\'e equations, \href{http://dx.doi.org/10.1088/0305-4470/30/22/029}{\textit{J.~Phys.~A:
  Math. Gen.}} \textbf{30} (1997), 7953--7966.

\bibitem{TTM:BBS}
Tokihiro T., Takahashi D., Matsukidaira J., Box and ball system as a
  realization of ultradiscrete nonautonomous {KP} equation, \href{http://dx.doi.org/10.1088/0305-4470/33/3/313}{\textit{J.~Phys.~A:
  Math. Gen.}} \textbf{33} (2000), 607--619.

\bibitem{TTMJ:ultradiscretization}
Tokihiro T., Takahashi D., Matsukidaira J., Satsuma J., From soliton equations
  to integrable cellular automata through a~limiting procedure, \href{http://dx.doi.org/10.1103/PhysRevLett.76.3247}{\textit{Phys.
  Rev. Lett.}} \textbf{76} (1996), 3247--3250.

\bibitem{Tsuda:QRT}
Tsuda T., Integrable mappings via rational elliptic surfaces,
  \href{http://dx.doi.org/10.1088/0305-4470/37/7/014}{\textit{J.~Phys.~A: Math. Gen.}} \textbf{37} (2004), 2721--2730.

\bibitem{Tsuda:Tropical2}
Tsuda T., Tropical {W}eyl group action via point conf\/igurations and
  {$\tau$}-functions of the {$q$}-{P}ainlev\'e equations, \href{http://dx.doi.org/10.1007/s11005-006-0052-z}{\textit{Lett. Math.
  Phys.}} \textbf{77} (2006), 21--30.

\end{thebibliography}
\end{document}